\documentclass[fleqn,usenatbib]{mnras}

\usepackage[T1]{fontenc}
\DeclareRobustCommand{\VAN}[3]{#2}
\let\VANthebibliography\thebibliography
\def\thebibliography{\DeclareRobustCommand{\VAN}[3]{##3}\VANthebibliography}

\usepackage{anyfontsize}        
\usepackage{graphicx}   	    
\usepackage{multirow}           
\usepackage{multicol}           
\usepackage{longtable}          
\usepackage{lscape}             
\usepackage{amsmath}            
\usepackage{amssymb}            
\usepackage{hyperref}           
\usepackage[hypcap=false]{caption}
\usepackage{txfonts}
\usepackage{wasysym}            
\usepackage{upgreek}            
\usepackage[svgnames]{xcolor}   
\usepackage{bm}                 

\newcommand{\gaia}{\textit{Gaia}}
\newcommand{\um}{$\mu$m}
\newcommand{\banyan}{\textsc{BANYAN}~$\Sigma$}
\newcommand{\kms}{km\,s$^{-1}$}

\newcommand{\logg}{\textit{log}(g)}
\newcommand{\teff}{T$_{\text{eff}}$}

\newcommand{\splat}{\texttt{SPLAT}}

\title[GUCDS VI. 51 \gaia{} UCDs]{The \gaia{} Ultracool Dwarf Sample -- VI. Spectral Types and Properties of 51 Ultracool Dwarfs}

\author[G. Cheng]{Gemma Cheng,$^{1}$\thanks{E-mail: g.j.cheng@herts.ac.uk}
H.R.A. Jones,$^{1}$
R.L. Smart,$^{1,2}$
Federico Marocco,$^{3}$
W.J. Cooper,$^{1,2}$
Adam Burgasser,$^{4}$
\newauthor
Juan Carlos Beamin,$^{5}$
D.J. Pinfield,$^{1}$
Jonathan Gagn\'{e},$^{6,7}$
and Leslie Moranta$^{6,7,8}$
\\
$^{1}$School of Physics, Astronomy and Mathematics, University of Hertfordshire, College Lane, Hatfield AL10 9AB, UK \\
$^{2}$Istituto Nazionale di Astrofisica, Osservatorio Astrofisico di Torino, Strada Osservatorio 20, I-10025 Pino Torinese, Italy \\
$^{3}$IPAC, Mail Code 100-22, Caltech, 1200 E. California Blvd., Pasadena, CA 91125, USA \\
$^{4}$Department of Astronomy \& Astrophysics, UC San Diego, 9500 Gilman Drive, La Jolla, CA 92093 \\
$^{5}$Fundaci\'on Chilena de Astronom\'ia, El Vergel 2252, Santiago, Chile \\
$^{6}$Plan\'{e}tarium Rio Tinto Alcan, Espace pour la Vie, 4801 av. Pierre-de Coubertin, Montr\'{e}al, Qu\'{e}bec, Canada \\
$^{7}$Institute for Research on Exoplanets, Universit\'{e} de Montr\'{e}al, D\'{e}partement de Physique, C.P. 6128 Succ. Centre-ville, Montr\'{e}al, QC H3C 3J7, Canada \\
$^{8}$Department of Astrophysics, American Museum of Natural History, Central Park West at 79th St, New York, NY 10024, USA \\
}

\date{Accepted XXX. Received YYY; in original form ZZZ}

\pubyear{\the\year{}}

\begin{document}
\label{firstpage}
\pagerange{\pageref{firstpage}--\pageref{lastpage}}
\maketitle

\begin{abstract}
Near-infrared spectra from the IRTF/SpeX and Blanco/ARCoIRIS telescope/instrument combinations are used for spectroscopic classification, to measure radial velocities and for the inference of astrophysical properties of 51 \gaia{}-selected nearby ultracool dwarfs. In this sample, 44 are newly classified in the near infrared. All but one of the UCDs are within 100\,pc, and 37 lie within 50\,pc. We find a total of 26 M-types, 24 L-types and one T-type in our sample. The positions of the majority of the UCDs on colour-magnitude diagrams and with evolutionary cooling track plots indicate that they are largely old and stellar in nature. There are a few UCDs of particular interest which lie away from expected trends, highlighting potential young, binary and thick disc/subdwarf UCDs. From spectral and kinematic analyses, we identify UCDs of particular interest for further investigation, including seven potentially young UCDs, three thick disc UCDs, one subdwarf, six wide binaries, and six unresolved binaries.

\end{abstract}


\section{Introduction}
\label{section:intro}
The \gaia{} third data release (hereafter DR3; \cite{vallenari2022}) was released on 13 June 2022, and contains five-parameter astrometric solutions for  over 1.46 billion (1.46$\times10^9$) sources. Ultracool dwarfs (hereafter UCDs) are defined as having spectral types of M7 or later (\citealt{kirkpatrick1997}). UCDs therefore straddle the boundary between stars and sub-stellar objects -- by studying such objects, we can learn more about the characteristics of, and physical processes in, UCDs.

The \gaia{} Ultracool Dwarf Sample (hereafter GUCDS) was introduced by \cite{smart2017}. The current GUCDS contains more than 8800 UCDs which have spectroscopic spectral type classifications and their associated photometric data from a number of infrared and optical surveys. Previous papers in this series have explored the known UCDs in the \gaia{} data releases (\citealt{smart2017}); improved upon main sequence modelling for UCDs (\citealt{smart2019}); identified new benchmark binary systems containing \gaia{}-detected UCDs (\citealt{marocco2020}); used GTC/OSIRIS optical spectra to determine astrophysical properties of 53 UCDs (\citealt{cooper2024}), and collated a catalogue of 278 multiple systems in the GUCDS with at least one spectroscopically confirmed UCD (\citealt{baig2024}).

Near-infrared (hereafter NIR) spectra are a valuable tool to determine the characteristics of UCDs, since they typically provide much better signal-to-noise ratio than optical observations due to the low temperatures and luminosities of UCDs (\citealt{patience2012}). In particular, NIR spectroscopy allows for the spectral type classification of UCDs, since many of the defining characteristics of L and T-dwarfs are features in their NIR spectra (e.g. methane in the \textit{J}-band of T-dwarfs; see \citealt{mclean2001, burgasser2002, geballe2002, kirkpatrick2005} among others).

This work targets UCDs in the GUCDS which have only photometrically determined spectral types, obtaining spectroscopic classifications and further investigating UCDs which have unusual characteristics. Although \gaia{}~DR3 contains BP and RP low-resolution spectra, the faint and red nature of UCDs means that BP spectra are only of use for the brightest and hottest UCDs. \cite{sarro2023} outline the difficulties in using RP spectra for UCD characterisation -- primarily these difficulties are the fairly bright \gaia{} \textit{G}-band limit of 15\,mag and the low resolution of the instrument results in spectral features which are blended and merged.

This paper is presented as follows: Section~\ref{section:data} outlines the candidate selection and spectral observations used for this work. Section~\ref{section:spt} presents the spectral type classification methods used for the UCDs in this work. Properties of the UCDs are derived from their spectra in Section~\ref{section:spec}, and further properties of the sample are derived in Sections~\ref{section:cmd}--\ref{section:rv}. UCDs of potential interest (such as young UCDs and subdwarfs) are identified in Section~\ref{section:ydh}. Finally, Section~\ref{section:indiv} investigates these UCDs in further detail, and the conclusions of this work are presented in Section~\ref{section:conc}.

\begin{table*}
    \caption{Entries for this table have been taken from the GUCDS input list and the \gaia{} archive. UCDs marked with a dagger $\dagger$ are those observed with SpeX Prism, those marked with a double dagger $\ddagger$ are observed with SpeX SXD, and those marked with an asterisk * are observed with ARCoIRIS. This data is also contained within the collated data table in Appendix~\ref{appendix:fulldata}.}
    \label{table:ucds}
	\centering
	\small
	\begin{tabular}{|c|c|c|c|c|c|c|c|c}
		\hline
            Object & RA & Dec & \gaia{}~DR3 & 2MASS & \gaia~DR3 & Parallax & Absolute \\
            Short Name & (degrees) & (degrees) & Source ID & \textit{J} (mag) & \textit{G} (mag) & (mas) & M$_G$ (mag) \\
		\hline
            J0508$+$3319 $\ddagger$ & 77.2282625 & 33.3214494 & 181724125038647040 & 14.217 $\pm$ 0.032 & 19.141 $\pm$ 0.003 & 53.026 $\pm$ 0.461 & 17.763 $\pm$ 0.019 \\
            J0515$+$0613 $\ddagger$ & 78.7951672 & 6.2331569 & 3240769806581623296 & 15.649 $\pm$ 0.064 & 20.312 $\pm$ 0.005 & 15.631 $\pm$ 0.806 & 16.282 $\pm$ 0.112 \\
            J0526$-$5026 * & 81.7491063 & -50.4383390 & 4796425760263724800 & 15.412 $\pm$ 0.066 & 20.685 $\pm$ 0.011 & 37.920 $\pm$ 0.859 & 18.579 $\pm$ 0.050 \\
            J0542$+$0041 $\ddagger$ & 85.6038451 & 0.6838717 & 3219457457305021568 & 15.418 $\pm$ 0.056 & 20.627 $\pm$ 0.008 & 41.525 $\pm$ 1.041 & 18.718 $\pm$ 0.055 \\
            J0723$+$4622 $\ddagger$ & 110.8896758 & 46.3784540 & 974639306231785344 & 16.027 $\pm$ 0.083 & 20.695 $\pm$ 0.012 & 17.906 $\pm$ 1.575 & 16.959 $\pm$ 0.191 \\
            J0808$+$3157 $\ddagger$ & 122.0077516 & 31.950797 & 901922932930351872 & 12.722 $\pm$ 0.019 & 16.743 $\pm$ 0.001 & 38.819 $\pm$ 0.086 & 14.689 $\pm$ 0.005 \\
            J0811$+$1855 $\ddagger$ & 122.7928110 & 18.9243897 & 669399515262792320 & 14.478 $\pm$ 0.027 & 19.220 $\pm$ 0.003 & 33.589 $\pm$ 0.414 & 16.851 $\pm$ 0.027 \\
            J0817$-$6155 * & 124.3735181 & -61.9161302 & 5278042880077383040 & 13.613 $\pm$ 0.024 & 20.035 $\pm$ 0.007 & 191.836 $\pm$ 0.419 & 21.450 $\pm$ 0.008 \\
            J0832$+$3538 $\dagger$ & 128.1977645 & 35.6477292 & 903765920576713856 & 12.522 $\pm$ 0.019 & 16.340 $\pm$ 0.001 & 32.328 $\pm$ 0.095 & 13.887 $\pm$ 0.006 \\
            J0850$-$0318 $\dagger$ & 132.5371964 & -3.3084429 & 5762038930728469888 & 14.247 $\pm$ 0.078 & 18.981 $\pm$ 0.002 & 28.433 $\pm$ 0.259 & 16.250 $\pm$ 0.020 \\
            J0900$+$5205 $\ddagger$ & 135.2161087 & 52.0862476 & 1017310351779243392 & 16.100 $\pm$ 0.108 &20.817 $\pm$ 0.009  & 11.056 $\pm$ 1.126 & 16.035 $\pm$ 0.221 \\
            J0911$+$1432 $\ddagger$ & 137.9024998 & 14.5454833 & 606673098451745920 & 15.447 $\pm$ 0.048 & 20.339 $\pm$ 0.005 & 14.914 $\pm$ 0.767 & 16.207 $\pm$ 0.112 \\
            J0916$-$1121 $\ddagger$ & 139.2384596 & -11.3519579& 5739240415392701440 & 14.095 $\pm$ 0.029 & 18.726 $\pm$ 0.002 & 32.950 $\pm$ 0.250 & 16.315 $\pm$ 0.017 \\
            J0941$+$3315A $\ddagger$ & 145.3006199 & 33.2519852 & 794031395948224640 & 13.622 $\pm$ 0.029 & 17.968 $\pm$ 0.001 & 30.3279 $\pm$ 0.163 & 15.377 $\pm$ 0.012 \\
            J0942$-$2551 $\ddagger$ & 145.6348537 & -25.8604361 & 5658599213249509120 & 15.860 $\pm$ 0.080 & 20.711 $\pm$ 0.008 & 20.006 $\pm$ 1.151 & 17.217 $\pm$ 0.125 \\
            J0948$+$5300 $\ddagger$ & 147.0526299 & 53.0102651 & 1020273539910037504 & 15.585 $\pm$ 0.064 & 20.465 $\pm$ 0.006 & 21.192 $\pm$ 0.798 & 17.096 $\pm$ 0.082 \\
            J1036$-$3441 * & 159.2208579 & -34.6960172 & 5444217638657849856 & 15.622 $\pm$ 0.048 & 20.850 $\pm$ 0.011 & 67.961 $\pm$ 1.595 & 20.012 $\pm$ 0.052 \\
            J1048$-$5254 * & 162.1143606 & -52.9048224 & 5353652721316104832 & 14.016 $\pm$ 0.030 & 18.887 $\pm$ 0.002 & 36.391 $\pm$ 0.183 & 16.692 $\pm$ 0.011 \\
            J1126$-$2706 $\ddagger$ & 171.6545247 & -27.1126095 & 3533077004345702656 & 14.253 $\pm$ 0.028 & 18.924 $\pm$ 0.002 & 30.016 $\pm$ 0.273 & 16.311 $\pm$ 0.020 \\
            J1143$+$5324 $\dagger$ & 175.8371376 & 53.4118116 & 840382713073253760 & 15.928 $\pm$ 0.075 & 20.620 $\pm$ 0.006 & 17.858 $\pm$ 0.772 & 16.879 $\pm$ 0.094 \\
            J1150$-$2914 $\ddagger$ & 177.6785418 & -29.2469265 & 3480771277705948928 & 15.040 $\pm$ 0.042 & 19.955 $\pm$ 0.004 & 23.338 $\pm$ 0.450 & 16.795 $\pm$ 0.042 \\
            J1152$+$5901 $\ddagger$ & 178.2461013 & 59.0185278 & 846350351086394752 & 15.886 $\pm$ 0.070 & 20.637 $\pm$ 0.008 & 18.609 $\pm$ 0.750 & 16.985 $\pm$ 0.088 \\
            J1158$-$0008 * & 179.6297159 & -0.1476202 & 3795026681770171264 & 16.149 $\pm$ 0.079 & 20.711 $\pm$ 0.009 & 13.644 $\pm$ 1.405 & 16.386 $\pm$ 0.224 \\
            J1158$+$3817 $\ddagger$ & 179.5707516 & 38.2877118 & 4034251549793554176 & 14.657 $\pm$ 0.035 & 18.930 $\pm$ 0.003 & 21.719 $\pm$ 0.241 & 15.614 $\pm$ 0.024 \\
            J1212$+$0206 * & 183.1413863 & 2.1072730 & 3699683699198329600 & 16.128 $\pm$ 0.128 & 20.715 $\pm$ 0.011 & 17.303 $\pm$ 1.668 & 16.906 $\pm$ 0.210 \\
            J1215$+$0042 * & 183.8261119 & 0.7156115 & 3698542165611096064 & 15.533 $\pm$ 0.073 & 20.486 $\pm$ 0.011 & 22.557 $\pm$ 1.135 & 17.252 $\pm$ 0.110 \\
            J1243$+$6001 $\dagger$ & 190.8838648 & 60.0239567 & -- & 18.37 $\pm$ 0.22 $^\text{a}$ & >23.233 & -- & -- \\
            J1250$+$0455 $\ddagger$ & 192.5644993 & 4.9186159 & 3705763723623660416 & 15.158 $\pm$ 0.060 & 20.460 $\pm$ 0.007 & 13.938 $\pm$ 1.140 & 16.181 $\pm$ 0.178 \\
            J1252$+$0347 * & 193.1586252 & 3.7929674 & 3704616555037021056 & 15.988 $\pm$ 0.081 & 20.709 $\pm$ 0.008 & 14.864 $\pm$ 1.427 & 16.570 $\pm$ 0.209 \\
            J1307$+$0246 * & 196.8911778 & 2.7666348 & 3692416064777472640 & 15.051 $\pm$ 0.037 & 19.720 $\pm$ 0.018 & 23.888 $\pm$ 0.569 & 16.611 $\pm$ 0.055 \\
            J1313$+$1404 $\ddagger$ & 198.4447057 & 14.0825798 & 3743732776411929472 & 15.793 $\pm$ 0.071 & 20.384 $\pm$ 0.006 & 16.186 $\pm$ 0.854 & 16.430 $\pm$ 0.115 \\
            J1315$+$3232 $\ddagger$ & 198.8293776 & 32.5338098 & 1466650487416101504 & 12.949 $\pm$ 0.023 & 16.912 $\pm$ 0.001 & 33.523 $\pm$ 0.098 & 14.539 $\pm$ 0.006 \\
            J1320$+$4238 $\ddagger$ & 200.1293042 & 42.6378205 & 1525546995989465344 & 12.839 $\pm$ 0.025 & 16.695 $\pm$ 0.001 & 37.454 $\pm$ 0.064 & 14.562 $\pm$ 0.004 \\
            J1420$+$3235 $\ddagger$ & 215.1086971 & 32.5981565 & 1477880589944207744 & 13.824 $\pm$ 0.026 & 17.774 $\pm$ 0.001 & 21.907 $\pm$ 0.141 & 14.477 $\pm$ 0.014 \\
            J1423$+$5146 $\ddagger$ & 215.7633280 & 51.7760046 & 1604901876901703168 & 11.885 $\pm$ 0.021 & 15.867 $\pm$ 0.001 & 57.802 $\pm$ 0.059 & 14.676 $\pm$ 0.002 \\
            J1441$+$4217 $\ddagger$ & 220.4522858 & 42.2970197 & 1490058849451967744 & 15.722 $\pm$ 0.061 & 20.439 $\pm$ 0.006 & 22.053 $\pm$ 0.626 & 17.156 $\pm$ 0.062 \\
            J1452$+$0931 * & 223.0056374 & 9.5260993 & 1174086386182157696 & 15.418 $\pm$ 0.070 & 20.114 $\pm$ 0.006 & 20.382 $\pm$ 0.668 & 16.660 $\pm$ 0.071 \\
            J1514$+$3547 $\ddagger$ & 228.7368646 & 35.7961237 & 1291186058168054016 & 16.097 $\pm$ 0.073 & 20.742 $\pm$ 0.008 & 18.902 $\pm$ 1.046 & 17.125 $\pm$ 0.120 \\
            J1536$+$0646 * & 234.2473725 & 6.7803897 & 4430701697214030464 & 15.609 $\pm$ 0.066 & 20.669 $\pm$ 0.009 & 16.451 $\pm$ 1.214 & 16.749 $\pm$ 0.161 \\
            J1544$-$0435 $\dagger$ & 236.1769024 & -4.5919062 & 4402274889228322944 & 15.901 $\pm$ 0.081 & 20.806 $\pm$ 0.008 & 19.245 $\pm$ 1.504 & 17.227 $\pm$ 0.170 \\
            J1544$+$3301 $\ddagger$ & 236.2299864 & 33.0294671 & 1370790111609798144 & 15.548 $\pm$ 0.057 & 20.653 $\pm$ 0.007 & 41.756 $\pm$ 0.775 & 18.756 $\pm$ 0.041 \\
            J1628$-$4652 * & 247.1506423 & -46.8816276 & 5942058396163925376 & 15.490 $\pm$ 0.011 & 20.132 $\pm$ 0.007 & 16.086 $\pm$ 1.035 & 24.100 $\pm$ 0.140 \\
            J1637$+$1813 $\dagger$ & 249.4715498 & 18.2284064 & 4562510841912952192 & 15.461 $\pm$ 0.049 & 20.312 $\pm$ 0.005 & 20.092 $\pm$ 0.698 & 23.797 $\pm$ 0.076 \\
            J1646$-$2115 $\dagger$ & 251.5970487 & -21.2533460 & 4126955670100614400 & 15.633 $\pm$ 0.064 & 20.698 $\pm$ 0.008 & 21.184 $\pm$ 1.858 & 24.068 $\pm$ 0.191 \\
            J1654$-$3819 $\ddagger$ & 253.5650997 & -38.3184147 & 5970493789759784192 & 12.084 $\pm$ 0.026 & 15.956 $\pm$ 0.001 & 39.666 $\pm$ 0.060 & 17.964 $\pm$ 0.003 \\
            J1700$-$4048 * & 255.2221268 & -40.8051656 & 5966993223906484992 & 16.252 $\pm$ 0.112 & 20.579 $\pm$ 0.010 & 4.397 $\pm$ 1.316 & 27.378 $\pm$ 0.650 \\
            J1713$-$3952 $\ddagger$ & 258.4192905 & -39.8712362 & 5972124644679705728 & 13.401 $\pm$ 0.026 & 18.133 $\pm$ 0.002 & 51.561 $\pm$ 0.178 & 19.571 $\pm$ 0.008 \\
            J1737$+$4705 $\ddagger$ & 264.4109622 & 47.0951605 & 1363482108789712000 & 14.572 $\pm$ 0.029 & 19.237 $\pm$ 0.003 & 26.803 $\pm$ 0.208 & 22.096 $\pm$ 0.017 \\
            J1847$-$3419 $\ddagger$ & 281.9459895 & -34.3260745 & 6735308241178655872 & 12.725 $\pm$ 0.029 & 16.737 $\pm$ 0.003 & 36.887 $\pm$ 0.088 & 18.902 $\pm$ 0.006 \\
            J1938$+$4321 $\ddagger$ & 294.7019820 & 43.3553497 & 2077988676275617408 & 12.708 $\pm$ 0.022 & 16.404 $\pm$ 0.001 & 37.376 $\pm$ 0.048 & 18.541 $\pm$ 0.003 \\
            J2019$+$2256 $\dagger$ & 304.9530278 & 22.9457810 & 1829571684884360832 & 13.820 $\pm$ 0.111 & 19.444 $\pm$ 0.003 & 33.938 $\pm$ 0.342 & 21.791 $\pm$ 0.022 \\
		\hline
            \multicolumn{5}{l}{$^\text{a}$ -- Estimated by \cite{faherty2021} using SpeX spectral data.} \\
	\end{tabular}
\end{table*}

\begin{table*}
    \caption{Basic data for each of the UCDs in this paper. Spectroscopic classifications (SpT) of the UCDs are as given by \splat, and photometric spectral types (PhT) are also shown. Where applicable, published NIR and optical spectroscopic classifications for each UCD are also shown. UCDs with entries in \textit{italics} are those which have spectra with sub-optimal signal-to-noise or poor telluric standard star availability, thus their derived properties are not definitive. Effective temperature and surface gravity estimate derivations are outlined in Section~\ref{section:temp}, and spectrophotometric distances are calculated in Section~\ref{section:distance}. UCDs marked with a dagger $\dagger$ are those observed with SpeX Prism, those marked with a double dagger $\ddagger$ are observed with SpeX SXD, and those marked with an asterisk * are observed with ARCoIRIS. This data is also contained within the collated data table in Appendix~\ref{appendix:fulldata}.}
    \label{table:simbad}
    \begin{tabular}{c c c c c c c c}
        \hline
        Object & SIMBAD & SpT & PhT & Published & Published & \teff & Spectrophotometric \\
        Short Name & Name & (this work) & (this work) & NIR SpT & Optical SpT & (K) & Distance (pc) \\
        \hline
        J0508$+$3319 $\ddagger$ & 2MASS J05085506$+$3319272 & L2 $\pm$ 0.4 & L1 $\pm$ 2.5 & -- & L2 $^{[1]}$ & 2000 $\pm$ 60 & 21.97 $\pm$ 1.42 \\
        J0515$+$0613 $\ddagger$ & \gaia{}~DR2 3240769806581623296 & M9 $\pm$ 0.4 & M9 $\pm$ 3.0 & -- & -- & 2800 $\pm$ 120 & 76.19 $\pm$ 1.48 \\
        \textit{J0526$-$5026 *} & \textit{2MASS J05265973$-$5026216} & \textit{L6 $\pm$ 0.4} & \textit{L8 $\pm$ 2.0} & -- & \textit{L3 $^{[2]}$} & \textit{1700 $\pm$ 90} & \textit{22.01 $\pm$ 0.88} \\
        \textit{J0542$+$0041 $\ddagger$} & \textit{\gaia{}~DR3 3219457457305021568} & \textit{L7 $\pm$ 1.2} & \textit{L9 $\pm$ 2.5} & -- & -- & \textit{1600 $\pm$ 180} & \textit{15.84 $\pm$ 1.15} \\
        J0723$+$4622 $\ddagger$ & \gaia{}~DR3 974639306231785344 & L1 $\pm$ 0.5 & L0 $\pm$ 4.5 & -- & -- & 2000 $\pm$ 200 & 74.95 $\pm$ 3.82 \\
        J0808$+$3157 $\ddagger$ & 2MASS J08080189$+$3157054 & M7 $\pm$ 0.4 & M7 $\pm$ 1.5 & -- & -- & 3000 $\pm$ 130 & 26.68 $\pm$ 0.36\\
        J0811$+$1855 $\ddagger$ & 2MASS J08111040$+$1855280 & L1 $\pm$ 0.4 & L3 $\pm$ 2.0 & -- & L1 $^{[3]}$ & 2600 $\pm$ 80 & 31.71 $\pm$ 0.45 \\
        J0817$-$6155 * & 2MASS J08173001$-$6155158 & T6 $\pm$ 0.4 & T7 $\pm$ 1.5 & T6 $^{[4]}$ & -- & 1200 $\pm$ 70 & 4.93 $\pm$ 0.31 \\
        J0832$+$3538 $\dagger$ & LP 258$-$34 & M6 $\pm$ 0.4 & M6 $\pm$ 2.0 & -- & -- & 2800 $\pm$ 140 & 26.79 $\pm$ 0.37 \\
        J0850$-$0318 $\dagger$ & 2MASS J08500913$-$0318305 & M8 $\pm$ 0.4 & M7.5 $\pm$ 1.5 & -- & -- & 2600 $\pm$ 100 & 44.02 $\pm$ 1.12 \\
        J0900$+$5205 $\ddagger$ & SDSS J090051.84$+$520512.1 & M9 $\pm$ 0.4 & L0 $\pm$ 1.0 & -- & L0 $^{[6]}$ & 2800 $\pm$ 160 & 92.68 $\pm$ 4.24 \\
        J0911$+$1432 $\ddagger$ & SDSS J091136.57$+$143244.4 & M9 $\pm$ 0.5 & M9 $\pm$ 2.5 & -- & L0 $^{[6]}$ & 2600 $\pm$ 80 & 58.50 $\pm$ 1.74 \\
        J0916$-$1121 $\ddagger$ & 2MASS J09165708$-$1120597 & M8 $\pm$ 0.4 & M7 $\pm$ 2.0 & M9 $^{[7]}$ & -- & 2600 $\pm$ 80 & 41.99 $\pm$ 1.02 \\
        J0941$+$3315A $\ddagger$ & \gaia{}~DR3 794031395948224640 & M8 $\pm$ 0.4 & M8 $\pm$ 2.0 & -- & -- & 2800 $\pm$ 100 & 38.85 $\pm$ 6.84 \\
        J0942$-$2551 $\ddagger$ & \gaia{}~DR3 5658599213249509120 & L1 $\pm$ 0.4 & L0 $\pm$ 2.0 & -- & -- & 2800 $\pm$ 130 & 60.40 $\pm$ 1.88 \\
        J0948$+$5300 $\ddagger$ & 2MASS J09481259$+$5300387 & L1 $\pm$ 0.4 & M9 $\pm$ 3.0 & L2 $^{[8]}$ & -- & 2000 $\pm$ 160 & 51.34 $\pm$ 2.53 \\
        J1036$-$3441 * & 2MASSW J1036530$-$344138 & L8 $\pm$ 0.4 & T0 $\pm$ 1.5 & L6.5 $^{[9]}$ & L6 $^{[10]}$ & 1700 $\pm$ 80 & 15.40 $\pm$ 0.22 \\
        J1048$-$5254 * & DENIS J104827.8$-$525418 & L1 $\pm$ 0.4 & M8.5 $\pm$ 1.0 & L0.5 $^{[11]}$ & L1.5 $^{[12]}$ & 2100 $\pm$ 150 & 25.06 $\pm$ 0.95 \\
        J1126$-$2706 $\ddagger$ & 2MASS J11263718$-$2706470 & M8 $\pm$ 0.4 & M7 $\pm$ 1.0 & -- & -- & 2700 $\pm$ 100 & 45.10 $\pm$ 0.97 \\
        J1143$+$5324 $\dagger$ & \gaia{}~DR3 840382713073253760 & L1 $\pm$ 0.4 & L1 $\pm$ 4.0 & -- & -- & 2400 $\pm$ 60 & 69.53 $\pm$ 1.33 \\
        J1150$-$2914 $\ddagger$ & \gaia{}~DR2 3480771277705948928 & L0 $\pm$ 0.4 & L0 $\pm$ 4.5 & -- & -- & 2000 $\pm$ 170 & 44.39 $\pm$ 1.76 \\
        J1152$+$5901 $\ddagger$ & \gaia{}~DR2 846350351086394752 & L0 $\pm$ 0.4 & M8.5 $\pm$ 3.0 & -- & -- & 2000 $\pm$ 200 & 64.88 $\pm$ 4.50 \\
        J1158$-$0008 * & 2MASS J11583104$-$0008491 & M9 $\pm$ 0.4 & M8 $\pm$ 3.0 & -- & -- & 2700 $\pm$ 100 & 91.78 $\pm$ 3.44 \\
        J1158$+$3817 $\ddagger$ & 2MASS J11581724$+$3817203 & M7 $\pm$ 0.4 & M7.5 $\pm$ 1.5 & -- & -- & 2700 $\pm$ 90 & 65.50 $\pm$ 1.39 \\
        J1212$+$0206 * & 2MASS J12123389$+$0206280 & L1 $\pm$ 0.4 & L1 $\pm$ 3.0 & L1 $^{[5]}$ & L1 $^{[6]}$ & 2600 $\pm$ 150 & 50.76 $\pm$ 5.36 \\
        J1215$+$0042 * & \gaia{}~DR3 3698542165611096064 & M9 $\pm$ 0.4 & L2 $\pm$ 2.5 & -- & -- & 2800 $\pm$ 160 & 72.98 $\pm$ 3.08 \\
        J1243$+$6001 $\dagger$ & \textit{WISE} J124332.17$+$600126.6 & L7 $\pm$ 0.4 & M8.5 $\pm$ 2.5 & L8 $^{[13]}$ & -- & 1600 $\pm$ 130 & 43.46 $\pm$ 12.28 \\
        J1250$+$0455 $\ddagger$ & 2MASS J12501559$+$0455065 & M9 $\pm$ 0.4 & L2 $\pm$ 1.5 & -- & -- & 2600 $\pm$ 110 & 58.53 $\pm$ 3.22 \\
        J1252$+$0347 * & 2MASS J12523834$+$0347351 & M8 $\pm$ 0.4 & M9 $\pm$ 2.5 & -- & M7 $^{[15]}$ & 2500 $\pm$ 70 & 82.67 $\pm$ 2.10 \\
        J1307$+$0246 * & 2MASS J13073376$+$0246048 & L1 $\pm$ 0.4 & L2 $\pm$ 5.0 & -- & -- & 2000 $\pm$ 90 & 54.08 $\pm$ 1.48 \\
        J1313$+$1404 $\ddagger$ & \gaia{}~DR2 3743732776411929472 & M8 $\pm$ 0.4 & L1 $\pm$ 2.5 & -- & -- & 2800 $\pm$ 130 & 86.18 $\pm$ 3.29 \\
        J1315$+$3232 $\ddagger$ & 2MASS J13151905$+$3232031 & M7 $\pm$ 0.4 & M6 $\pm$ 2.0 & -- & -- & 2900 $\pm$ 110 & 30.90 $\pm$ 0.99 \\
        J1320$+$4238 $\ddagger$ & LP 218$-$81 & M7 $\pm$ 0.4 & M6 $\pm$ 1.0 & -- & -- & 3000 $\pm$ 130 & 29.41 $\pm$ 0.99 \\
        J1420$+$3235 $\ddagger$ & \gaia{}~DR2 1477880589944207744 & M7 $\pm$ 0.4 & M6 $\pm$ 2.5 & -- & -- & 2900 $\pm$ 110 & 45.04 $\pm$ 0.97 \\
        J1423$+$5146 $\ddagger$ & LP 134$-$7 & M7 $\pm$ 0.4 & M6 $\pm$ 2.0 & -- & M7 $^{[14]}$ & 2900 $\pm$ 130 & 18.64 $\pm$ 0.44 \\
        J1441$+$4217 $\ddagger$ & \gaia{}~DR3 1490058849451967744 & L3 $\pm$ 0.6 & L0 $\pm$ 1.5 & -- & -- & 1600 $\pm$ 140 & 42.65 $\pm$ 1.83 \\
        J1452$+$0931 * & 2MASS J14520130$+$0931372 & M8 $\pm$ 0.4 & M9 $\pm$ 2.5 & -- & L0 $^{[15]}$ & 2700 $\pm$ 90 & 75.64 $\pm$ 2.28 \\
        J1514$+$3547 $\ddagger$ & \gaia{}~DR3 1291186058168054016 & L1 $\pm$ 0.5 & L1 $\pm$ 3.0 & -- & -- & 2600 $\pm$ 60 & 59.63 $\pm$ 4.50 \\
        J1536$+$0646 * & 2MASS J15365938$+$0646507 & L0 $\pm$ 0.4 & L2 $\pm$ 3.0 & -- & -- & 2500 $\pm$ 70 & 65.82 $\pm$ 2.63 \\
        J1544$-$0435 $\dagger$ & \gaia{}~DR2 4402274889228322944 & L1 $\pm$ 0.4 & L1 $\pm$ 1.5 & -- & -- & 1700 $\pm$ 100 & 55.66 $\pm$ 2.54 \\
        J1544$+$3301 $\ddagger$ & 2MASS J15445518$+$3301447 & L2 $\pm$ 0.4 & L1.5 $\pm$ 1.5 & -- & L6 $^{[16]}$ & 1500 $\pm$ 110 & 44.47 $\pm$ 5.83 \\
        J1628$-$4652 * & \gaia{}~DR3 5942058396163925376 & M8 $\pm$ 0.4 & L1.5 $\pm$ 3.0 & -- & -- & 2800 $\pm$ 100 & 81.69 $\pm$ 2.46 \\
        J1637$+$1813 $\dagger$ & \gaia{}~DR2 4562510841912952192 & M7 $\pm$ 0.4 & M6.5 $\pm$ 1.5 & -- & -- & 2200 $\pm$ 100 & 84.94 $\pm$ 5.82 \\
        J1646$-$2115 $\dagger$ & 2MASS J16462325$-$2115064 & sdL2 $\pm$ 0.4 & L2 $\pm$ 3.0 & -- & -- & 2300 $\pm$ 40 & 49.11 $\pm$ 4.14 \\
        \textit{J1654$-$3819 $\ddagger$} & \textit{2MASS J16541588$-$3818593} & \textit{M6 $\pm$ 0.4} & \textit{M6 $\pm$ 1.0} & -- & -- & \textit{2600 $\pm$ 200} & \textit{26.17 $\pm$ 4.03} \\
        \textit{J1700$-$4048 *} & \textit{\gaia{}~DR3 5966993223906484992} & \textit{L4 $\pm$ 0.4} & \textit{L5.5 $\pm$ 4.0} & -- & -- & \textit{1600 $\pm$ 220} & \textit{50.23 $\pm$ 3.84} \\
        J1713$-$3952 $\ddagger$ & \gaia{}~DR3 5972124644679705728 & L1 $\pm$ 0.4 & L2 $\pm$ 1.0 & -- & -- & 2500 $\pm$ 80 & 20.15 $\pm$ 1.26 \\
        J1737$+$4705 $\ddagger$ & 2MASS J17373855$+$4705511 & M8 $\pm$ 0.4 & M8.5 $\pm$ 1.5 & -- & -- & 1200 $\pm$ 100 & 48.78 $\pm$ 1.16 \\
        J1847$-$3419 $\ddagger$ & 2MASS J18474700$-$3419345 & M7 $\pm$ 0.4 & M7 $\pm$ 1.5 & -- & -- & 2900 $\pm$ 120 & 26.98 $\pm$ 0.55 \\
        J1938$+$4321 $\ddagger$ & LSPM J1938$+$4321 & M6 $\pm$ 0.4 & M6 $\pm$ 2.5 & -- & -- & 2900 $\pm$ 120 & 29.50 $\pm$ 0.36 \\
        J2019$+$2256 $\dagger$ & \gaia{}~DR2 1829571684884360832 & L2 $\pm$ 0.4 & L1.5 $\pm$ 1.5 & -- & -- & 1800 $\pm$ 80 & 21.84 $\pm$ 1.74 \\
        \hline
    \end{tabular}
    \vspace{-2ex}
    \begin{tabular}{l l l l}
        \textbf{References} & $^{[1]}$ -- \cite{kirkpatrick2016} & $^{[2]}$ -- \cite{reid2008} & $^{[3]}$ -- \cite{schmidt2015} \\
        $^{[4]}$ -- \cite{artigau2010} & $^{[5]}$ -- \cite{faherty2009} & $^{[6]}$ -- \cite{kiman2019} & $^{[7]}$ -- \cite{kirkpatrick2014} \\
        $^{[8]}$ -- \cite{kellogg2017} & $^{[9]}$ -- \cite{burgasser2010} & $^{[10]}$ -- \cite{gizis2002} & $^{[11]}$ -- \cite{folkes2012} \\
        $^{[12]}$ -- \cite{phanbao2008} & $^{[13]}$ -- \cite{faherty2021} & $^{[14]}$ -- \cite{kirkpatrick2024} & $^{[15]}$ -- \cite{zhang2010} \\
        $^{[16]}$ -- \cite{schmidt2014}\\
    \end{tabular}
\end{table*}

\section{Data Collection and Reduction}
\label{section:data}
\subsection{Candidate Selection}
\label{section:data.cand}
The UCDs selected for this investigation were initially found using \gaia{}~DR1 and DR2. The selection criteria were determined empirically from known UCDs and exploited the following external photometric datasets: 

\begin{itemize}
    \item \vskip -0.2cm Panoramic Survey Telescope and Rapid Response System \textit{i}, \textit{z} and \textit{y} magnitudes (hereafter Pan-STARRS; \citealt{chambers2016}).
    \item Two Micron All Sky Survey \textit{J}, \textit{H} and \textit{Ks} magnitudes (hereafter 2MASS; \citealt{skrutskie2006}).
    \item AllWISE or CatWISE2020 \textit{W1} and \textit{W2} magnitudes, depending on availability at time of candidate selection (AllWISE is a combination of \textit{Wide-field Infrared Survey Explorer}, hereafter \textit{WISE}, and NEOWISE data; \citealt{wright2010, mainzer2011}; and CatWISE2020 expands the baseline to six years, improving accuracy and precision; \citealt{marocco2021}).
\end{itemize}

The selection process was conservative enough to retain all real UCDs, but also results in a number of false positives. To remove false positives, each target was visually checked in different sky surveys for inconsistencies, e.g. colour, proper motion, crowding.

The \gaia{} data releases within 100\,pc have significant numbers of objects with incorrect parallaxes (e.g. for DR3; \citealt{lindegren2018,smart2019}), and many of the UCDs in this sample are faint, thus quality assurance flags are not used since they are unreliable. The final selection criteria for the UCDs in the GUCDS selection therefore do not rely solely on absolute \textit{G}-band magnitude and instead use estimated spectral types and colour relations. In order to estimate the spectral types, we used known literature UCDs to find relations between median absolute \textit{G} magnitude and various colour relations. 

The adopted selection criteria are: 
\begin{itemize}
    \item \vskip -0.2cm No published spectroscopic spectral classification.
    \item Has parallax, \textit{G}, \(G_{RP}\) and at least one 2MASS magnitude.
    \item M$_\textit{G}$-based spectral type later than M6 (nearest subclass rounded to 0.5).
    \item Median colour-derived spectral type later than M6 (median of nearest subclasses for available colours).
\end{itemize}

Table~\ref{table:ucds} contains \gaia{}~DR3 data for the UCDs, including their \textit{G}- and \textit{J}-band magnitudes and parallaxes. The \gaia{} source ID for each UCD is cross-referenced with SIMBAD (\citealt{wenger2000}), in order to find any published classifications for the UCDs.

Many of the sources of interest have SIMBAD entries, and we included some additional sources with spectroscopic spectral type classifications (Table~\ref{table:simbad}; \citealt{wenger2000}). This is useful for the purposes of this paper, since we can check that our derived spectral types are consistent with those published, and thus we can gauge the reliability of our spectral type classifications. We included these UCDs with literature spectral types (such as J0817$-$6155) to use as comparison objects. We also included J1243$+$6001 in our sample -- although this UCD does not have a published \gaia{} parallax, \cite{faherty2021} ascertain it is in a wide binary system with BD$+$60 1417. We therefore adopt the parallax of the primary for the UCD.

\subsection{Observations of UCDs}
\label{section:data.obs}
The UCDs selected for investigation in this work were observed with the Astronomy Research using the Cornell Infra Red Imaging Spectrograph instrument (hereafter ARCoIRIS; \citealt{schlawin2014}) at the Blanco 4\,m telescope in Chile and the SpeX instrument at the NASA Infrared Telescope Facility in Hawaii (hereafter IRTF; \citealt{rayner2003}).

We used A-B-B-A observing patterns with both telescopes, so that sky subtraction could be carried out in order to optimise the signal from the UCDs. For most observations, the slit was aligned to the paralactic angle, in order to keep the effects of atmospheric distortion to a minimum. After each candidate observation, we observed a nearby A0V star to use for telluric corrections, taken from the  Tycho-2 Catalogue (\citealt{hog2000}). The standards used for telluric correction in the reduction process were cross-referenced with SIMBAD to check the spectral types used for the telluric correction of the spectral data. No uncertainties for the standards' spectral classifications are published, so an uncertainty of ${\pm}$1 sub-type is applied -- all of the telluric correction standard stars used in this work are within the uncertainties of an A0V-type. Appendix~\ref{appendix:log} contains the observation log, which also includes details of the standard stars used for the telluric corrections of each candidate.

During some of the observing runs, the conditions were too poor for UCDs to be observed (e.g. poor seeing or excessive cloud cover), so non-UCDs were observed instead. Appendix~\ref{appendix:nonucd} discusses these UCDs, along with the objects that were observed and classified as non-UCDs.

\subsubsection{Blanco Observations}
\label{section:data.obs.blanco}
We observed 13 UCDs in 2018 using the Blanco ARCoIRIS instrument under program 2018A-0910 (PI: Beamin). ARCoIRIS is composed of a cross-dispersed spectrograph, producing six spectral orders between the wavelengths of 0.8--2.4\,\um{} (\citealt{james2015}). The instrument has a resolution of R${\approx}$3000, making the spectra obtained using ARCoIRIS the highest resolution spectra in our sample.

\subsubsection{IRTF Observations}
\label{section:data.obs.spex}
Between 2018 and 2021, 38 UCDs were observed using the IRTF SpeX instrument under programs 2018A067, 2020A069, 2020B057 and 2021A023 (PI: Smart), with a total of eight observations using the prism configuration and 30 observations using the short-wavelength cross-dispersion (hereafter SXD) configuration. The prism configuration creates one order of spectral data between the wavelengths of 0.70--2.52\,\um{}, while the SXD grating covers spectral orders 3--9 between the wavelengths of 0.70--2.55\,\um{} (\citealt{rayner2003}). The resolutions of the prism and SXD configurations are R$\approx$200 and R$\approx$2000, respectively. Due to the low resolution of the prism, this configuration was only used when observing conditions were poor (e.g. clouds, wind).

\begin{figure}
\centering
\includegraphics[width=\linewidth]{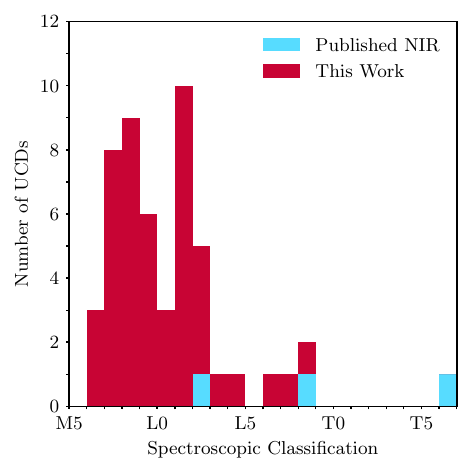}
\caption{Histogram of previously published NIR spectroscopic classifications for the UCDs presented in this work (blue) and the contribution of the new NIR spectral type classifications of UCDs from this work (red).}
\label{figure:histogram}
\end{figure}

\subsection{Data Reduction with \textsc{Spextool}}
\label{section:data.red}
The data reduction process was completed with the \textsc{Spextool} program (\citealt{cushing2004}). The reduction process for the SpeX data is generally the same for both the prism and SXD data, with the exception of the telluric correction methods: the prism data use a convolution kernel created using the Instrument Profile method, while the convolution kernels for the SXD data are constructed through the deconvolution method using the Pa\,$\updelta$ line which lies around 1.005\,\um{} (\citealt{paschen1908}). The reduction for the ARCoIRIS data utilised a version of \textsc{Spextool} adapted for ARCoIRIS spectra by K. Allers\footnote{The adaptation of \textsc{Spextool} for ARCoIRIS data is publicly available from: \url{https://www.dropbox.com/sh/wew08hcqluib8am/AAC6HiCeAVoiDPSe-MXRU2Cda?dl=0}}, and thus follows the same general reduction process as the SpeX SXD data.

Perhaps the most critical part of the reduction process is the telluric correction for each UCD, specifically the selection of the standard star to be used for correcting each science object's spectrum. The primary method of choosing the standard star was finding the best airmass-match for each object, so that the observing conditions are as similar as possible for the science target and the standard. In most cases, the best airmass-matched telluric standard also has the smallest time difference between when the observations were taken. The observing schedule was designed so that each UCD observation was followed by an observation of a A0V standard star, minimising the effects of the changing sky conditions. When the airmass difference between a UCD and the following telluric standard star exceeds 0.100, the next-closest time-matched standard star is tested in the telluric correction process and the result judged by visual inspection. This process is repeated until the standard star with the optimal balance of time-match and airmass difference is found: the telluric standard with the smallest residuals is selected.

The spectra are also flux calibrated during the telluric correction process, using the \textit{B} and \textit{V} magnitudes of the A0-type standard stars. These magnitudes are used by the \texttt{xtellcor} program for \textsc{IDL} to scale a high-resolution model spectrum of Vega to use for the absolute flux calibration (\citealt{vacca2003}). The \textit{B} and \textit{V} magnitudes for scaling the Vega model spectrum are included in the observing log in Appendix~\ref{appendix:log}.

\section{Spectral Classification}
\label{section:spt}
\subsection{Spectroscopic Classification using \splat}
\label{section:spt.splat}
The \texttt{SpeX Prism Library Analysis Toolkit}\footnote{{The \splat{} module for \textsc{Python} is publicly available from: \url{https://github.com/aburgasser/splat}}} \textsc{Python} module (hereafter \splat; \citealt{burgasser2017}) was used to classify the spectral type of each UCD. By using the \texttt{classifyByStandard} routine, the spectroscopic type of each of the UCDs can be found (presented in Table~\ref{table:simbad}). The \texttt{classifyByStandard} classification method uses a spectral standard of each spectral type, M0 through T9, from the SpeX Prism Library (\citealt{burgasser2006,kirkpatrick2010,cushing2011}) and compares it with the spectrum of interest by minimising chi-squared values. \splat{} then returns the best-fitting spectral classification for the UCD, along with associated uncertainties calculated using the distribution of chi-squared values for each spectral type fit.

First, we used the \texttt{classifyByStandard} built-in \texttt{Kirkpatrick} method to classify the spectral type of each UCD by fitting the standard spectra to only the 0.9--1.4\,\um{} band (\citealt{kirkpatrick2010}). In most cases, this gives a spectral type which matches the overall shape of the spectrum. To verify these classifications, we also used the \splat{} \texttt{classifyByStandard} routine to match each object's spectrum with the SpeX Library standards over the entire spectrum, with the spectral types derived using the two methods agreeing with each other within $\pm$\,1.0 spectral types. We take the adopted spectroscopic classification as the weighted mean of the Kirkpatrick-method and whole-spectrum fits, and the associated uncertainties are calculated as the standard errors of the weighted means. When the weighted mean gives a classification in-between spectral types, the two spectral types are compared by-eye and the best-fitting classification is selected.

Overall, comparisons with published spectral types for these UCDs show that our classifications are consistent with published spectral types (Table~\ref{table:simbad}). For some UCDs, the spectral classifications are equivalent to those published, while other UCDs' spectral types vary by a few subtypes (such as J0526$-$5026), likely due to poor observing conditions leading to spectra with poor signal-to-noise ratios or the UCD being a non-single source (see Section~\ref{section:ydh.unres}). Fig.~\ref{figure:histogram} shows a histogram of previously spectroscopically classified UCDs in the GUCDS sample (blue) and the new UCD classifications contributed by this work (red).

\splat{} returns L0 as the best-fitting spectral classification for J1646$-$2115, however visual inspection and comparison with its photometric data suggest that it is more likely to be an early-L subdwarf. Using the \texttt{std\_class = "subdwarf"} option, \splat{} gives an sdL0 classification, which is closer to our suspected early-L subdwarf classification. Based on the combination of our by-eye classification, \splat's spectroscopic classification and the photometric classification (see Section~\ref{section:spt.phot}), we take the spectral type of J1646$-$2115 to be sdL2. The potential of J1646$-$2115 being a subdwarf is discussed in further detail in Section~\ref{section:indiv.sd}.

\subsection{Photometric Classification}
\label{section:spt.phot}
The spectral classification method outlined by \cite{skrzypek2015} can be used to find the spectral types of the UCDs photometrically, as a means of checking the spectral types given by \splat. For this work, we use Pan-STARRS \textit{i}, \textit{z} and \textit{y} magnitudes; 2MASS \textit{J}, \textit{H} and \textit{Ks} photometry, and CatWISE2020 \textit{W1} and \textit{W2} magnitudes (\citealt{eisenhardt2020, marocco2021}) of published UCDs to create a template relation between spectral classification and a range of colours. Appendix~\ref{appendix:template} outlines the template used for the photometric classification. Not all of the 51 UCDs have complete sets of 2MASS and CatWISE2020 magnitudes, so these were classified using the photometry which is available. By calculating $\chi^2$ values for each of the UCDs, the best-fitting photometric spectral type can be found. The photometric classifications are shown alongside the spectroscopic types in Table~\ref{table:simbad}.

\subsection{Comparison of Spectroscopic and Photometric Spectral Types}
\label{section:spt.comp}
The spectroscopic classifications that we obtained with \splat{} can be compared with those derived from photometric data, in order to ensure consistency between the different classification methods. This also allows us to identify any UCDs with differing spectroscopic and photometric classifications, since these may indicate non-solar metallicities or unusual surface gravity values. Table~\ref{table:simbad} shows a comparison between the two sets of classifications, verifying that the spectral types given by the spectroscopic and photometric methods are fairly consistent. The majority of the photometric spectral types lie within the uncertainties of the spectroscopic classifications, with only a few UCDs which do not have overlapping photometric and spectroscopic classification uncertainties: J1048$-$5254, J1243$+$6001 and J1250$+$0455. These UCDs will be investigated in more detail in Section~\ref{section:indiv}.

\subsection{Comparison of NIR and Optical Classifications}
\label{section:spt.niropt}
NIR and optical observations of UCDs probe different depths of the objects' atmospheres, and can provide clarification when an object is suspected to be an unresolved binary. Nine of our UCDs have published optical spectral classifications (Table~\ref{table:niropt}). The robustness of our NIR spectroscopic types is verified by the optical classifications, with most of the NIR and optical types falling within the uncertainties of each other, although there are two exceptions: J1036$-$3441 and J1536$+$0646.

J1036$-$3441 has a NIR spectral classification of L8, while its optical spectral type is given as L6 by \cite{gizis2002}. This 2-subtype difference could be due to a number of reasons, such as the UCD's age or the UCD being an unresolved binary. Further investigation of J1036$-$3441's potential binarity is presented in Section~\ref{section:indiv.1036}.

J1536$+$0646 has an L0 NIR spectral type and a L2-type optical photometric classification given by \cite{zhang2010}. As with J1036$-$3441, this difference in spectral classifications could be caused by the UCD's age or the UCD being an unresolved binary. Section~\ref{section:indiv.1536} contains further investigation of the properties of J1536$+$0646.

\begin{table}
\centering
    \caption{Comparison of NIR SpT classifications (this work) and optical spectral types (published). Where available, spectral type uncertainties are also included.}
    \label{table:niropt}
    \begin{tabular}{c|c|c|c}
        \hline
        Short Name & NIR SpT & Optical SpT & Optical PhT \\
        \hline
        J0508$+$3319 & L2 $\pm$ 0.4 & L2 \footnotesize{$^{[1]}$} & -- \\
        J0526$-$5026 & L6 $\pm$ 0.4 & L3 \footnotesize{$^{[2]}$} & -- \\
        J0811$+$1855 & L1 $\pm$ 0.4 & L1 \footnotesize{$^{[3]}$} & -- \\
        J1036$-$3441 & L8 $\pm$ 0.4 & L6 \footnotesize{$^{[4]}$} & -- \\
        J1048$-$5254 & L1 $\pm$ 0.4 & L1.5 \footnotesize{$^{[5]}$} & -- \\
        J1158$-$0008 & M9 $\pm$ 0.4 & -- & M9.5 \footnotesize{$^{[6]}$} \\
        J1212$+$0206 & L1 $\pm$ 0.4 & L1 \footnotesize{$^{[7]}$} & -- \\
        J1452$+$0931 & M8 $\pm$ 0.4 & L0 \footnotesize{$^{[6]}$} & -- \\
        J1536$+$0646 & L0 $\pm$ 0.4 & -- & L2 \footnotesize{$^{[6]}$} \\
        \hline
    \end{tabular}
    \begin{tabular}{l|l}
        \textbf{References} & \footnotesize{$^{[1]}$} -- \cite{kirkpatrick2016} \\
        \footnotesize{$^{[2]}$} -- \cite{reid2008} & \footnotesize{$^{[3]}$} -- \cite{schmidt2015} \\
        \footnotesize{$^{[4]}$} -- \cite{gizis2002} & \footnotesize{$^{[5]}$} -- \cite{phanbao2008} \\
        \footnotesize{$^{[6]}$} -- \cite{zhang2010} & \footnotesize{$^{[7]}$} -- \cite{kiman2019} \\
        \footnotesize{$^{[8]}$} -- \cite{faherty2009} \\
    \end{tabular}
\end{table}

\begin{figure*}
    \centering
    \includegraphics[width=\linewidth]{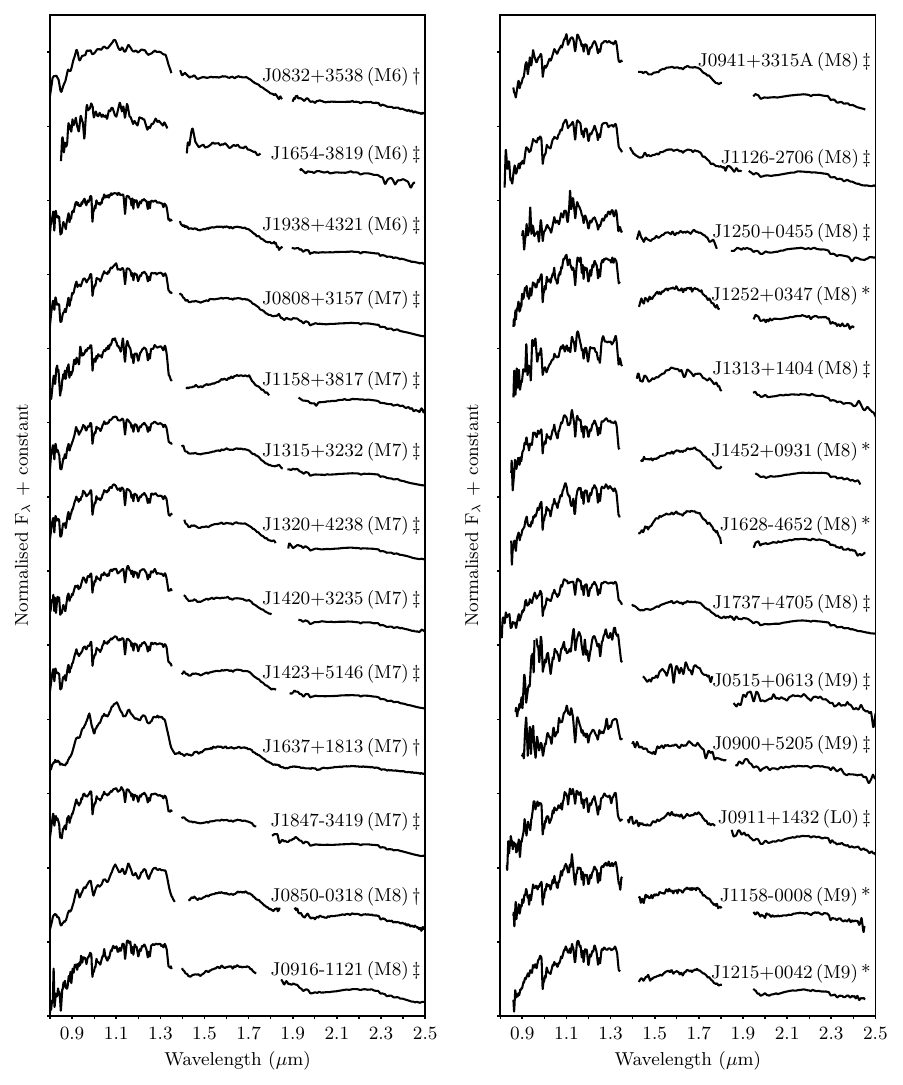}
    \caption{Left: plots of the UCD spectra, sorted by spectral classification (M6--M8). Right: plots of the UCD spectra, sorted by spectral classification (M8--M9). Each spectrum is normalised at 1.27\,\um{} and vertically offset by equal flux increments. Noisy areas around the H$_2$O bands at ${\sim}$1.3\,\um{} and ${\sim}$1.9\,\um{} have been removed to make plots clearer. UCDs marked with a dagger $\dagger$ are those observed with SpeX Prism, those marked with a double dagger $\ddagger$ are observed with SpeX SXD, and those marked with an asterisk * are observed with ARCoIRIS.}
\label{figure:spec1}
\end{figure*}

\begin{figure*}
    \centering
    \includegraphics[width=\linewidth]{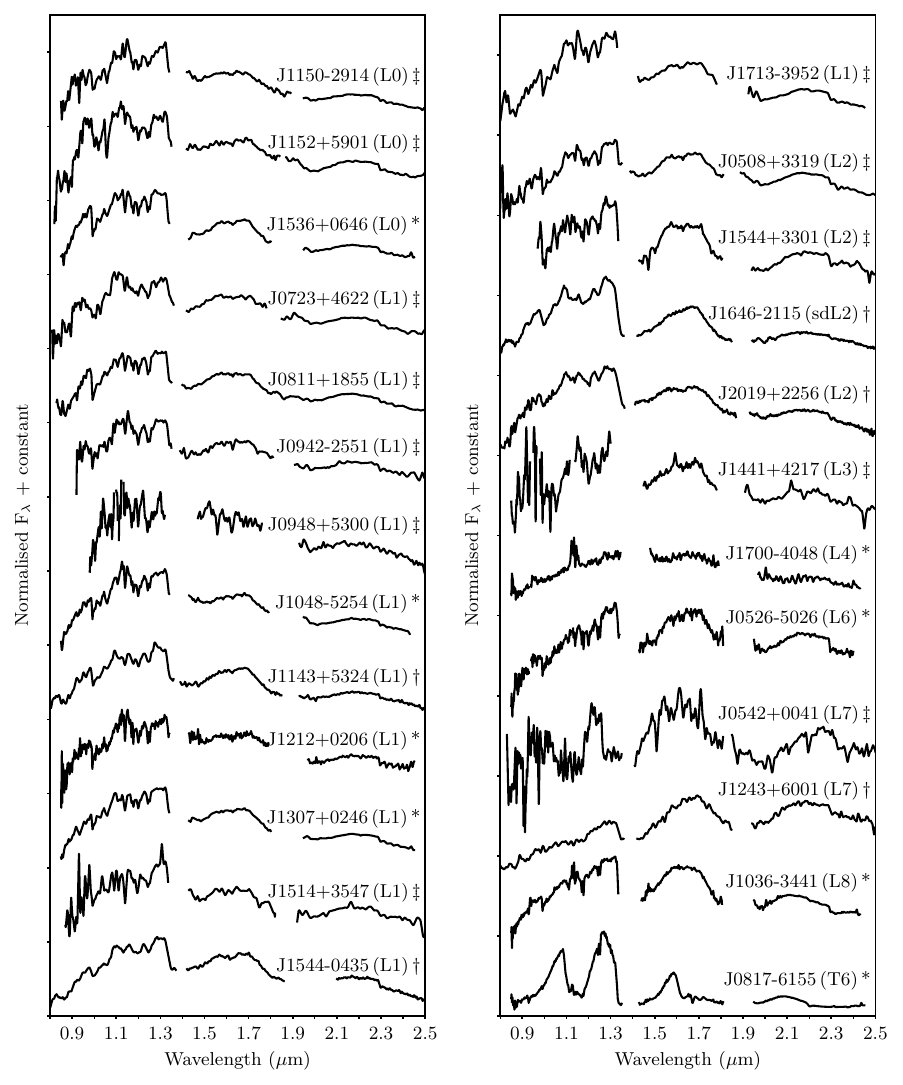}
    \caption{Left: plots of the UCD spectra sorted by spectral classification (L0--L1). Right: plots of the UCD spectra, sorted by spectral classification (L1--T6). Each spectrum is normalised at 1.27\,\um{} and vertically offset by equal flux increments. Noisy areas around the H$_2$O bands at ${\sim}$1.3\,\um{} and ${\sim}$1.9\,\um{} have been removed to make plots clearer. UCDs marked with a dagger $\dagger$ are those observed with SpeX Prism, those marked with a double dagger $\ddagger$ are observed with SpeX SXD, and those marked with an asterisk * are observed with ARCoIRIS.}
\label{figure:spec2}
\end{figure*}

\section{Spectral Analysis}
\label{section:spec}
\subsection{Spectral Plots}
\label{section:spec.plots}
By creating stacked plots of the spectra sorted by spectral type, it is possible to check that our spectral classifications are logical and follow the expected trend as spectral classifications progress from late-M to T-type (see \citealt{geballe2002,kirkpatrick2005,burgasser2009}, among others). By inspecting these plots, we are also able to identify any interesting features within the spectra. As can be seen from the spectral plots in Figs.~\ref{figure:spec1}~\&~\ref{figure:spec2}, the low resolution of the SpeX prism compared to the other instruments used in this work is clear. The later-type UCDs have much noisier spectra, so the highest and lowest wavelength ends of the spectra have been removed from the plots since there is too much noise to make use of the flux these regions.

With simple visual inspection, it can be seen that most of the UCD spectra follow the sequence that would be expected from the progression from M-type through T-type classifications, though there are a few UCDs with unusual spectral shapes: J1143$+$5324 has a relatively triangular-shaped \textit{H}-band; J1243$+$6001 has an overall positive gradient in its spectrum, and J1646$-$2115 has a very blue NIR spectrum. There are also UCDs with unusual spectral features: J0811$+$1855 and J1637$+$1813 have deep FeH lines in the \textit{J}-band, and J0542$+$0041 and J0811$+$1855 have strong TiO features in the \textit{J}-band. Each of these UCDs will be investigated individually in Section~\ref{section:indiv}.

Some of the spectra appear to be interesting, but this is caused by poor signal-to-noise ratios due to inadequate observing conditions. J0900$+$5205 has an unusual shape to its \textit{J}-band and comparatively flat \textit{H} and \textit{K}-bands, however the overall shape of the spectrum and strengths of the spectral features are not dissimilar to that of a typical late-M UCD. J0948$+$5300 has a relatively flat \textit{H}-band, and J1700$-$4048 has a spectrum which is flatter and smoother than expected in its entirety (although this may be a consequence of poor observing conditions rather than its intrinsic spectral characteristics). All of these spectra are likely to be victims of poor conditions during the observations: adverse weather conditions such as cloud cover or wind gusts can cause poor signal-to-noise ratios in observations, hence causing the spectra observed to be difficult to characterise once the noise is smoothed out.

\subsection{Comparisons with Other Available NIR Spectra}
\label{section:spec.comp}
\begin{figure}
    \centering
    \includegraphics[width=\linewidth]{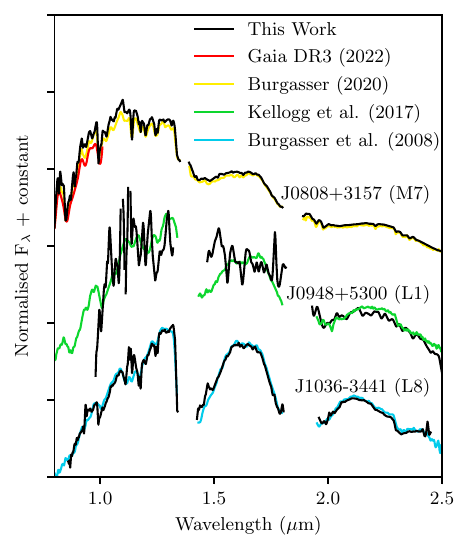}
    \caption{Comparison of the spectra obtained in this work with previously published NIR spectra of the same UCDs.}
\label{figure:compare}
\end{figure}

Using the GUCDS Data Browser (\citealt{cooper2023}) and the SIMPLE Database (\citealt{cruz2023}), any available spectra for the UCDs can be found and compared with those in this work. There are a handful of UCDs with published NIR spectra: J0808$+$3157, J0948$+$5300 and J1036$-$3441. Fig.~\ref{figure:compare} shows a comparison between the spectra obtained in this work with those previously published. Generally, all of the previously published spectra match those in Figs.~\ref{figure:spec1}~\&~\ref{figure:spec2}.

J0808$+$3157 was observed in 2020 (program 2020B057, PI: Smart), and although the spectrum has not been included in any publications it can be found in the GUCDS Data Browser and NASA/IPAC Infrared Science Archive\footnote{A summary of the observation can be found at \url{https://irsa.ipac.caltech.edu/data/IRTF/2020B/20201124/summary/2020B057/sbd_20201124_140156/summary.html}}. This spectrum was obtained using the SpeX prism configuration, thus it is directly comparable to the spectrum presented in this work which was taken with the SpeX SXD configuration. There is also \gaia{}~DR3 XP spectral data available for J0808$+$3157, which reaches up to 1.05\,\um{} (\citealt{prusti2016}). As can be seen in Fig.~\ref{figure:compare}, all of the observations show the same general spectral shape expected of an M7-type UCD, with only minor deviations which likely arose due to small differences in the observing conditions and reduction processes.

J0948$+$5300 has a spectrum that was previously obtained using the SpeX prism configuration in 2015 and published by \cite{kellogg2017}. A visual comparison of our spectrum and the one obtained by \cite{kellogg2017} immediately highlights some differences. While the general shapes of the \textit{J}-bands are similar, the \textit{H}-band has significant deviations. The \textit{H}-band of the \cite{kellogg2017} spectrum has a much more pronounced arched shape than that of the spectrum obtained for this work. A major contributor to the differences in the two spectra is likely to be the poor quality of the observation obtained for this work, since the observing conditions for our SpeX SXD observation were poor, resulting in a spectrum which is much noisier than desired. Due to the poor signal-to-noise ratio of our spectrum for J0948$+$5300, it is likely that the spectrum presented by \cite{kellogg2017} is a better representation of UCD's spectral shape.

J1036$-$3441 was observed in 2010 with the SpeX prism configuration, and the results were published by \cite{burgasser2010}. The two spectra are incredibly similar, aside from the differences in resolutions of the instruments and minor deviations due to differences in the reduction methods. This similarity between the spectra strengthens our L9 spectral classification for this UCD.

\begin{figure}
    \centering
    \includegraphics[width=\linewidth]{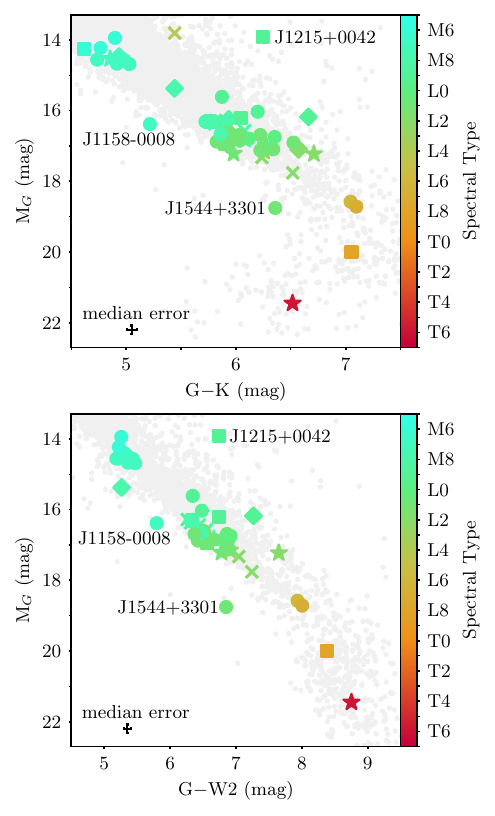}
    \caption{Top: CMD of M$_G$ versus \textit{G}$-$\textit{Ks} for the UCDs. Bottom: CMD of M$_G$ versus \textit{G}$-$\textit{W2} for the UCDs. The colour corresponds to the spectral type of the UCD: blue objects are earlier types and red objects are later types. The median uncertainties are plotted as error bars in the lower left corner of each plot. Grey points are objects from the GUCDS Master list, showing the expected distribution of UCDs on the CMD. Labelled UCDs lie away from the expected distribution and are investigated in Section~\ref{section:indiv}. UCDs plotted as a star $\bigstar$ are potentially young (see Section~\ref{section:ydh.young}), and those plotted as a cross $\bm{\times}$ are thick disc UCDs and subdwarfs (see Section~\ref{section:ydh.sd}). Diamonds $\Diamondblack$ indicate wide binary UCDs (Section~\ref{section:ydh.wide}) and squares $\blacksquare$ denote potential unresolved binaries (Section~\ref{section:ydh.unres}).}
    \label{figure:cmd}
\end{figure}

\section{Colour-Magnitude Diagrams}
\label{section:cmd}
Plotting colour-magnitude diagrams (hereafter CMDs) allows for the redness (or blueness) of each UCD to be visualised and compared to the expectations for UCDs with similar absolute \textit{G}-band magnitudes. This, in turn, means that any of our UCDs which have unusual colours or absolute \textit{G}-band magnitudes can be identified so that they can be investigated in more depth in Section~\ref{section:indiv}.

\subsection{Absolute \textit{G}-band magnitude versus \textit{G}\texorpdfstring{$-$}\textit{Ks} colour}
\label{section:cmd.gk}
By plotting absolute \textit{G}-band magnitude against \textit{G}$-$\textit{Ks} colour in a CMD, a clear relation can be seen (Fig.~\ref{figure:cmd}, top panel). As with the photometric classification outlined in Section~\ref{section:spt.phot}, \gaia{} \textit{G} and 2MASS \textit{Ks} magnitude values are used. Two of our UCDs do not have \textit{Ks} magnitude data available, so are not plotted on this CMD: J1243$+$6001 and J1628$-$4652.

As expected, the brighter UCDs have bluer colours than the fainter UCDs. The shape of the distribution of our UCDs resembles that seen in fig.~10 of \cite{smart2019}, although J1215$+$0042 is brighter than would be expected for a UCD of the same colour. J1158$-$0008 and J1544$+$3301 both lie slightly below the grey points, suggesting that they may be subdwarfs. Section~\ref{section:indiv} discusses these UCDs individually in more detail.

\subsection{Absolute \textit{G}-band magnitude versus \textit{G}\texorpdfstring{$-$}\textit{W2} colour}
\label{section:cmd.gw2}
Using CatWISE2020 \textit{W2} magnitude values, the absolute \textit{G}-band magnitude can be plotted against the \textit{G}$-$\textit{W2} colour in a CMD (Fig.~\ref{figure:cmd}, bottom panel). CatWISE2020 \textit{W2} magnitudes are not available for some of the UCDs: J0850$-$0318, J1628$-$4652, J1700$-$4048 and J2019$+$2256.

This CMD is similar to fig.~11 of \cite{smart2019}, and the trends are comparable, following the expected relation of brighter UCDs having bluer \textit{G}$-$\textit{W2} colours than fainter UCDs. J1215$+$0042 again is brighter than would be expected. J1158$-$0008 and J1544$+$3301 again lie below the grey points, furthering that they could be subdwarfs. These UCDs will be discussed in more detail in Section~\ref{section:indiv}.

\begin{figure}
    \centering
    \includegraphics[width=\linewidth]{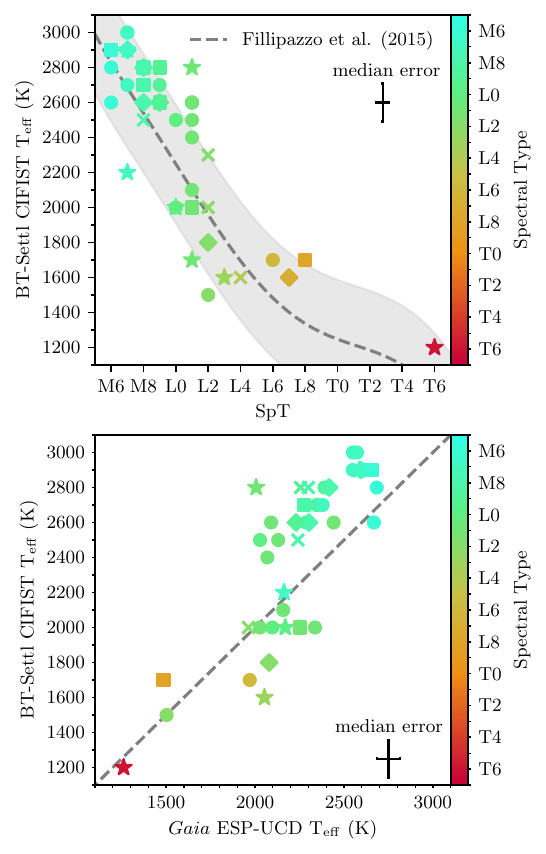}
    \caption{Top: Plot showing estimated effective temperature against spectral type for the UCDs in this paper. The grey dashed line shows the spectral type-temperature relation from \citet{filippazzo2015}, and the shaded region corresponds to their 1$\sigma$ uncertainty. Bottom: comparison of ESP-UCD effective temperature calculations and our effective temperature estimates. The grey dashed line represents the 1:1 ratio of ESP-UCD and BT-Settl CIFIST effective temperature.The median uncertainties are plotted as error bars in the corner of each plot. Colours and symbols are the same as in Fig.~\ref{figure:cmd}.}
    \label{figure:temperature}
\end{figure}

\section{Spectrophotometric Distances}
\label{section:distance}
Using the spectral type-absolute magnitude relations defined by \cite{dupuy2012}, we have been able to calculate spectrophotometric distances for the UCDs in our sample. Since the majority of our UCDs have 2MASS and CatWISE2020 photometry (only five UCDs are missing some photometric data), we calculated spectrophotometric distances using 2MASS \textit{J}, \textit{H} and \textit{K}-band magnitudes as well as CatWISE2020 \textit{W1} and \textit{W2}-band photometry. We take the spectrophotometric distance to be the mean of the distance values, and the uncertainties to be the standard deviation of the values.  (Table~\ref{table:simbad}). These distances can then be compared to the parallactic distances derived from the \gaia{} parallax measurements, thus highlighting any over-bright UCDs (suggestive of unresolved binarity).

\section{Estimating Effective Temperatures}
\label{section:temp}
Effective temperatures are a valuable tool for constraining the properties of UCDs, particularly in categorising potential planetary objects and identifying brown dwarfs. By performing a chi-squared fit of the BT-Settl CIFIST models from \cite{baraffe2015} to each UCD (as outlined by \citealt{marocco2013}), the effective temperature of the UCD can be deemed to be that corresponding to the smallest $\chi^2$. It was possible to use this fitting method to also find an estimate for the surface gravity of the UCD, as \cite{baraffe2015} provides models containing variations in both temperature and \logg, however the surface gravity was kept constant at \logg{} = 5.0\,dex throughout the fitting process to ensure consistency in the fits (and because of the known degeneracy between effective temperatures and \logg; \citealt{kirkpatrick2005}). Uncertainties are derived from the standard deviations of the results of the chi-squared fitting process for each UCD. Table~\ref{table:simbad} shows the effective temperatures for each UCD from the chi-squared fitting.

The top plot in Fig.~\ref{figure:temperature} shows the relation between our estimated temperatures using the BT-Settl CIFIST grid and our spectral classification of each UCD (as in fig.~4 by \citealt{ravinet2024}). The general shape distribution of our estimated temperatures follows the relation from \cite{filippazzo2015}, with the majority of our effective temperature estimates laying within 3$\sigma$ of their relation once effective temperature uncertainties are taken into account (\citealt{filippazzo2015} state that their uncertainty for their spectral type-effective temperature relation is $\sigma = 113$\,K, with this 3$\sigma$ uncertainty shown as the shaded grey region in the top panel of Fig.~\ref{figure:temperature}).

The bottom plot in Fig.~\ref{figure:temperature} shows a comparison between our estimated effective temperatures and those given by the \gaia{} Astrophysical Parameters Inference System (hereafter Apsis; \citealt{bailerjones2013}) module for UCDs called the Extend Stellar Parametrizer - UltraCool Dwarfs (hereafter ESP-UCD). The grey dashed line corresponds to a 1:1 relation between the ESP-UCD effective temperature and our estimates. As can be seen from the plot, our estimated temperatures for the hottest UCDs ($>$\,2000\,K) are higher than those determined by the Apsis module, and there is significant scatter around ${\sim}$2000\,K. A similar trend is seen by \cite{ravinet2024}, with their estimated effective temperatures also being higher than the ESP-UCD temperatures for the hottest UCDs. This difference at the hotter temperatures likely arises due to the correction factor that the ESP-UCD module uses to correct for over-estimations of effective temperatures at in the hotter regime (\citealt{ulla2022}).

\begin{table}
    \centering
    \caption{Spectral lines used in the measurements of radial velocity with both line-centring and cross-correlation methods of \texttt{rvfitter}.}
    \label{table:rvlines}
    \begin{tabular}{c|c}
        \hline
        Line & Wavelength (\um{}) \\
        \hline
        Na\,I-a & 0.8133 \\
        Na\,I-b & 0.8195 \\
        Na\,I-c & 1.1381 \\
        Na\,I-d & 1.1404 \\
        K\,I-a & 1.1690 \\
        K\,I-b & 1.1773 \\
        K\,I-c & 1.2432 \\
        K\,I-d & 1.2522 \\
        \hline
    \end{tabular}
\end{table}

\begin{table}
    \centering
    \caption{Published radial velocity values for UCDs in our sample, along with the difference between our measured radial velocity and those previously published. J0900$+$5205 has no uncertainties associated with its published radial velocity.}
    \label{table:rvpub}
    \begin{tabular}{c|c|c|c}
        \hline
        Object & Published & Measured & Radial Velocity \\
        Short & Radial Velocity & Radial Velocity & Difference \\
        Name & (\kms) & (\kms) & (\kms) \\
        \hline
        \multicolumn{4}{c}{SpeX SXD} \\
        \hline
        J0808$+$3157 & 31.1 $\pm$ 0.06 \footnotesize{$^{[1]}$} & 36.5 $\pm$ 12.9 & 5.4 \\
        J0811$+$1855 & 31.0 $\pm$ 10.6 \footnotesize{$^{[2]}$} & 26.5 $\pm$ 14.6 & 4.5 \\
        J0900$+$5205 & -14.5 \footnotesize{$^{[2]}$} & 5.1 $\pm$ 17.5 & 19.6 \\
        J1423$+$5146 & 28.9 $\pm$ 0.04 \footnotesize{$^{[1]}$} & 33.4 $\pm$ 12.8 & 4.5 \\
        J1544$+$3301 & -32.1 $\pm$ 16.5 \footnotesize{$^{[2]}$} & -8.6 $\pm$ 16.3 & 23.5 \\
        \hline
        \multicolumn{4}{c}{ARCoIRIS} \\
        \hline
        J1536$+$0646 & 41.7 $\pm$ 13.7 \footnotesize{$^{[2]}$} & 32.6 $\pm$ 17.9 & 9.1 \\
        \hline
    \end{tabular}
    \begin{tabular}{c|c|c}
        \textbf{References} &  \footnotesize{$^{[1]}$} -- \cite{jonsson2020} &  \footnotesize{$^{[2]}$} -- \cite{kiman2019} \\
    \end{tabular}
\end{table}

\begin{table}
    \centering
    \caption{The radial velocity values measured using the cross-correlation and line-centring methods for the SXD and ARCoIRIS observations. We take the adopted radial velocity to be the weighted mean of the two methods. Na\,I-a and Na\,I-b lines were not used in the radial velocity measurements for UCDs in \textit{italics} due to excessive noise in the short-wavelength part of the spectra. The adopted radial velocity data is also contained within the collated data table in Appendix~\ref{appendix:fulldata}.}
    \label{table:rv}
    \begin{tabular}{c|c|c|c}
        \hline
        Object & Cross & Line & Adopted \\
        Short & Correlation & Centring & Radial Velocity \\
        Name & (km s$^{-1}$) & (km s$^{-1}$) & (km s$^{-1}$) \\
        \hline
        J0508$+$3319 & 51.5 $\pm$ 18.3 & 46.3 $\pm$ 20.3 & 49.4 $\pm$ 13.6 \\
        \textit{J0515$+$0613} & \textit{74.1 $\pm$ 19.5} & \textit{75.5 $\pm$ 20.3} & \textit{74.8 $\pm$ 14.1} \\
        J0526$-$5026 & 3.5 $\pm$ 25.7 & 0.0 $\pm$ 25.6 & 1.8 $\pm$ 18.2 \\
        \textit{J0542$+$0041} & \textit{-5.5 $\pm$ 23.2} & \textit{-53.7 $\pm$ 17.7} & \textit{-35.9 $\pm$ 14.1} \\
        \textit{J0723$+$4622} & \textit{23.4 $\pm$ 18.5} & \textit{24.4 $\pm$ 19.1} & \textit{23.9 $\pm$ 13.3} \\
        J0808$+$3157 & 39.4 $\pm$ 18.2 & 33.4 $\pm$ 18.4 & 36.5 $\pm$ 12.9 \\
        J0811$+$1855 & 31.1 $\pm$ 21.2 & 22.4 $\pm$ 20.3 & 26.5 $\pm$ 14.6 \\
        \textit{J0817$-$6155} & \textit{-21.9 $\pm$ 25.7} & \textit{-22.9 $\pm$ 18.4} & \textit{-22.7 $\pm$ 4.9} \\
        \textit{J0900$+$5205} & \textit{4.7 $\pm$ 22.4} & \textit{5.6 $\pm$ 27.9} & \textit{5.1 $\pm$ 17.5} \\
        J0911$+$1432 & 15.5 $\pm$ 18.1 & 24.9 $\pm$ 18.5 & 20.1 $\pm$ 13.0 \\
        J0916$-$1121 & -9.6 $\pm$ 18.0 & -12.6 $\pm$ 18.2 & -11.1 $\pm$ 12.8 \\
        \textit{J0941$+$3315A} & \textit{3.5 $\pm$ 17.8} & \textit{-5.6 $\pm$ 18.2} & \textit{-1.0 $\pm$ 12.7} \\
        J0942$-$2551 & 3.5 $\pm$ 18.0 & 5.3 $\pm$ 18.9 & 4.3 $\pm$ 13.0 \\
        \textit{J0948$+$5300} & \textit{37.3 $\pm$ 28.7} & \textit{33.7 $\pm$ 28.6} & \textit{35.5 $\pm$ 20.2} \\
        \textit{J1036$-$3441} & \textit{13.7 $\pm$ 25.2} & \textit{-0.7 $\pm$ 25.6} & \textit{6.6 $\pm$ 18.0} \\
        J1126$-$2706 & 13.8 $\pm$ 19.0 & 7.7 $\pm$ 18.5 & 10.7 $\pm$ 13.3 \\
        J1150$-$2914 & 16.9 $\pm$ 18.4 & -8.4 $\pm$ 18.7 & 4.5 $\pm$ 13.1 \\
        J1152$+$5901 & -1.4 $\pm$ 19.2 & 7.2 $\pm$ 19.8 & 2.7 $\pm$ 13.8 \\
        J1158$-$0008 & 22.4 $\pm$ 25.2 & 21.1 $\pm$ 25.1 & 21.8 $\pm$ 17.8 \\
        \textit{J1158$+$3817} & \textit{20.0 $\pm$ 18.0} & \textit{21.1 $\pm$ 18.2} & \textit{20.4 $\pm$ 12.8} \\
        J1212$+$0206 & -21.5 $\pm$ 25.7 & -17.6 $\pm$ 25.9 & -19.5 $\pm$ 18.3 \\
        \textit{J1215$+$0042} & \textit{-2.8 $\pm$ 25.3} & \textit{-9.4 $\pm$ 25.5} & \textit{-6.1 $\pm$ 17.9} \\
        \textit{J1250$+$0455} & \textit{13.5 $\pm$ 25.8} & \textit{12.0 $\pm$ 28.7} & \textit{12.9 $\pm$ 19.2} \\
        J1252$+$0347 & 18.7 $\pm$ 25.2 & 13.1 $\pm$ 25.3 & 15.9 $\pm$ 17.9 \\
        J1307$+$0246 & 14.0 $\pm$ 25.1 & 8.6 $\pm$ 25.2 & 11.4 $\pm$ 17.8 \\
        J1313$+$1404 & -21.3 $\pm$ 26.0 & -60.8 $\pm$ 21.0 & -45.2 $\pm$ 16.3 \\
        J1315$+$3232 & 36.0 $\pm$ 18.3 & 35.0 $\pm$ 18.1 & 35.5 $\pm$ 12.9 \\
        J1320$+$4238 & -30.2 $\pm$ 18.0 & -37.7 $\pm$ 18.3 & -33.9 $\pm$ 12.9 \\
        J1420$+$3235 & -11.7 $\pm$ 17.8 & -21.6 $\pm$ 18.5 & -16.4 $\pm$ 12.8 \\
        J1423$+$5146 & 35.5 $\pm$ 18.1 & 31.2 $\pm$ 18.2 & 33.4 $\pm$ 12.8 \\
        \textit{J1441$+$4217} & \textit{-23.5 $\pm$ 19.5} & \textit{-24.9 $\pm$ 19.8} & \textit{-24.2 $\pm$ 13.9} \\
        J1452$+$0931 & -12.2 $\pm$ 25.0 & -8.7 $\pm$ 25.2 & -10.4 $\pm$ 17.7 \\
        \textit{J1514$+$3547} & \textit{1.2 $\pm$ 20.3} & \textit{15.8 $\pm$ 19.3} & \textit{8.9 $\pm$ 14.0} \\
        J1536$+$0646 & 32.9 $\pm$ 25.1 & 32.3 $\pm$ 25.5 & 32.6 $\pm$ 17.9 \\
        \textit{J1544$+$3301} & \textit{-16.3 $\pm$ 22.2} & \textit{0.4 $\pm$ 24.1} & \textit{-8.6 $\pm$ 16.3} \\
        J1628$-$4652 & -21.7 $\pm$ 25.2 & -22.4 $\pm$ 25.2 & -22.0 $\pm$ 17.8 \\
        J1713$-$3952 & -34.0 $\pm$ 18.6 & -44.5 $\pm$ 19.7 & -39.0 $\pm$ 13.5 \\
        \textit{J1737$+$4705} & \textit{0.8 $\pm$ 19.4} & \textit{-3.3 $\pm$ 18.9} & \textit{-1.3 $\pm$ 13.6} \\
        J1847$-$3419 & -47.9 $\pm$ 20.1 & -52.9 $\pm$ 24.1 & -49.9 $\pm$ 15.4 \\
        J1938$+$4321 & 28.1 $\pm$ 13.9 & 2.5 $\pm$ 18.2 & 21.1 $\pm$ 12.8 \\
        \hline
    \end{tabular}
\end{table}

\begin{figure}
    \centering
    \includegraphics[width=\linewidth]{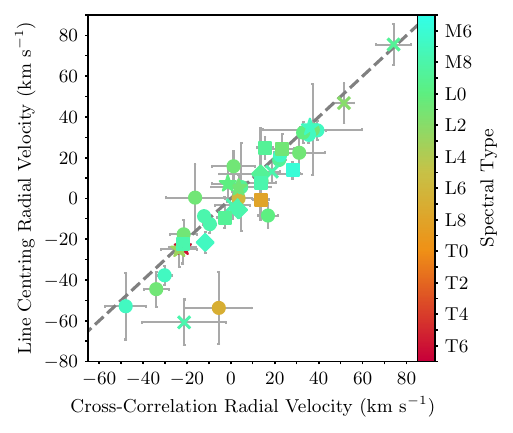}
    \caption{Comparison of line centring radial velocity and cross-correlation radial velocity measurements. The grey dashed line represents the 1:1 ratio of line centring and cross-correlation values. Colours and symbols are the same as in Fig.~\ref{figure:cmd}.}
    \label{figure:rv}
\end{figure}

\section{Calculating Radial Velocities}
\label{section:rv}
Using the \textsc{Python} module \texttt{rvfitter}\footnote{{The \texttt{rvfitter} module for \textsc{Python} is publicly available from: \url{https://github.com/Will-Cooper/rvfitter}}} (\citealt{cooper2022}), the radial velocities of the UCDs observed with the higher-resolution instruments (i.e. SpeX SXD and ARCoIRIS observations) can be measured with two methods: line-centring and cross-correlation.

The lines used to measure the radial velocity all lie in the \textit{J}-band due to the shorter wavelengths having the best signal-to-noise in our observations. The lines are shown in Table~\ref{table:rvlines}. All measurements of radial velocity used the Na\,I-c, Na\,I-d, K\,I-a, K\,I-b, K\,I-c and K\,I-d lines, while the Na\,I-a and Na\,I-b lines were only used when the signal of the spectrum was sufficient to clearly distinguish the features from the continuum (detections were deemed to be sufficient by-eye, approximating a 3$\sigma$ detection). UCDs with radial velocities measured without using the Na\,I-a and Na\,I-b lines are italicised in Table~\ref{table:rv}. The line-centring routine fits a continuum to the spectrum around the spectral line, and a Gaussian is fitted to the manually-identified feature, thus giving a radial velocity measurement for each line based on its shift from its position in air (taken from the NIST Atomic Spectra Database; \citealt{kramida2023}). The cross-correlation measurements are made in a similar way, finding the best-fitting radial velocity-shifted BT-Settl CIFIST spectrum by eye for each spectral line. Comparing the measured radial velocity values reveals that the two methods are generally consistent with each other (Fig.~\ref{figure:rv}).

The \texttt{rvfitter} module gives an uncertainty for each of the radial velocity measurements, which is calculated using inverse variance weighting equations (see section~4.3.3 of \citealt{cooper2024}). This uncertainty is added in quadrature with the median difference in order to account for systematic uncertainties, such as instrumental drift and undersampling. The median differences for the SpeX SXD and ARCoIRIS instruments are 17.7\,\kms{} and 25.0\,\kms, respectively (see Table~\ref{table:rvpub} for the published radial velocities). This is comparable to the ${\sim}$20\,\kms radial velocity uncertainty associated with the typical signal-to-noise ratio of ${\sim}$30 in our observations.

Barycentric corrections are applied to the measured radial velocities: calculating a correction factor using the \texttt{Astropy}\footnote{{The \texttt{Astropy} module for \textsc{Python} is publicly available from: \url{https://docs.astropy.org/en/stable/install.html}}} module for \textsc{Python} and adding it multiplicatively. Equation~(\ref{equation:bary}) shows how the correction factor is added to the radial velocity measurements; $v_{r}$ is the barycentric-corrected radial velocity, $v_{meas}$ is the measured radial velocity, $v_{corr}$ is the correction factor and $c$ is the speed of light in a vacuum.

\begin{equation}
    v_{r} = v_{meas} + v_{corr} + \frac{v_{meas} \cdot v_{corr}}{c}
    \label{equation:bary}
\end{equation}

Table~\ref{table:rv} shows the values measured using the two methods. Overall, the values measured using the two methods are consistent with each other, with only two UCDs lying away from the 1:1 line of Figure~\ref{figure:rv}. These UCDs are J0542$+$0041 and J1313$+$1404, both of which have their \textit{J}-band spectral features hidden by noise -- this is likely the cause of the inconsistencies between the radial velocity values measured using the two methods.

The radial velocity measurements presented in Table~\ref{table:rv} can be used in Section~\ref{section:indiv} to help confirm young moving group memberships, as well as for calculating the space velocities of potential thick disc and halo objects.

\begin{figure}
    \centering
    \includegraphics[width=\linewidth]{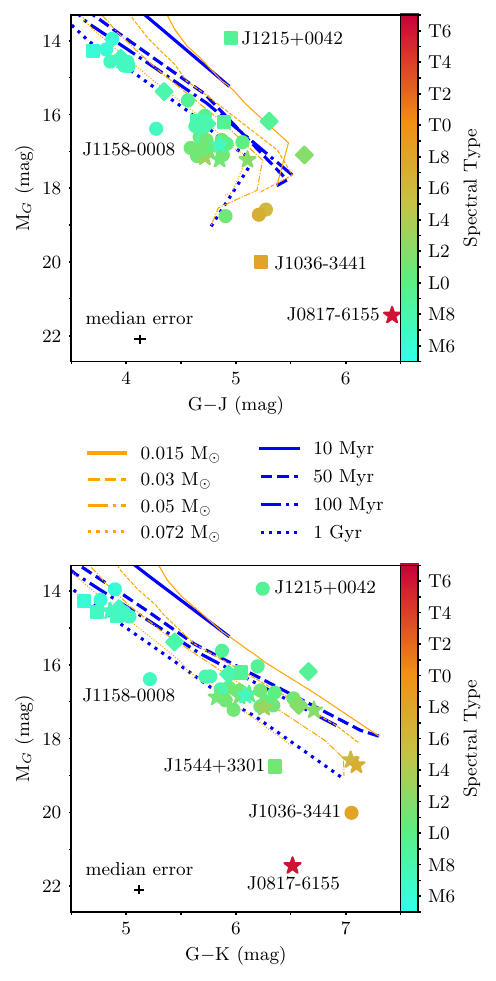}
    \caption{Top: cooling track plot showing BHAC15 mass (orange) and age (blue) relations for M$_G$ versus \textit{G}$-$\textit{J} (\citealt{baraffe2015}). Bottom: cooling track plot showing BHAC15 mass (orange) and age (blue) relations for M$_G$ versus \textit{G}$-$\textit{Ks} for the UCDs (\citealt{baraffe2015}). The median uncertainties are plotted as error bars in the corner of each plot. Suspected thick disc and subdwarf UCDs (see Section~\ref{section:ydh.sd}) are not plotted, since the BHAC15 isochrones assume solar metallicities. Colours and symbols are the same as in Fig.~\ref{figure:cmd}.}
    \label{figure:cooling}
\end{figure}

\section{Young, Thick Disc, Subdwarf or Binary?}
\label{section:ydh}
\subsection{Identification of Young Objects}
\label{section:ydh.young}
The \banyan{} tool (\citealt{gagne2018}) can be used to give a classification of young group membership for each of our UCDs. Table~\ref{table:ydh} shows the most probable classification given by \banyan{} for each UCD, along with the probability of the UCD being a member of the young moving group. Most of the UCDs are classified as being field objects, i.e. not likely to be a member of any of the young groups that the \banyan{} tool considers. The probabilities assigned by \banyan{} to each nearby moving group for a UCD are used to classify its membership into one of six categories based on a probability threshold. If the assigned probability exceeds 90\,per\,cent, the UCD's group membership is classified as Bona Fide. This classification additionally requires a strong kinematic match, complete three-dimensional velocity information, and independent indicators of youth, such as X-ray emission, lithium depletion signatures, or a position on the colour-magnitude diagram consistent with the estimated age of the moving group (see section~7 of \citealt{gagne2018} for further details). Using this threshold probability, we find seven UCDs to be likely members of young moving groups: J0942$-$2551, J1152$+$5901, J1441$+$4217, J1544$-$0435, J1544$+$3301, J1637$+$1813 and J2019$+$2256.

As outlined by \cite{sarro2023}, low gravity is another indicator of young objects. \splat{} can be used to determine each UCD's gravity classification, making use of the index-based method outlined by \cite{allers2013}: classifying the gravity of the UCD from its VO-band, FeH-bands, Na\,I lines, K\,I lines and H-band continuum indices. This \cite{allers2013} method was created for spectral types M5--L6, so any UCDs outside of this spectral type range do not have surface gravity classifications. A low surface gravity classification (hereafter VL-G) is associated with objects with ages ${\sim}$10--30\,Myr, while a classification of intermediate surface gravity (hereafter INT-G) is associated with objects with ages ${\sim}$50--200\,Myr. Table~\ref{table:ydh} shows the UCDs with non-field gravity classifications of INT-G and VL-G: J0900$+$5205, J0948$+$5300, J1152$+$5901, J1315$+$3232, J1544$-$0435, J1646$-$2115 and J1700$-$4048.

As discussed by \cite{allers2013}, visual inspection of a UCD's spectrum can help verify the UCD's youth -- young UCDs have a triangular \textit{H}-band and a positive slope in the \textit{Ks}-band. In our sample, there are two UCDs with such spectral shapes: J1243$+$6001 and J1544$-$0435. More detailed inspection of the spectra can reveal more features of youth: weak FeH bands at 0.99, 1.20 and 1.55\,\um{} (\citealt{allers2013}) and strong TiO features at ${\sim}$0.78 and ${\sim}$0.82\,\um{} in the \textit{J}-band (\citealt{mcgovern2004}) are indicators of very young objects, with J1544$-$0435 and J2019$+$2256 having such features.

Cooling tracks are useful tools for distinguishing stellar UCDs from sub-stellar UCDs. Using the BHAC15 isochrones from \cite{baraffe2015}, evolutionary age and mass tracks can be plotted on CMDs, showing the distribution of the UCDs across the BHAC15 models. All of the UCDs have 2MASS \textit{J}-band photometry available, and only two UCDs do not have 2MASS \textit{Ks}-band photometry (J1243$+$6001 and J1628$-$4652, as mentioned in Section~\ref{section:cmd.gk}). There are no available BHAC15 isochrones for CatWISE2020 photometry, hence a cooling track plot for \textit{G}$-$\textit{W2} (which would allow for a more comprehensive comparison with the CMDs in Fig.~\ref{figure:cmd}) cannot be plotted. The suspected thick disc and subdwarf objects (see Section~\ref{section:ydh.sd}) are not included in this plot, since the BHAC15 isochrones assume solar metallicities and we suspect these three UCDs to be located in the thick disc (see Section~\ref{section:ydh.sd}).

The BHAC15 isochrones are valid for temperatures ${\sim}$2000--6000\,K, thus cooler UCDs with later spectral types (i.e. mid-late L-dwarfs and T-dwarfs) cannot be accurately represented by the cooling tracks. Our sample is comprised primarily of late-M and early-L type UCDs, thus the BHAC15 isochrones are suitable for characterising the majority of our UCDs.

As can be seen in Fig.~\ref{figure:cooling}, the majority of the UCDs in this work are old and stellar in nature, while the few dim red UCDs in the lower right corners of the plots are almost certain to be sub-stellar objects. These UCDs are J0817$-$6155 and J1036$-$3441, which have late-L and T-type classifications. J1215$+$0042 is located above the cooling tracks in Fig.~\ref{figure:cooling}, suggesting that it could be an unresolved binary. J1158$-$0008 and J1544$+$3301 both lie further to the left of the isochrones than any of the other UCDs, suggesting that they may be binaries since they are more massive than would be expected based on their spectral classifications, or subdwarfs with relatively old ages. As with the UCDs highlighted in the CMDs, all of these UCDs will be investigated in more detail in Section~\ref{section:indiv}.

By combining the \splat{} gravity classifications with visual inspection and positions on the cooling track plots of Fig.~\ref{figure:cooling}, the UCDs with convincing characteristics suggestive of youth will be investigated further in Section~\ref{section:indiv.young}: J0526$-$5026, J0817$-$6155, J0942$-$2551, J1152$+$5901, J1315$+$3232, J1441$+$4217, J1544$-$0435, J1637$+$1813 and J1847$-$3419.

\begin{figure*}
\includegraphics[width=0.95\linewidth]{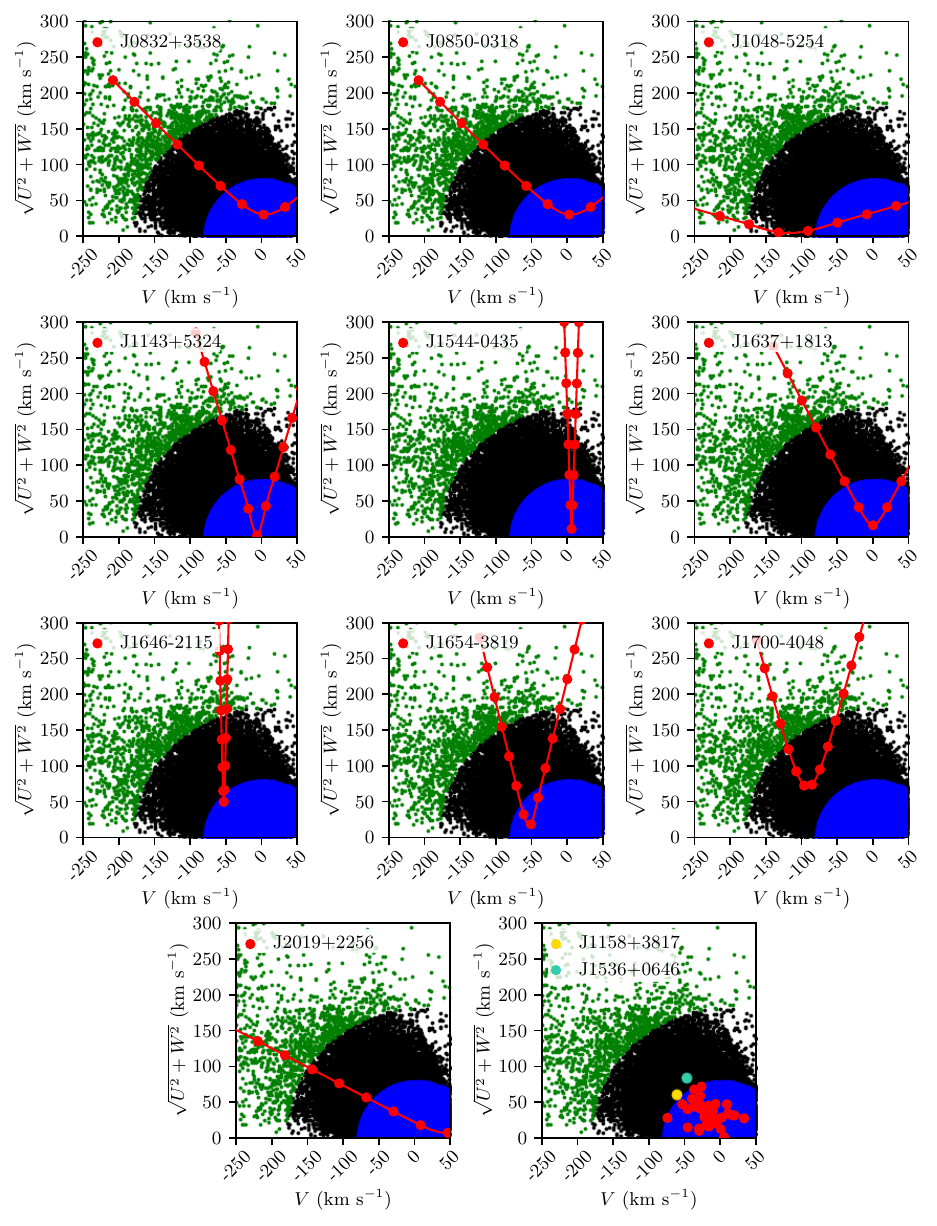}
    \caption{Toomre diagrams for our UCDs sample. UCDs without radial velocity measurements are plotted individually by allowing radial velocities between $\pm$300\,\kms. The final panel shows the UCDs with measured radial velocities. Red objects are likely thin disc members, while other colours indicate thick disc candidates. Background sources are from GCNS: thin disc members are plotted in blue, thick disc in black and halo in green (\citealt{smart2021}).}
    \label{figure:toomre-all}
\end{figure*}

\subsection{Identification of Thick Disc and Subdwarf Objects}
\label{section:ydh.sd}
Subdwarfs are low-metallicity Population~II objects, crucial to the understanding of interior structures at the sub-stellar mass limit and star formation in the early Universe. As outlined by \cite{zhang2017}, subdwarfs include both the lowest mass metal-poor stars and brown dwarfs with subsolar metallicities. Subdwarfs can be characterised by their NIR spectral features: weak metal oxides, strong metal hydrides and enhanced collision-induced H$_2$ absorption (see \citealt{bates1952, borysow1989, borysow2002, abel2012, saumon2012, zhang2013a}, among others). Potential subdwarf objects can be identified by visually comparing each spectrum with fig.~7 from \cite{greco2019}. The old nature of subdwarfs means that they belong to the Milky Way's thick disc and halo and can be identified by their surface gravity classification: thick disc objects and subdwarfs often have higher surface gravity than younger thin disc UCDs (\citealt{martin2017}). This means that the \splat{} gravity classifications shown in Table~\ref{table:ydh} can be used to aid the identification of thick disc and subdwarf objects.

Space velocities are also an indicator of an object being located in the galactic disc. Using the criteria outlined by \cite{nissen2004}, UCDs which are likely to be located in the thick disc (v$_\text{tot}\,>$ 85\,\kms) and halo (v$_\text{tot}\,>$ 180\,\kms) can be identified using their total space velocities. Our radial velocity measurements can be used to calculate $UVW$ space velocities for the UCDs observed with the SpeX SXD and ARCoIRIS instruments, while the UCDs without radial velocity measurements have space velocities estimated by allowing their radial velocity to vary between -300\,\kms{} and 300\,\kms, as by \cite{ravinet2024}. Fig.~\ref{figure:toomre-all} shows the positions of each UCD on a Toomre plot of the sources in the \gaia{} Catalogue of Nearby Stars (hereafter GCNS; \citealt{smart2021}). UCDs without radial velocity measurements are plotted individually in red; all of these UCDs have inconclusive positions on the Toomre diagram, with the exception of J1700$-$4048 which is likely to be a member of the thick disc or galactic halo. This is supported by the UCD's relatively high tangential velocity (112.75 $\pm$ 33.82\,\kms). Of the UCDs with radial velocity measurements, there are two which lie within the thick disc region of the Toomre diagram in Fig.~\ref{figure:toomre-all}: J1158$-$3817 and J1536$+$0646.

Using these criteria, and the positions of the UCDs on the CMDs of Fig.~\ref{figure:cmd}, four objects in our sample can be further investigated as being a potential thick disc UCD or subdwarf in Section~\ref{section:indiv.sd}: J0515$+$0613, J1536$+$0646 and J1700$-$4048.

\begin{figure*}
\centering
    \includegraphics[width=\linewidth]{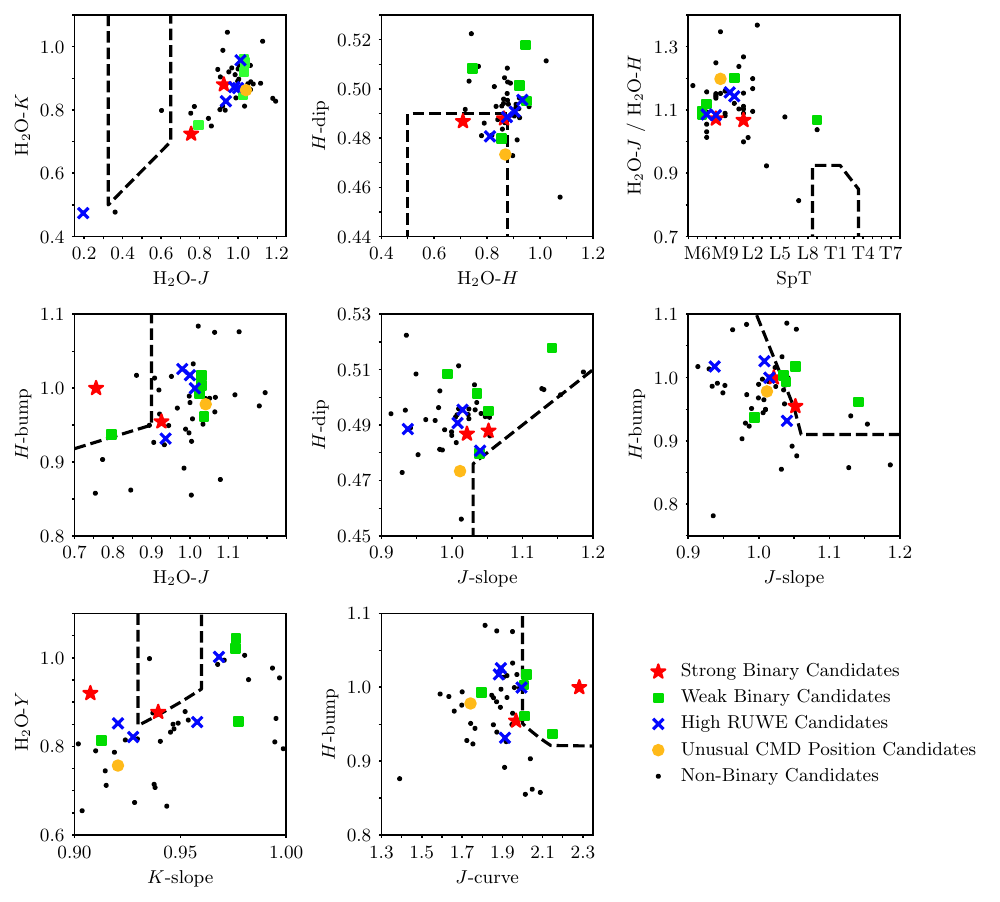}
    \caption{Spectral index relations for unresolved binary candidate selection. Dashed lines indicate the selection areas, red stars \textcolor{red}{$\bigstar$} indicate strong objects and green squares \textcolor{LimeGreen}{$\blacksquare$} indicate weak objects. Blue crosses \textcolor{blue}{$\bm{\times}$} denote objects with high RUWE values, and yellow circles \textcolor{Gold}{$\CIRCLE$} are objects with unusual positions on the CMDs of Fig.~\ref{figure:cmd}.}
    \label{figure:specindex}
\end{figure*}

\subsection{Identification of Wide Binary Objects}
\label{section:ydh.wide}
Using the methods outlined by \cite{smart2019}, we found that some of the UCDs have secondary (or tertiary) associated \gaia{} objects. We used the criteria outlined by \cite{baig2024} to identify systems with common proper motion, allowing for perturbations invoked by orbital motion. We also use the false positive probability equations as laid out in section 2.4.1 of \cite{baig2024} to confirm the companionships of our binary systems.

The UCDs in our sample that we identify as being in wide binaries are: J0850$-$0318, J0941$+$3315A, J1243$+$6001, J1250$+$0455, J1420$+$3235, and J2019$+$2256. A summary of the UCDs and their wide companions can be found in Table~\ref{table:cpm}.

\subsection{Identification of Unresolved Binary Objects}
\label{section:ydh.unres}
The \gaia{} re-normalised unit weight error value (hereafter RUWE) can be used as an indication of whether the UCD is likely to be an unresolved binary (for a single-star source, the RUWE value is expected to be 1.0, while a significantly larger value is often an indication that the source is non-single). In this sample, we use a RUWE cut-off value of 1.2 to identify potential binary objects (\citealt{tofflemire2021}). There are a handful of potential binary UCDs with RUWE values above 1.2 in our sample: J0817$-$6155, J1036$-$3441, J1126$-$2706 and J1423$+$5146. In addition, there is a brighter-than-expected UCD which may also be considered to be a potential unresolved binary candidate: J1215$+$0042.

\cite{bardalezgagliuffi2014} outlines a method which makes use of spectral indices to identify potential unresolved binary objects. Using the indices and selection criteria they present (Tables~\ref{table:specind}~\&~\ref{table:bincrit}), the spectral index relations can be plotted and any potential unresolved binaries can be identified (Fig.~\ref{figure:specindex}). With these indices, we identified two strong candidates (which satisfy three or more of the selection criteria) and five weak candidates (which satisfy two of the selection criteria; see Table~\ref{table:binary} for number of spectral indices satisfied by each unresolved binary candidate).

Using methods similar to those outlined by \cite{burgasser2010} and \cite{marocco2015}, we created a set of binary template spectra from the M0--T9 standard from the library of spectral standards in the SpeX Prism Library (hereafter SPL, \citealt{burgasser2014}). Some of these standards are known to be variable or binary systems, in these cases we used alternative objects with consistent NIR and optical spectral classifications to create our templates. The objects used in the creation of our template binary spectra are presented in Table~\ref{table:standards}. We performed absolute flux calibration of the standard spectra using 2MASS \textit{J}-band magnitudes: first interpolating the 2MASS \textit{J}-band filter curve to match each standard UCD spectrum and calculating a weighted flux before integrating under the filter curve to give the new absolute-calibrated standard spectra. We then combined these standard spectra to create our binary templates.

By performing a chi-squared fit of the template binary spectra to the spectra of the UCDs, we can find the most probable combination of objects which make up the unresolved binary system. We perform our fits using the wavelengths 0.9--1.35\,\um, 1.45--1.8\,\um{} and 2.0--2.35\,\um{} in order to avoid noise introduced by water bands. The chi-squared values of the binary template fits are shown in Table~\ref{table:binary}.

We also calculated a one-sided F-test for each unresolved binary candidate following the method presented by \cite{burgasser2010}. Equation~(\ref{equation:ftest}) is the distribution statistic, where we assume $\nu_{\text{single}} = \nu_{\text{binary}} = \nu$, and we take $\nu$ to be the number of data points used in the chi-squared fit minus one (as in \citealt{burgasser2010}). In order to achieve a 90\,per\,cent confidence level in the binary fit, we require $\eta_{\text{SB}}\,>\,1.1$. The values of $\eta_{\text{SB}}$ are presented in Table~\ref{table:binary}.

\begin{equation}
    \eta_{\text{SB}} = \frac{{\chi_{\text{single,min}}^2}/{\nu_{\text{single}}}}{{\chi_{\text{binary,min}}^2}/{\nu_{\text{binary}}}}
    \label{equation:ftest}
\end{equation}

The potential binarity of each of these unresolved binary candidates is investigated in Section~\ref{section:indiv.unres}.

\begin{table}
    \centering
    \caption{Objects used in the creation of the template binary spectra. Objects are taken from the SPL with known variables and binaries are substituted for alternative objects.}
    \label{table:standards}
    \begin{tabular}{l|c}
        \hline
        \multicolumn{1}{c}{Object} & SpT \\
        \hline
        Gl270 & M0 \\
        IRAS F04196$+$2638 & M1 \\
        Gl91 & M2 \\
        Gl752A & M3 \\
        2MASS J03452021$+$3217223 & M4 \\
        2MASS J01294256$-$0823580 & M5 \\
        2MASS J00130931$-$0025521 & M6 \\
        vB 8 & M7 \\
        2MASS J00552554$+$4130184 & M8 \\
        LHS 2924 & M9 \\
        2MUCD 20165 & L0 \\
        SIPS J2130$-$0845 & L1 \\
        2MASS J00062250$+$1300451 & L2 \\
        2MASSW J1506544$+$132106 & L3 \\
        2MASS J21580457$-$1550098 & L4 \\
        SDSS J083506.16$+$195304.4 & L5 \\
        2MASSI J1010148$-$040649 & L6 \\
        SDSS J104409.43$+$042937.6 & L7 \\
        2MASSW J1632291$+$190441 & L8 \\
        2MASS J02550357$-$4700509 & L9 \\
        2MASS J00164364$+$2304262 & T0 \\
        SDSSp J083717.22$-$000018.3 & T1 \\
        WISE J045746.08$-$020719.2 & T2 \\
        2MASS J01383648$-$0322181 & T3 \\
        2MASS J08195820$-$0335266 & T4 \\
        2MASS J00454267$+$3611406 & T5 \\
        SDSSp J162414.37$+$002915.6 & T6 \\
        2MASSI J0727182$+$171001 & T7 \\
        2MASSI J0415195$-$093506 & T8 \\
        UGPS J072227.51$-$054031.2 & T9 \\
        \hline
    \end{tabular}
\end{table}

\begin{table}
    \centering
    \caption{Unresolved binary candidates from index-based selection, objects with RUWE values $>$\,1.2, and objects with unusual CMD positions.}
    \label{table:binary}
    \begin{tabular}{c|c|c|c|c|c}
        \hline
        Object & Single & N$^\text{\underline{o}}$ of Index & Best-fitting & Binary & $\eta_{\text{SB}}$ \\
        Short Name & SpT & Criteria & Binary & $\chi^2$ Value & Value \\
        \hline
        \multicolumn{6}{c}{Strong Candidates} \\
        \hline
        J1544$+$3301 & L2 & 3 & L0 $+$ L2 & 2.253 & 0.744 \\
        J1628$-$4652 & M8 & 3 & M9 $+$ L9 & 3.720 & 1.133 \\
        \hline
        \multicolumn{6}{c}{Weak Candidates} \\
        \hline
        J0832$+$3538 & M6 & 2 & M6 $+$ L6 & 7.479 & 0.302 \\
        J0911$+$1432 & M9 & 2 & M8 $+$ L3 & 2.445 & 1.076 \\
        J1036$-$3441 & L8 & 2 & L5 $+$ T1 & 1.676 & 1.875 \\
        J1320$+$4238 & M7 & 2 & M6 $+$ M7 & 3.263 & 1.025 \\
        J1938$+$5146 & M6 & 2 & M7 $+$ M9 & 3.512 & 0.839 \\
        \hline
        \multicolumn{6}{c}{High RUWE Objects} \\
        \hline
        J0723$+$4622 & L1 & 1 & M8 $+$ L7 & 1.574 & 2.229 \\
        J0817$-$6155 & T6 & 1 & T0 $+$ T6 & 9.142 & 0.115 \\
        J1126$-$2706 & M8 & 0 & M8 $+$ L1 & 1.073 & 1.244 \\
        J1423$+$5146 & M7 & 0 & M6 $+$ M7 & 4.494 & 0.974 \\
        J1536$+$0646 & L0 & 1 & L1 $+$ T4 & 2.048 & 0.550 \\
        \hline
        \multicolumn{6}{c}{Unusual CMD Position Objects} \\
        \hline
        J1215$+$0042 & M9 & 1 & M8 $+$ L9 & 3.601 & 1.022 \\
        \hline
    \end{tabular}
\end{table}

\begin{table*}
    \centering
    \caption{Spectral indices used for identification of potential unresolved binary objects.}
    \label{table:specind}
    \begin{tabular}{c|c|c|c|c}
        \hline
        Spectral & Numerator & Denominator & Spectral & \multirow{2}{4em}{Reference} \\
        Index & Range (\um) & Range (\um) & Feature & -- \\
        \hline
        H$_2$O-\textit{Y} & 1.04--1.07 & 1.14--1.17 & 1.15\,\um{} H$_2$O & 1 \\
        H$_2$O-\textit{J} & 1.14--1.165 & 1.26--1.285 & 1.15\,\um{} H$_2$O & 2 \\
        H$_2$O-\textit{H} & 1.48--1.52 & 1.56--1.60 & 1.4\,\um{} H$_2$O & 2 \\
        H$_2$O-\textit{K} & 1.975--1.995 & 2.08--2.10 & 1.9\,\um{} H$_2$O & 2 \\
        \textit{K} / \textit{J} & 2.06--2.10 & 1.25--1.29 & \textit{J}-\textit{K} colour & 2 \\
        \textit{J}-slope & 1.27--1.30 & 1.30--1.33 & 1.28\,\um{} flux peak shape & 1 \\
        \textit{J}-curve & 1.04--1.07 $+$ 1.26--1.29\,\footnotesize{$^{a}$} & 1.14--1.17 & Curvature across \textit{J}-band & 1 \\
        \textit{H}-bump & 1.54--1.57 & 1.66--1.69 & Slope across \textit{H}-band peak & 1 \\
        \textit{H}-dip & 1.61--1.64 & 1.56--1.59 $+$ 1.66--1.69\,\footnotesize{$^{a}$} & 1.634\,\um{} FeH/CH$_4$ & 3 \\
        \textit{K}-slope & 2.06--2.10 & 2.10--2.14 & \textit{K}-band shape/CIA H$_2$ & 4 \\
        \hline
    \end{tabular}
    \vspace{-2ex}
    \begin{tabular}{l l l l l}
        \textbf{Notes} & \multicolumn{4}{l}{\footnotesize{$^{a}$} -- Average of the two wavelength ranges} \\
        \textbf{References} & 1 -- \cite{bardalezgagliuffi2014} & 2 -- \cite{burgasser2006} & 3 -- \cite{burgasser2010} & 4 -- \cite{burgasser2002} \\
    \end{tabular}
\end{table*}

\begin{table*}
    \centering
    \caption{Selection criteria used for identification of potential unresolved binary objects in different parameter spaces.}
    \label{table:bincrit}
    \begin{tabular}{c|c|l|c}
        \hline
        Abscissa ($x$) & Ordinate ($y$) & \multicolumn{1}{c}{Limits} & N$^\text{\underline{o}}$ Sources \\
        \hline
        H$_2$O-\textit{J} & H$_2$O-\textit{K} & Intersection of: \(y = 0.615 x + 0.3\), \(x = 0.325\) and \(x = 0.65\) & 1 \\
         &  & Select points in upper middle &  \\
        H$_2$O-\textit{H} & \textit{H}-dip & Intersection of: \(y = 0.49\), \(x = 0.5\) and \(x = 0.875\) & 12 \\
         &  & Select points in lower middle &  \\
        SpT & H$_2$O-\textit{J}/H$_2$O-\textit{H} & Intersection of: \(y = 0.95\), \(y = -0.0375 x + 1.731\), \(x = \text{L}8.5\) and \(x = \text{T3.5}\) & 0 \\
         &  & Where L0 = 10, T0 = 20 etc. Select points in lower middle &  \\
        H$_2$O-\textit{J} & H-bump & Intersection of: \(y = 0.16 x + 0.806\) and \(x = 0.90\) & 4 \\
         &  & Select points in upper left corner &   \\
        \textit{J}-slope & \textit{H}-dip & Intersection of: \(y = 0.20x + 0.27\) and \(x = 1.03\) & 0 \\
         &  & Select points in lower right corner &  \\
        \textit{J}-slope & \textit{H}-bump & Intersection of: \(y = -2.75x + 3.84\) and \(y = 0.91\) & 12 \\
         &  & Select points in upper right corner &  \\
        \textit{K}-slope & H$_2$O-\textit{Y} & Intersection of: \(y = 12.036x^2 - 20.000x + 8.973\), \(x = 0.93\) and \(x = 0.96\); \(\sigma = 0.064\) & 3 \\
         &  & Select points above 1$\sigma$ curve in upper middle &  \\
        \textit{J}-curve & \textit{H}-bump & Intersection of: \(y = 0.269x^2 - 1.326x + 2.479\), \(y = 2.00\) and \(x = 0.92\); \(\sigma = 0.048\) & 6 \\
         &  & Select points above 1$\sigma$ curve in upper right corner &  \\
        \hline
    \end{tabular}
\end{table*}

\section{Analysis of Specific Objects}
\label{section:indiv}
Most UCDs in our sample have RUWE values $<$\,1.2 and v$_\text{tan}\,<$\,85\,\kms, typical of solo objects located in the thin disc, however some UCDs can be discussed in a little more detail; here we look at the UCDs which have been identified as being interesting in Sections~\ref{section:spec}--\ref{section:ydh}, mainly by considering their \banyan{} classification, \splat{} gravity classification, spectral features, RUWE and tangential and radial velocities. By systematically inspecting various features (summarised in Table~\ref{table:ydh}) the characteristics of these UCDs can be determined. The majority of the UCDs have field \banyan{} and \splat{} gravity classifications, with spectral shapes and features typical of their spectral types.

\subsection{Potentially Young Objects}
\label{section:indiv.young}
\subsubsection{J0526$-$5026 (L6)}
\label{section:indiv.0526}
J0526$-$5026 has a fairly prominent CO feature at ${\sim}$2.3\,\um{}, suggesting that it could have high metallicity. Performing a chi-squared fitting of the \cite{witte2011} models at the effective temperature value determined in Section~\ref{section:temp}, the metallicity of J0526$-$5026 is estimated to be [M/H]$ = +0.3$\,dex. The spectroscopic classification of L6 is also notably different from the photometric spectral type of L8, suggesting that the UCD is fainter and redder than would typically be expected of a L6-type UCD. The high metallicity, along with the weak FeH lines in the \textit{J}-band and spectral classification differences, suggest that J0526$-$5026 is likely to be a young UCD, although the field surface gravity classification and photometric faintness suggests that youth is unlikely. The TiO features in the \textit{J}-band are also weak, further suggesting that J0526$-$5026 is not a young UCD. Ultimately, it is unlikely that this is a young UCD, and further studies will be required to determine the cause of the discrepancy between the spectroscopic and photometric spectral type classifications.

\subsubsection{J0942$-$2551 (L1)}
\label{section:indiv.0942}
J0942$-$2551 is classified as likely being a member of the Carina-Near moving group by the \banyan{} tool, with a probability of 0.939. Carina-Near is a group of co-moving stellar objects with ages of ${\sim}$200\,Myr at a distance of around 30 $\pm$ 20\,pc (\citealt{zuckerman2006, gagne2018}), which is consistent with the distance of J0942$-$2551. Using the measured radial velocity of 4.3 $\pm$ 13.0\,\kms, we calculate the space velocity of J0942$-$2551 to be \(U = -21.8\)\,\kms, \(V = -4.4\)\,\kms, \(W = -11.3\)\,\kms. This is somewhat similar to the average space velocity of the Carina-Near moving group as defined by \cite{zuckerman2006} (\(U = -25.9\)\,\kms, \(V = -18.1\)\,\kms, \(W = -2.3\)\,\kms), so it is likely that J0942$-$2551 is a member of the young moving group.

We identify \gaia{}~DR3 5661603972370860800 as a potential wide companion to J0942$-$2551, however a calculation of the false positive probability yields a value of 0.939, thus it is unlikely that this is a wide binary.

\subsubsection{J1152$+$5901 (L0)}
\label{section:indiv.1152}
J1152$+$5901 is classified as a member of the Crius 198 moving group by the \banyan{} tool, with a probability of 0.983. Crius 198 is a young moving group at an average distance of around 47\,pc and an age of 100--700\,Myr (\citealt{moranta2022}). J1152$+$5901 has a parallactic distance of 53.7 $\pm$ 2.1\,pc, so it is reasonable that it could be part of the Crius 198 group. Using our measured radial velocity of 2.7 $\pm$ 13.8\,\kms, we can calculate the space velocity of J1152$+$5901: \(U = -33.6\)\,\kms, \(V = -12.9\)\,\kms, \(W = -7.0\)\,\kms. \cite{moranta2022} defines the average space velocity of the Crius 198 group as \(U = -34.3\)\,\kms, \(V = -12.9\)\,\kms, \(W = -11.0\)\,\kms. \banyan{} uses multiple input parameters to determine young moving group memberships, thus the discrepancy between the W-components of the space velocities is minor in comparison to the agreement between the other parameters, and we can consider J1152$+$5901 to be a member of the Crius 198 young moving group.

\begin{figure}
\centering
\includegraphics[width=\linewidth]{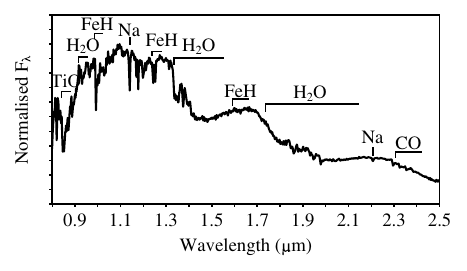}
\caption{Spectrum of J1315$+$3232 with key spectral features labelled. Note the shallow \textit{J}-band FeH features and strong TiO lines -- a telltale sign of youth in UCDs.}
\label{figure:1315}
\end{figure}

\subsubsection{J1315$+$3232 (M7)}
\label{section:indiv.1315}
J1315$+$3232 has a surface gravity classification of INT-G, with fairly weak FeH lines and relatively strong TiO features in the \textit{J}-band, suggesting that it could be a young UCD. Its tangential velocity of 11.702 $\pm$ 0.04\,\kms{} is consistent with J1315$+$3232 being young, as is its radial velocity of 35.5 $\pm$ 12.9\,\kms. Using this radial velocity measurement, we calculate the space velocities of J1315$+$3232 to be: \(U = 6.8\)\,\kms, \(V = -14.5\)\,\kms, \(W = -43.8\)\,\kms. These low space velocity values place the UCD in the thin disc region of the Toomre diagram of Fig.~\ref{figure:toomre-all}, further supporting the theory that J1315$+$3232 may be a young UCD. J1315$+$3232 has only been previously published as a potential UCD by \cite{reyle2018}.

\subsubsection{J1441$+$4217 (L3)}
\label{section:indiv.1441}
The \banyan{} tool identifies J1441$+$4217 as being a likely member of the Crius 198 moving group, which is consistent with its distance of 45.3 $\pm$ 1.3\,pc (the Crius 198 group has an average distance of ${\sim}$47\,pc; \citealt{moranta2022}). The radial velocity of J1441$+$4217 is measured to be -24.2 $\pm$ 13.9\,\kms, which can be used to calculate the UCD's space velocity (\(U = -35.2\)\,\kms, \(V = -11.5\)\,\kms, \(W = -5.6\)\,\kms). This space velocity is very similar to that of the Crius 198 moving group (\(U = -34.3\)\,\kms, \(V = -12.9\)\,\kms, \(W = -11.0\)\,\kms; \citealt{moranta2022}). Only the surface gravity classification of FLD-G suggests that J1441$+$4217 is not a young UCD, though the 100--700\,Myr age of the Crius 198 group (\citealt{moranta2022}) means that a FLD-G gravity classification is plausible for a member of this group (similar to some of the surface gravity classifications of the sample presented by \citealt{manjavacas2019}).

\subsubsection{J1544$-$0435 (L1)}
\label{section:indiv.1544-}
The \banyan{} tool classifies J1544$-$0435 as likely to be a member of the Ursa Major Corona, the peripheral regions of the Ursa Major moving group. The Ursa Major group has an average distance of \(25.4^{+0.8}_{-0.7}\)\,pc and an average age of 414\,Myr (\citealt{jones2015, gagne2018}). J1544$-$0435 has a distance of 52.0 $\pm$ 3.8\,pc, so it is unlikely that it is a member of the core of this young moving group, however it is plausible that it could be a member of the extended Ursa Major Corona, as suggested by the \banyan{} tool. We did not measure the radial velocity of J1544$-$0435 since it was observed with the SpeX prism configuration, however estimating space velocities by allowing radial velocities between $\pm$300\,\kms{} suggests that it is likely to be located in the thin disc (see Fig.~\ref{figure:toomre-all}), consistent with being a younger UCD.

\subsubsection{J1637$+$1813 (M7)}
\label{section:indiv.1637}
J1637$+$1813 has TiO features which are deeper than usual for an M7-type UCD, suggesting that this could be a young UCD. Its tangential velocity is 16.132 $\pm$ 0.59\,\kms, consistent with J1637$+$1813 being a young UCD. The \banyan{} tool identifies J1637$+$1813 as being a member of the Ursa Major Corona. \cite{dopcke2019} outlines that the kinematics of the extended Ursa Major Corona suggests that the members of the extended group are found over a wider range of distances than the core of the moving group (which has a distance of \(25.4^{+0.8}_{-0.7}\)\,pc; \citealt{gagne2018}). J1637$+$1813 has a distance of 49.8 $\pm$ 1.7\,pc, so it is possible that it is a member of the Ursa Major Corona, though we do not deem it to be a young UCD. \cite{reyle2018} lists J1637$+$1813 as a UCD.

\subsubsection{J1847$-$3419 (M7)}
\label{section:indiv.1847}
J1847$-$3419 is classified as being a potential member of the Ursa Major moving group by the \banyan{} tool, and the moving group's distance is consistent with the UCD's distance of 27.1 $\pm$ 0.1\,pc. The radial velocity of J1847$-$3419 is measured to be -49.9 $\pm$ 15.4\,\kms, which is a much larger radial velocity than that of the Ursa Major Corona stars (\citealt{chupina2006}), hence it is unlikely that J1847$-$3419 is a member of the young moving group. Since the \banyan{} classification is the only indicator of youth for this UCD, it is most likely that J1847$-$3419 is not a young UCD.

\begin{figure}
\centering
\includegraphics[width=\linewidth]{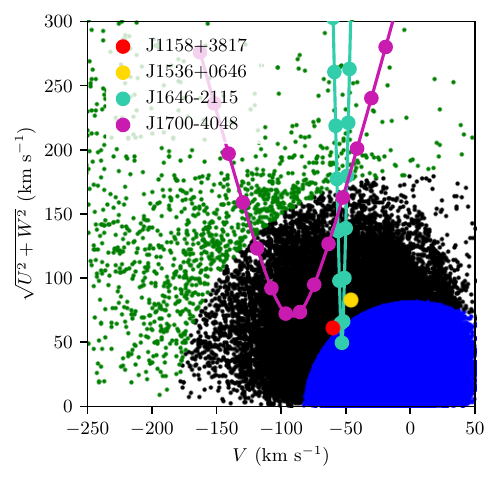}
\caption{Toomre diagram showing potential thick disc and subdwarf UCDs identified in Section~\ref{section:ydh.sd}. Curves for UCDs without measured radial velocities assume radial velocities between $\pm$\,300\,\kms{} with markers plotted at intervals of 50\,\kms. Background points are stars taken from GCNS with known radial velocities (\citealt{smart2021}). Blue points are thin disc stars, black are thick disc and green are halo stars.}
\label{figure:toomre}
\end{figure}

\subsection{Thick Disc and Subdwarf Objects}
\label{section:indiv.sd}
\subsubsection{J0515$+$0613 (M9)}
\label{section:indiv.0515}
J0515$+$0613 has a measured radial velocity of 74.8 $\pm$ 14.1\,\kms, thus we can calculate the space velocity of J0515$+$0613 to be \(U = -86.0\)\,\kms, \(V = -42.1\)\,\kms, \(W = -25.7\)\,\kms. Plotting the position of J0515$+$0613 on a Toomre diagram reveals that, based on its space velocities, J0515$+$0613 is likely to be a member of the thick disc (Fig.~\ref{figure:toomre}). There are no other indications of J0515$+$06713 being a subdwarf, so overall, it is likely that J0515$+$0613 is simply a normal UCD located in the thick disc.

\subsubsection{J1646$-$2115 (sdL2)}
\label{section:indiv.1646}
J1646$-$2115 has a best-fitting \splat{} spectral type of L0, however its photometric spectral type and position on the CMDs in Fig.~\ref{figure:cmd} indicate that it has a spectral classification of L2. Visual inspection of the spectrum compared to the \splat{} L2 and T0 standards show that the spectrum is more similar to that of the L2 UCD. There are, however, deviations from the standard in the spectrum of J1646$-$2115 which suggest that it may be a subdwarf.

\begin{figure}
\centering
\includegraphics[width=\linewidth]{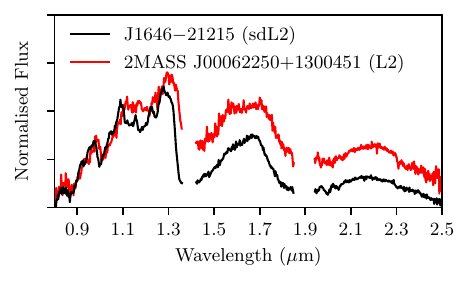}
\caption{Spectrum of J1646$-$2115 compared with that of 2MASS J00062250+1300451, the L2-type UCD which we have adopted for use as the class standard in this work. Our UCD's spectrum is clearly dimmer in the \textit{H} and \textit{K}-bands, and shows a lack of CH$_4$ features and strong FeH features, characteristic of a subdwarf.}
\label{figure:1646}
\end{figure}

Fig.~\ref{figure:1646} shows a comparison of the spectrum of J1646$-$2115 with that of our adopted L2 standard, 2MASS J00062250+1300451. As is evident from this comparative plot, J1646$-$2115 is fainter and bluer in the \textit{H} and \textit{K}-bands, and has stronger FeH features in the \textit{J}-band. The lack of CH$_4$ features at 1.6\,\um{} and 2.2\,\um{} can also be seen in the plot. This plot looks remarkedly similar to that of \textit{WISE} J1013$-$7246 in the top plot of fig.~67 by \cite{kirkpatrick2016}, furthering that J1646$-$2115 is a sdL2-type object.

The photometry for J1646$-$2115 is typical of a L2-type UCD, and the UCD's position on the CMDs does not suggest that it is a subdwarf. The position of J1646$-$2115 on the Toomre diagram of Fig.~\ref{figure:toomre} shows that it is likely to be a thick disc or halo object, although the range of potential space velocity values extends into the thin disc region, so thick disc membership cannot be confirmed using this method.

\subsubsection{J1700$-$4048 (L4)}
\label{section:indiv.1700}
J1700$-$4048 has a \splat{} surface gravity classification of VL-G, but its spectrum is too noisy to identify any spectral shapes indicative of young UCDs, and the poor signal-to-noise of our spectrum means that the VL-G surface gravity classification should not be taken as definite. The tangential velocity of J1700$-$4048 is 112.745 $\pm$ 33.819\,\kms, and plotting a Toomre diagram using radial velocities varying between $\pm$\,300\,\kms{} shows that J1700$-$4048 is likely to be a thick disc object (Fig.~\ref{figure:toomre}). It has the faintest absolute \gaia{} \textit{G}-band magnitude (M$_G$) of any of the UCDs in our sample (27.763 $\pm$ 0.650), suggesting that it may be a subdwarf. Overall, J1700$-$4048 is likely to be located within the thick disc, but further spectroscopic observations will be required to confirm whether it is a subdwarf. J1700$-$4048 has not been included in any previous publications.

\subsection{Wide Binary Objects}
\label{section:indiv.wide}

\begin{table*}
    \centering    
    \caption{Key information about the UCDs in wide binary systems, including the companion object spectral type (if available) and false positive probability, as calculated following the method outlined by \citet{baig2024}. Some UCDs have more than one wide companion.}
    \label{table:cpm}
    \begin{tabular}{c|c|c|c|c|c|c|c}
        \hline
        UCD & Spectral & Companion & Companion & Separation & Projected & False Positive \\
        Short Name & Type & \gaia{}~DR3 ID & SpT & (arcsec) & Separation (AU) & Probability \\
        \hline
        J0850$-$0318 & M8 & 5762038930729097728 & M1 \footnotesize{$^{[1]}$} & 8.84 & 310.9 & $3.86\times10^{-8}$ \\
        J0941$+$3315A & M8 & 794031395950867968 & M5 \footnotesize{$^{[2]}$} & 7.34 & 242.0 & $5.17\times10^{-10}$ \\
        J1243$+$6001 & L7 & 1579929906250111232 & K0 \footnotesize{$^{[3]}$} & 35.19 & -- & $2.92\times10^{-14}$ \\
        J1250$+$0455 & M9 & 3705763723623026304 & -- & 10.45 & 449.92 & $6.90\times10^{-7}$ \\
        J1420$+$3235 & M7 & 1477880761742899456 & -- & 129.87 & 286.76 & $3.69\times10^{-5}$ \\
        J2019$+$2256 & L2 & 1829571684890816384 & M3 $+$ M0 \footnotesize{$^{[4]}$} & 11.02 & 324.7 & \multirow{2}{4.8em}{$5.70\times10^{-8}$} \\
         &  & 1829571684890816512 & M3.5 \footnotesize{$^{[4]}$} & 9.84 & 289.9 &  \\
        \hline
    \end{tabular}
    \vspace{-2ex}
    \begin{tabular}{l l l l l}
        \textbf{References} & \footnotesize{$^{[1]}$} -- \cite{gaidos2014} & \footnotesize{$^{[2]}$} -- This work (see Appendix~\ref{appendix:nonucd.spt}) & \footnotesize{$^{[3]}$} -- \cite{faherty2021} & \footnotesize{$^{[4]}$} -- \cite{cifuentes2021} \\
    \end{tabular}
\end{table*}

\subsubsection{J0850$-$0318 (M8) $+$ SCR J0850$-$0318 (M1)}
\label{section:indiv.0850}
As outlined in Section~\ref{section:ydh.wide}, the methods defined by \cite{smart2019} were used to identify potential wide companions for the UCDs. J0850$-$0318 was identified to have a wide companion: SCR J0850$-$0318, at a separation of 8.84\,arcsec (a projected separation of 310.9\,AU at the distance of the UCD). SCR J0850$-$0318 is a well-characterised M1 star published by \cite{gaidos2014}, with an effective temperature of 3881 $\pm$ 78\,K, radius of 0.56 $\pm$ 0.04\,R$_{\odot}$ and mass of 0.60 $\pm$ 0.07\,M$_{\odot}$.

This system is included in the Ultracool Dwarf Binary Catalogue (hereafter UCDC; \citealt{baig2024}), where J0850$-$0318 and SCR J0850$-$0318 are confirmed to be a binary system with reasonable projected separations, comparable parallaxes and common proper motion. The system has a false positive probability of $3.68\times10^{-8}$, confirming the companionship of the system.

\subsubsection{J0941$+$3315A (M8) $+$ J0941$+$3315B (M5)}
\label{section:indiv.0941}
J0941$+$3315A is identified as a companion to J0941$+$3315B. Both of these objects were observed for this work, and J0941$+$3315B is omitted from the UCD sample since we determine it to have an M5 spectral classification (see Appendix~\ref{appendix:nonucd}). The J0941$+$3315AB system is not included in the UCDC, however using the criteria from \cite{baig2024} we find the pair to have appropriate projected separation, and consistent parallaxes and proper motions. Using the equations outlined by \cite{baig2024}, we calculate the false positive probability for this system to be $5.17\times10^{-10}$, confirming the companionship of this binary. \cite{seli2021} identifies J0941$+$3315AB as a co-moving pair, with J0941$+$3315A being listed as one of their TRAPPIST-1 analogue systems.

\subsubsection{J1243$+$6001 (L7) $+$ BD$+$60~1417 (K0)}
\label{section:indiv.1243}
J1243$+$6001 only has limiting magnitudes in \gaia{} and 2MASS photometry, so is not plotted on the CMDs in Fig.~\ref{figure:cmd} or the cooling track plots in Fig.~\ref{figure:cooling}. The \splat{} surface gravity tool gives a classification of VL-G, and visual inspection of the spectrum reveals a triangular \textit{H}-band and positive slope in the \textit{K}-band, indicating that this is likely to be a young UCD. Performing a chi-squared fit of the \cite{witte2011} models at \teff{}$ = 1600$\,K and \logg{}$ = 5.0$\,dex reveals a best-fitting metallicity of [M/H]$ = +0.3$\,dex. This is consistent with J1243$+$6001 being a young UCD. \cite{faherty2021} also conclude that J1243$+$6001 is a young UCD, with an estimated age of 10--150\,Myr.

J1243$+$6001 is also identified as a wide companion to BD$+$60~1417, which has been previously published by \cite{faherty2021} and \cite{calamari2024}. The system is not published in the UCDC, although this is likely due to the lack of \gaia{} detection for J1243$+$6001. Our calculated spectrophotometric distance ($43.5 \pm 12.3$\,pc) is consistent with the parallactic distance of BD$+$60~1417 ($45.0 \pm 0.0$\,pc, based on its \gaia~DR3 parallax measurement). Calculating a false positive probability for this pair in the same way as \cite{baig2024}, we find a false positive probability value of $2.92\times10^{-14}$, thus confirming the companionship of the system.

\subsubsection{J1250$+$0455 (M9) $+$ \gaia{}~DR3 3705763723623026304}
\label{section:indiv.1250}
J1250$+$0455 has a RUWE value of 1.5097 -- the highest RUWE value for any UCD within the sample in this work. J1250$+$0455 is included in the list of L-dwarfs by \cite{skrzypek2015}, with a photometric classification of L1. The discrepancy between the photometric and spectroscopic classifications is reflected in the position of J1250$+$0455 on the CMDs and cooling tracks: the UCD is unusually red for an UCD of its spectral type (Fig.~\ref{figure:cmd}~\&~\ref{figure:cooling}). In addition to the redness in the photometry, the spectrum of J1250$+$0455 also displays suggestions that there may be hidden companion, since there are strong H$_2$O features at ${\sim}$0.9\,\um{} and ${\sim}$1.1\,\um{} and weak FeH features at ${\sim}$1.0\,\um{} and ${\sim}$1.2\,\um{}. Baig et al. (sub.) resolve this UCD as a binary.

Our methods also identify \gaia{}~DR3 3705763723623026304 as a wide companion to J1250$+$0455, at a separation of 10.45\,arcsec (a projected separation of 449.92\,AU at the distance of the UCD). \cite{baig2024} includes J1250$+$0455 in the UCDC, as a wide binary with \gaia{}~DR3 3705763723623026304, and also identifies the UCD as having a potential hidden companion using blended photometry methods (see section~3.2 of their paper). As with J0850$-$0318, the inclusion of J1250$+$0455 in the UCDC means that the binary has a sensible projected separation and consistent parallaxes and proper motions, thus confirming the companionship of the pair.

\subsubsection{J1420$+$3235 (M7) $+$ \gaia{}~DR3 1477880761742899456}
\label{section:indiv.1420}
J1420$+$3235 is identified to have one potential wide companion: \gaia{}~DR3 1477880761742899456, an M-dwarf identified by \cite{morrell2019}. The two objects are separated by 129.87\,arcsec, corresponding to a projected separation of 286.76\,AU at the distance of the UCD. J1420$+$3235 is included in the UCDC, meaning that the projected separation of the components is reasonable, and there is common proper motion and consistent parallaxes between the two objects, thus confirming the binarity.

Some of the properties of \gaia{}~DR3 1477880761742899456 are estimated by \cite{morrell2019}: \teff{} = 3500\,K, \logg{} = 5.5\,dex, R = 0.43\,R$_{\odot}$. These values have a poor goodness of fit from their model-fitting methods, so further observations of this pair will be required to confirm their properties.

\subsubsection{J2019$+$2256 (L2) $+$ LP 395-8 AB (M3 $+$ M0 $+$ M3.5)}
\label{section:indiv.2019}
\begin{figure}
    \centering
    \includegraphics[width=\linewidth]{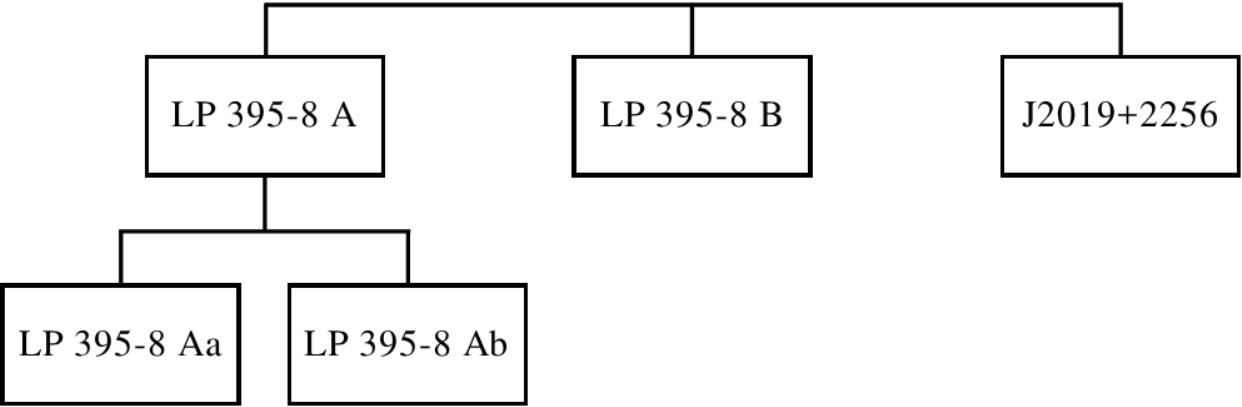}
    \caption{Diagram showing the hierarchical structure of the J2019$+$2256 $+$ LP 395-8 AB system described by \citep{cifuentes2021}.}
    \label{figure:2019sys}
\end{figure}

J2019$+$2256 is identified to have two wide companions: LP 395-8 A and LP 395-8 B. LP 395-8 B is a known M3.5-type companion to the M-dwarf binary system comprised of LP 395-8 Aa and LP 395-8 Ab (M3 and M0 types respectively; \citealt{cifuentes2021}). J2019$+$2256 is  identified by \cite{cifuentes2021} as a M9-type companion to the hierarchical triple system, listing the UCD as LP 395-8 C in their table~1. Fig.~\ref{figure:2019sys} shows a diagram depicting the \cite{cifuentes2021} description of the hierarchical nature of the system.

The J2019$+$2256 $+$ LP 395-8 AB system is also listed in the UCDC. This verifies that the system has reasonable projected separation, consistent parallaxes and common proper motion. \cite{baig2024} gives a false positive probability of $5.70\times10^{-8}$, confirming the membership of J2019$+$2256 in this hierarchical system.

\subsection{Unresolved Binary Objects}
\label{section:indiv.unres}
\subsubsection{J0723$+$4622 (M8 $+$ L7)}
\label{section:indiv.0723}
J0723$+$4622 has a \gaia{} RUWE value of 1.2470, which suggests that this is likely to be a binary object; although there were no Image Parameters Determination (hereafter IPD) windows with more than one peak (a double peak means that the detection may be a resolved double star, i.e. a visual double or real binary), so it most likely that this is an unresolved binary.

J0723$+$4622 only satisfies one of the spectral index criteria laid out in Section~\ref{section:ydh.unres}, but its spectrophotometric distance suggests that it is over-luminous for a single L1-type UCD (the parallactic distance is 55.8 $\pm$ 4.5\,pc and the spectrophotometric distance is 74.9 $\pm$ 3.8\,pc). The significant difference in the parallactic and spectrophotometric distance values suggests that it may not be a single source. A chi-squared fitting of the binary spectra created using the library of standard spectra included in \splat{} reveals the best-fitting binary to be one with M8 and L7 components. Plotting this binary template spectrum over the spectrum of J0723$+$4622 shows that the binary spectrum is more closely matched with the UCD spectrum (Fig.~\ref{figure:binary}), and the $\eta_{\text{SB}}$ value is 2.229, consistent with a $>\,90$\,per\,cent confidence that this spectrum is better-fitted by a binary template. This is consistent with the over-luminosity suggested by the spectrophotometric distance, thus J0723$+$4622 is likely to be an unresolved binary with M8 and L7-type components.

\subsubsection{J0817$-$6155 (T6)}
\label{section:indiv.0817}
\begin{figure*}
\centering
    \includegraphics[width=0.88\linewidth]{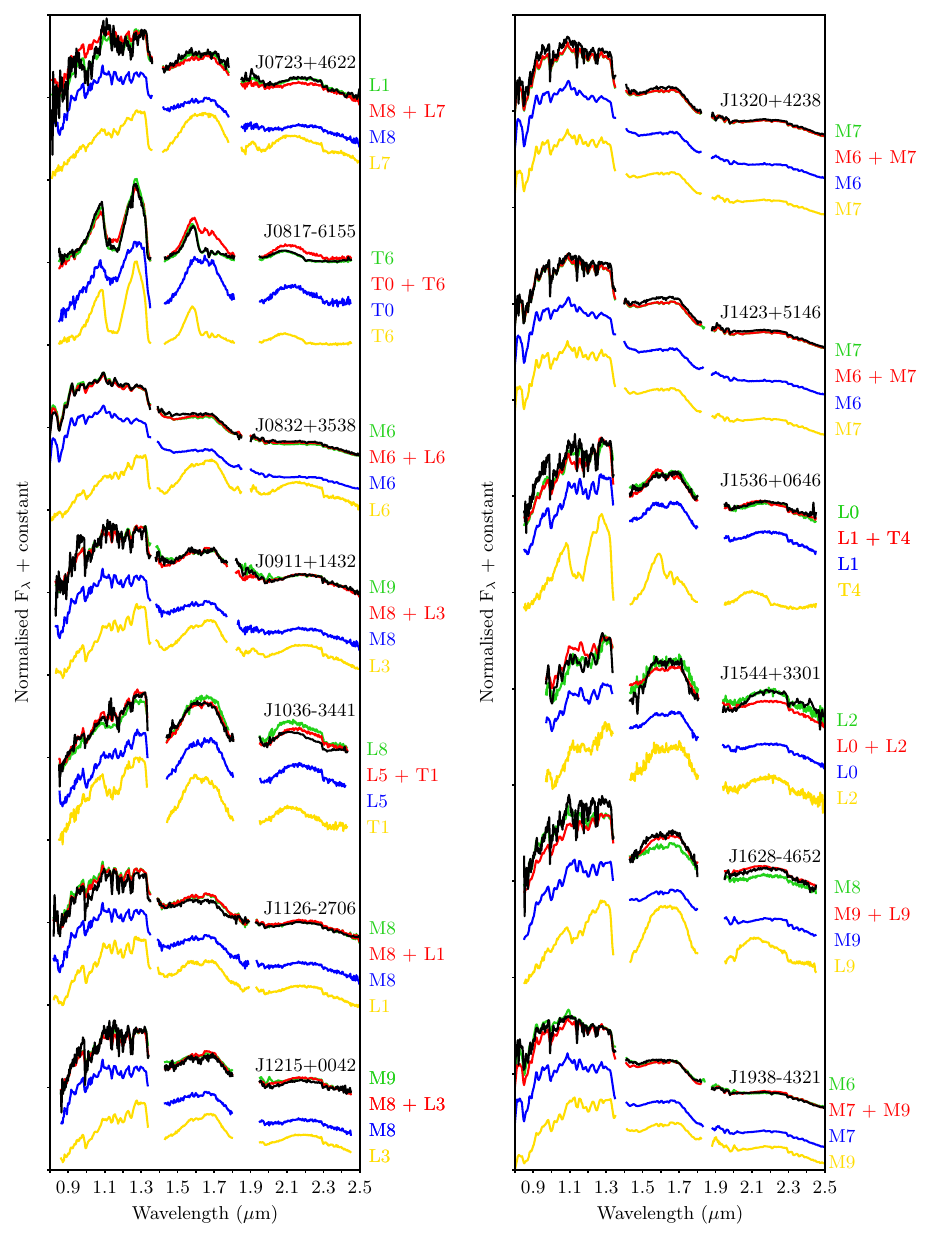}
    \caption{Spectral deconvolution of the unresolved binary UCDs. The spectrum of the UCD is plotted (black), along with the best-fitting single standard spectrum (green), the best-fitting binary template spectrum (red), and the two components of the binary spectrum (blue and yellow). Plotted spectra are normalised at 1.27\,\um. Spectra used to create the binary templates are listed in Table~\ref{table:standards}.}
    \label{figure:binary}
\end{figure*}

J0817$-$6155 lies in the region of the CMDs which would be expected for a sub-stellar object. Its RUWE value is 1.2429, so it is possible that this is a binary object, although the spectral index criteria do not suggest that J0817$-$6155 is an unresolved binary. A chi-squared fit of the binary template spectra gives a best-fitting binary with T0 and T6 components. The $\eta_{\text{SB}}$ value of 0.115 means that we can safely conclude the single-source spectral fit is better than the binary, which is consistent with a by-eye comparison of the spectral fits (Fig.~\ref{figure:binary}), so it is mostly likely that J0817$-$6155 is a single UCD.

The sub-stellar nature of J0817$-$6155 is furthered by the \splat{} surface gravity classification of INT-G. The \gaia{}~DR3 tangential velocity of 27.437 $\pm$ 0.062\,\kms{} and our measured radial velocity of -22.9 $\pm$ 18.4\,\kms{} are also consistent with J0817$-$6155 being a young object. This UCD has been included in a number of works: \cite{kiman2019, schneider2016, kirkpatrick2011}. J0817$-$6155 is discussed in detail by \cite{artigau2010}, and our findings largely agree with theirs.

\subsubsection{J0832$+$3538 (M6)}
\label{section:indiv.0832}
J0832$+$3538 is identified as a weak unresolved binary candidate using the spectral index criteria, with the best-fitting binary template being that of an M6 $+$ L6 binary. The $\eta_{\text{SB}}$ value is 0.302, thus the single-source spectrum can be deemed a better fit than that of the binary. When comparing the two spectral fits by-eye, the single-source spectrum matches the spectrum of J0832$+$3538 more closely than the binary spectrum, furthering that J0832$+$3538 is a single UCD. The spectrophotometric distance is similar to the parallactic distance (26.8 $\pm$ 0.4\,pc and 30.9 $\pm$ 0.1\,pc respectively), so it is most likely that it is a single M6-type UCD.

\subsubsection{J0911$+$1432 (M8 $+$ L3)}
\label{section:indiv.0911}
J0911$+$1432 is a strong unresolved binary candidate based on the spectral index criteria. The best-fitting binary template spectrum is that of a M8 $+$ L3 binary, and when compared by-eye to the spectrum of J0911$+$1432, as shown in Fig.~\ref{figure:binary}, the binary spectrum is more similar than that of the M9 standard (J0911$+$1432 was spectroscopically classified as a M9-type UCD in Section~\ref{section:spt.splat}). This is reflected by the $\eta_{\text{SB}}$ value of 1.076, which is consistent with an 80\,per\,cent confidence of it being a binary. The possibility of J0911$+$1432 being an unresolved binary is further supported by the discrepancy between the parallactic distance (67.1 $\pm$ 3.3\,pc) and spectrophotometric distance (58.5 $\pm$ 1.7\,pc). This means that it is likely that J0911$+$1432 is an unresolved binary with M8 and L3 constituent components, rather than a single M9-type UCD.

\subsubsection{J1036$-$3441 (L5 $+$ T1)}
\label{section:indiv.1036}
J1036$-$3441 has a RUWE value of 1.2900, meaning that is likely that this could be a binary object. There are no potential wide companions, so this is likely to be an unresolved binary, a possibility supported by the 3$-$subtype difference in optical and NIR spectral classifications. The spectral index criteria in Section~\ref{section:ydh.unres} indicate that J1036$-$3441 is a weak binary candidate, and it was published by \cite{bardalezgagliuffi2014} as a weak binary candidate with a best-fitting binary spectrum composed of L5 $+$ T1.5 components.

Comparing the L5 $+$ T1 template binary and single-source spectra to that of J1036$-$3441 by-eye reveals that the binary spectrum better matches the spectrum of the UCD, so it is likely that it is a unresolved binary UCD (Fig.~\ref{figure:binary}). This is supported by the $\eta_{\text{SB}}$ value of 1.875, which corresponds to a $>$\,90\,per\,cent confidence that the binary spectral fit is better than that of the single-UCD spectrum.

J1036$-$3441 has been previously published as a high proper motion object by \cite{schneider2016}. It has also been published as a UCD by \cite{gizis2002}, with a spectral classification of L6.

\subsubsection{J1126$-$2706 (M8 $+$ L1)}
\label{section:indiv.1126}
J1126$-$2706 has a RUWE value of 1.3152, so it is likely to be a binary. Its spectral indices do not suggest that it is an unresolved binary, but there are also no potential wide companions. A chi-squared fit suggests that J1126$-$2706 could be a M8 $+$ L1 binary, and comparing the spectrum of J1126$-$2706 with the binary and standard spectra by-eye reveals that the binary spectrum matches that of the UCD more closely in the \textit{J}-band, although the spectral types comprising the binary template are similar so the differences between the spectra are minimal. The $\eta_{\text{SB}}$ value of 1.244 gives a $>90$\,per\,cent confidence that the binary template is a better spectral fit for J1126$-$2706.

When compared to the parallactic distance (33.3 $\pm$ 0.3\,pc), the spectrophotometric distance (45.1 $\pm$ 1.0\,pc) suggests that J1126$-$2706 is brighter than expected. This suggests that J1126$-$2706 is likely to be an unresolved binary with M8 and L1-type components. \cite{reyle2018} lists J1126$-$2706 as a UCD.

\subsubsection{J1215$+$0042 (M8 $+$ L3)}
\label{section:indiv.1215}
J1215$+$0042 lies in the top right corners of the CMDs (i.e. it is brighter than would be expected considering its redness). It also lies above the cooling tracks in Fig.~\ref{figure:cooling}, and has a considerable difference between its spectroscopic and photometric spectral type classifications. This difference is reflected in the parallactic and spectrophotometric distance discrepancy (the parallactic distance is 44.3 $\pm$ 2.1\,pc and the spectrophotometric distance is 73.0 $\pm$ 3.1\,pc).

Performing a chi-squared fit of template binary spectra gives a best-fitting binary with M8 and L3-type components. A by-eye comparison of the binary template spectrum with that of the M8 standard and the UCD, reveals that the binary spectrum is the better match. The $\eta_{\text{SB}}$ value of 1.022 shows that there is not a $>$\,90\,per\,cent confidence that the binary fit is better than the single-spectrum fit, although all other indicators suggest that J1215$+$0042 is a binary. It is thus more likely that J1215$+$0042 is an unresolved binary with M8 and L3-type components. J1215$+$0042 has been previously published as a UCD with a photometric spectral type of L1.5 by \cite{skrzypek2015}.

\subsubsection{J1320$+$4238 (M7)}
\label{section:indiv.1320}
J1320$+$4238 is a weak unresolved binary candidate based on its spectral indices, with a best-fitting binary template of a M6 $+$ M7 system. This binary template is not a closer match to our observed spectrum than that of a single M7-type UCD when compared by-eye. This is further supported by the $\eta_{\text{SB}}$ value of 1.025, indicating a $<$\,90\,per\,cent confidence on the binary spectral fit being better than the single-source fit.

The single-source nature of J1320$+$4238 is further supported by the agreement between then parallactic and spectrophotometric distances (26.7 $\pm$ 0.0\,pc and 29.4 $\pm$ 1.0\,pc, respectively). We can thus conclude that J1320$+$4238 is most likely to be a single M7-type UCD.

\subsubsection{J1423$+$5146 (M7)}
\label{section:indiv.1423}
J1423$+$5146 is not identified as an unresolved binary candidate based on its spectral indices, but it has a RUWE value of 1.2084, and we have not found any potential wide companions for J1423$+$5146, so it is possible that it is an unresolved binary. A chi-squared fit of binary template spectra results in a best-fitting binary comprised of M6 $+$ M7 components, which is almost identical to the single-source spectral template. The $\eta_{\text{SB}}$ value of 0.974 shows that there is a $<$\,90\,per\,cent confidence that the binary template is a better-fit to the spectrum of J1423$+$5146, thus this is most likely a single M7-type UCD. This is further supported by the consistency between the parallactic distance (17.3 $\pm$ 0.0\,pc) and spectrophotometric distance (18.6 $\pm$ 0.4\,pc). J1423$+$5146 is therefore unlikely to be an unresolved binary.

\subsubsection{J1536$+$0646 (L0)}
\label{section:indiv.1536}
J1536$+$0646 has a RUWE value of 1.2198, suggesting that this could be a binary object. J1536$+$0646 has been published as a UCD by \cite{zhang2010}, and has a photometric spectral type of L2 by \cite{skrzypek2015}. The small discrepancy between their photometric spectral type and our spectroscopic classification of L0 further suggests that it may be a binary. This is furthered by the 2-subtype difference between the optical and NIR spectroscopic classifications.

A chi-squared fit of binary template spectra reveals a L1 $+$ T4 to be the best-fitting binary spectrum; however, the single-source spectrum of the L0 standard better represents the spectrum of J1536$+$0646 when making a comparison by-eye. There is a $<$\,90\,per\,cent confidence in the binary fit being better than the single-source fit, as shown by the $\eta_{\text{SB}}$ value of 0.550. The possibility of the UCD being a single-object source is further supported by the consistency between the parallactic distance (60.8 $\pm$ 4.2\,pc) and spectrophotometric distance (65.8 $\pm$ 2.6\,pc); thus it is mostly likely that this is a single UCD.

J1536$+$0646 is also identified as a potential thick disc object in Section~\ref{section:ydh.sd}, with a radial velocity of 74.8 $\pm$ 14.1\,\kms and space velocities of \(U = -65.2\)\,\kms, \(V = -36.3\)\,\kms, \(W = -18.6\)\,\kms. This places the UCD in the thick disc region of the Toomre diagram, although the spectrum is not indicative of a subdwarf; thus it is likely that J1536$+$0646 is a single UCD located in the thick disc.

\subsubsection{J1544$+$3301 (L2)}
\label{section:indiv.1544+}
J1544$+$3301 is a strong unresolved binary candidate, based on the spectral index criteria laid out in Section~\ref{section:ydh.unres}. A chi-squared fit of the binary template spectra reveals an L0 $+$ L2 binary to be the best-fitting binary template, and plotting a comparison between this binary spectrum and that of the L2 standard 2MASSW J2130446$-$084520 shows that the single source is a better match to the spectrum of J1544$+$3301 when compared by-eye. The $\eta_{\text{SB}}$ value of 0.744 reflects this, showing that there is a $<$\,90\,per\,cent confidence that the binary template is a better fit.

This suggestion of J1544$+$3301 being a single source is inconsistent with the discrepancy between the parallactic (23.9 $\pm$ 0.4\,pc) and spectrophotometric distances (44.5 $\pm$ 5.8\,pc), since the spectrophotometric distance suggests that J1554$+$3301 is brighter than a typical L2-type UCD. This could be explained by J1544$+$3301 being a young UCD, since the \banyan{} tool identifies J1544$+$3301 as being a potential member of the Rhea 300 moving group, a young moving group at an average distance of around 24\,pc. This distance is very similar to J1544$+$3301's parallactic distance of ${\sim}$23.9\,pc, so it is possible that the UCD is a member of the young moving group, although distance alone is insufficient for determining a UCD's membership to a young moving group. Further investigation will be required to confirm the youth of J1544$+$3301.

\subsubsection{J1628$-$4652 (M9 $+$ L9)}
\label{section:indiv.1628}
J1628$-$4652 is a strong unresolved binary candidate, based on its spectral indices (Section~\ref{section:ydh.unres}). The best-fitting binary template is a M9 $+$ L9 system, which matches our observed spectrum better than that of a single M8-type UCD when compared by-eye. The $\eta_{\text{SB}}$ value of 1.133 indicates a $>$\,90\,per\,cent confidence that the binary template is a better fit to the UCD's spectrum than the single-source model. 

This binarity is further supported by the difference between the parallactic (62.2 $\pm$ 3.8\,pc) and spectrophotometric distances (81.7 $\pm$ 2.5\,pc). It is therefore likely that J1628$-$4652 is an unresolved M9 $+$ L9 binary.

\subsubsection{J1938$+$4321 (M6)}
\label{section:indiv.1938}
J1938$+$4321 is a weak unresolved binary candidate, satisfying two of the spectral index selection criteria outlined in Section~\ref{section:ydh.unres}. A $\chi^2$ fitting of binary templates reveals a M7 $+$ M9 to be the best-fitting binary template. A by-eye comparison of the binary and single-source spectral fits reveals that the single M6-type spectrum is more similar to the spectrum of J1938$+$4321. This is supported by the $\eta_{\text{SB}}$ value of 0.839, since there is a $<$\,90\,per\,cent confidence that the binary template is a better fit than the single-source spectrum. The spectrophotometric distance of 29.5 $\pm$ 0.4\,pc is consistent with the parallactic distance of 26.8 $\pm$ 0.3\,pc, thus is it likely that J1938$+$4321 is not an unresolved binary.

\section{Conclusions}
\label{section:conc}
In this paper, 51 \gaia{} UCDs have been observed with the Blanco ARCoIRIS and IRTF SpeX instruments, in order to classify the spectral types of each UCD. 44 of these are new classifications, while three already have published NIR spectroscopic types. These previously published spectral classifications are consistent with our NIR spectral types. All but one of the UCDs are within 100\,pc: 37 lie within 50\,pc, and the more distant UCD lies at ${\sim}$227\,pc.

The \textsc{Python} module \splat{} was used to classify the spectra, identifying the spectral types of the UCDs by comparing them with standards from the SpeX Prism Library. Our sample is comprised of a total of 26 M-types, 24 L-types and one T-type UCD, and the spectral type consistency has been confirmed using the photometric classification method outlined by \cite{skrzypek2015}.

The photometry for the UCDs follow the general trends that would be expected, based on \cite{smart2019}, and plotting the UCDs on CMDs and cooling track plots reveals that the majority of our UCDs are old and stellar in nature. There are, however, a few UCDs of interest that lie away from the general trends, highlighting potential young, binary and thick disc/subdwarf UCDs.

In addition, our spectra allow for more detailed analysis of the UCDs, namely their effective temperatures and radial velocities can be estimated. By comparing our UCDs' spectra with the BT-Settl CIFIST spectra, we can find the best-fitting temperature for each UCD, and doing so gives a temperature range between 1200\,K and 3000\,K for the UCDs in our sample. Using the \texttt{rvfitter} module for \textsc{Python}, we were able to measure the radial velocities for the higher resolution observations (i.e. the ARCoIRIS and SpeX SXD observations).

By combining published data with measurements of radial velocities using our spectra, we find a number of UCDs which have properties similar to young UCDs, as well as some thick disc and subdwarf UCDs and a plethora of wide and unresolved binary UCDs. Looking at these UCDs individually, we find a total of seven young, three thick disc, one subdwarf, six wide binary and six unresolved binary UCDs, though further observations and deeper investigation will be required to confirm the full properties of these UCDs.

\onecolumn
\begin{landscape}
\centering
    \begin{longtable}[c]{c c c c c c c c c c c c c}
        \caption{Indications of youth, thick disc, halo and binary UCDs given by sensitive features for each UCD in this work, including \banyan{} group membership, \splat{} surface gravity classification, spectral features and \gaia{}~DR3 data.}
        \label{table:ydh} \\
        \hline
        Object & \multirow{2}{2em}{SpT} & CMD & Cooling & \banyan & \banyan & Surface & Spectral & FeH/TiO & Toomre & \gaia{}~DR3 & \gaia{}~DR3 & Young, Thick Disc, \\
        Short Name & -- & Position & Track & Membership & Probability & Gravity & Shape & Features & Diagram & RUWE & v$_\text{tan}$ (\kms) & or Binary? \\
        \hline
        \endfirsthead

        \caption[]{\raggedright{\textit{-- continued}}} \\
        \hline
        Object & \multirow{2}{2em}{SpT} & CMD & Cooling & \banyan & \banyan & Surface & Spectral & FeH/TiO & Toomre & \gaia{}~DR3 & \gaia{}~DR3 & Young, Thick Disc, \\
        Short Name & -- & Position & Track & Membership & Probability & Gravity & Shape & Features & Diagram & RUWE & v$_\text{tan}$ (\kms) & or Binary? \\
        \hline
        \endhead

        \hline
        \endfoot
        
        \hline
        \endlastfoot
        
        J0508$+$3319 & L2 & -- & -- & Field & 1.000 & FLD-G & -- & -- & -- & 1.0803 & 58.428 $\pm$ 0.512 & -- \\
        J0515$+$0613 & M9 & -- & -- & Field & 0.970 & FLD-G & -- & -- & Thick disc & 1.0773 & 17.956 $\pm$ 0.969 & Thick disc \\
        J0526$-$5026 & L6 & -- & -- & Field & 0.988 & FLD-G & -- & -- & -- & 1.1563 & 28.615 $\pm$ 0.686 & -- \\
        J0542$+$0041 & L7 & -- & -- & Field & 0.993 & VL-G & Young & Young & -- & 1.1348 & 6.276 $\pm$ 0.208 & -- \\
        J0723$+$4622 & L1 & -- & -- & Field & 1.000 & FLD-G & -- & -- & -- & 1.2470 & 43.728 $\pm$ 3.890 & UB \\
        J0808$+$3157 & M7 & -- & -- & Field & 1.000 & FLD-G & -- & -- & -- & 1.0626 & 17.167 $\pm$ 0.041 & -- \\
        J0811$+$1855 & L1 & -- & -- & Field & 0.830 & FLD-G & -- & Young & -- & 1.1226 & 12.953 $\pm$ 0.173 & -- \\
        J0817$-$6155 & T6 & VLM & VLM & Field & 1.000 & INT-G & -- & -- & -- & 1.2429 & 27.437 $\pm$ 0.062 & Young \\
        J0832$+$3538 & M6 & -- & -- & Field & 1.000 & FLD-G & -- & -- & -- & 1.0809 & 52.493 $\pm$ 0.155 & -- \\
        J0850$-$0318 & M8 & -- & -- & Field & 1.000 & FLD-G & -- & -- & -- & 1.0475 & 30.226 $\pm$ 0.282 & WB \\
        J0900$+$5205 & M9 & -- & -- & Field & 1.000 & INT-G & -- & -- & -- & 1.0702 & 47.448 $\pm$ 4.877 & -- \\
        J0911$+$1432 & M9 & -- & -- & Field & 1.000 & FLD-G & -- & -- & -- & 1.1326 & 23.698 $\pm$ 1.276 & UB \\
        J0916$-$1121 & M8 & -- & -- & Field & 1.000 & FLD-G & -- & -- & -- & 1.1117 & 64.271 $\pm$ 0.489 & -- \\
        J0941$+$3315A & M8 & -- & Young & Field & 1.000 & FLD-G & -- & -- & -- & 0.9854 & 71.002 $\pm$ 0.383 & WB \\
        J0942$-$2551 & L1 & -- & -- & CARN & 0.939 & FLD-G & -- & -- & -- & 1.2074 & 24.561 $\pm$ 1.482 & Young \\
        J0948$+$5300 & L1 & -- & -- & Field & 1.000 & VL-G & -- & -- & -- & 1.1384 & 23.938 $\pm$ 0.933 & -- \\
        J1036$-$3441 & L8 & VLM & VLM & Field & 1.000 & INT-G & -- & -- & -- & 1.2900 & 31.890 $\pm$ 0.759 & UB \\
        J1048$-$5254 & L1 & -- & -- & CARN & 0.753 & FLD-G & -- & -- & -- & 1.0817 & 31.793 $\pm$ 0.163 & -- \\
        J1126$-$2706 & M8 & -- & -- & Field & 1.000 & FLD-G & -- & Young & -- & 1.3152 & 19.522 $\pm$ 0.184 & UB \\
        J1143$+$5324 & L1 & -- & -- & Field & 1.000 & FLD-G & Young & Young & -- & 1.1220 & 6.603 $\pm$ 0.381 & -- \\
        J1150$-$2914 & L0 & -- & -- & CARN & 0.538 & FLD-G & -- & -- & -- & 1.0407 & 35.029 $\pm$ 0.686 & -- \\
        J1152$+$5901 & L0 & -- & -- & Cri198 & 0.983 & INT-G & -- & -- & -- & 1.0817 & 36.561 $\pm$ 1.517 & Young \\
        J1158$-$0008 & M9 & Old & Old & Field & 1.000 & FLD-G & -- & -- & -- & 0.9978 & 50.337 $\pm$ 5.224 & -- \\
        J1158$+$3817 & M7 & -- & -- & Field & 1.000 & FLD-G & Young & -- & -- & 1.0730 & 76.462 $\pm$ 0.849 & -- \\
        J1212$+$0206 & L1 & -- & -- & Field & 1.000 & FLD-G & -- & -- & -- & 1.1962 & 33.893 $\pm$ 3.321 & -- \\
        J1215$+$0042 & M9 & Binary & Binary & Field & 0.999 & FLD-G & -- & -- & -- & 1.1442 & 21.441 $\pm$ 1.123 & UB \\
        J1243$+$6001 & L7 & -- & -- & Cri227 & 0.439 & VL-G & Young & -- & -- & -- & -- & Young, WB \\
        J1250$+$0455 & M9 & -- & -- & Field & 1.000 & FLD-G & -- & -- & -- & 1.5097 & 46.326 $\pm$ 3.845 & WB \\
        J1252$+$0347 & M8 & -- & -- & Field & 1.000 & FLD-G & -- & -- & -- & 1.1078 & 85.143 $\pm$ 8.211 & -- \\
        J1307$+$0246 & L1 & -- & -- & Field & 1.000 & FLD-G & -- & -- & -- & 1.0784 & 65.477 $\pm$ 1.569 & -- \\
        J1313$+$1404 & M8 & -- & -- & Field & 0.997 & FLD-G & -- & -- & -- & 1.1653 & 41.188 $\pm$ 2.220 & -- \\
        J1315$+$3232 & M7 & -- & -- & Field & 1.000 & INT-G & -- & -- & -- & 0.9841 & 11.702 $\pm$ 0.038 & Young \\
        J1320$+$4238 & M7 & -- & -- & Field & 1.000 & FLD-G & -- & -- & -- & 0.9897 & 25.281 $\pm$ 0.044 & -- \\
        J1420$+$3235 & M7 & -- & -- & Field & 1.000 & FLD-G & -- & -- & -- & 1.1934 & 13.828 $\pm$ 0.096 & WB \\
        J1423$+$5346 & M7 & -- & -- & Field & 1.000 & FLD-G & -- & -- & -- & 1.2084 & 33.692 $\pm$ 0.035 & -- \\
        J1441$+$4217 & L3 & -- & -- & Cri198 & 0.957 & FLD-G & -- & -- & -- & 1.1242 & 34.567 $\pm$ 1.002 & Young \\
        J1452$+$0931 & M8 & -- & -- & Field & 1.000 & FLD-G & Young & -- & -- & 1.1047 & 52.494 $\pm$ 1.733 & -- \\
        J1514$+$3547 & L1 & -- & -- & Field & 1.000 & FLD-G & -- & -- & -- & 0.9870 & 30.576 $\pm$ 1.731 & -- \\
        J1536$+$0646 & L0 & -- & -- & Field & 1.000 & FLD-G & -- & -- & -- & 1.2198 & 24.587 $\pm$ 1.895 & Thick disc \\
        J1544$-$0435 & L1 & -- & -- & UMC & 0.988 & INT-G & Young & Young & -- & 1.0325 & 12.626 $\pm$ 1.114 & Young \\
        J1544$+$3301 & L2 & Old & Old & OCTN & 0.996 & FLD-G & -- & -- & -- & 1.0985 & 8.965 $\pm$ 0.212 & -- \\
        J1628$-$4652 & M8 & -- & -- & Field & 1.000 & FLD-G & -- & -- & -- & 1.2326 & 37.572 $\pm$ 2.468 & UB \\
        J1637$+$1813 & M7 & -- & -- & UMC & 0.993 & FLD-G & -- & -- & -- & 1.0922 & 16.132 $\pm$ 0.589 & -- \\
        J1646$-$2115 & sdL2 & -- & -- & Field & 1.000 & INT-G & Subdwarf & Subdwarf & Thick disc & 1.1187 & 72.582 $\pm$ 6.387 & Subdwarf \\
        J1654$-$3819 & M6 & -- & -- & Field & 1.000 & FLD-G & -- & -- & -- & 1.0395 & 54.161 $\pm$ 0.082 & -- \\
        J1700$-$4048 & L4 & -- & Young & Field & 0.989 & VL-G & -- & -- & Thick disc & 1.0781 & 112.745 $\pm$ 33.819 & Thick disc \\
        J1713$-$3952 & L1 & -- & -- & Field & 1.000 & FLD-G & Flat & -- & -- & 1.0167 & 28.122 $\pm$ 0.101 & -- \\
        J1737$+$4705 & M8 & -- & -- & Field & 1.000 & FLD-G & -- & -- & -- & 1.0364 & 86.863 $\pm$ 0.676 & -- \\
        J1847$-$3419 & M7 & -- & -- & Field & 0.998 & FLD-G & -- & -- & -- & 1.1031 & 5.651 $\pm$ 0.022 & -- \\
        J1938$+$4321 & M6 & -- & -- & Field & 1.000 & FLD-G & -- & -- & -- & 1.1077 & 34.835 $\pm$ 0.046 & -- \\
        J2019$+$2256 & L2 & -- & -- & CARN & 0.855 & FLD-G & -- & Young & -- & 1.0464 & 19.938 $\pm$ 0.207 & WB \\
    \end{longtable}
    \begin{tabular}{l|p{22cm}}
    \vspace{-5ex} \\
        \textbf{Notes.} & \footnotesize{The young association full names are: Cassiopeia-Taurus (Cas-Tau), Octans-Near (OCTN), Carina-Near (CARN), Rhea 449 (Rhea449), Crius 198 (Cri198), Ursa Major Corona (UMC) and Rhea 300 (Rhea300).} \\
         & \footnotesize{WB = wide binary; UB = unresolved binary.}
    \end{tabular}
\end{landscape}

\twocolumn

\section *{Acknowledgements}
GC is funded by a University of Hertfordshire studentship. RLS has been supported by a STSM grant from COST Action CA18104: MW-Gaia. GC, HRAJ, WJC, DJP recognise the computing infrastructure provided via STFC grant ST/R000905/1 at the University of Hertfordshire.

Based in part on observations made at the Infrared Telescope Facility, which is operated by the University of Hawaii under contract 80HQTR19D0030 with the National Aeronautics and Space Administration, and observations made at Cerro Tololo Inter-American Observatory at NSF’s NOIRLab (NOIRLab Prop. ID 2018A-0910; PI: J. Beamin), which is managed by the Association of Universities for Research in Astronomy (AURA) under a cooperative agreement with the National Science Foundation. This paper makes use of observations made at Infrared Telescope Facility, which is operated by the University of Hawaii under contract 80HQTR19D0030 with the National Aeronautics and Space Administration.

This paper makes use of data from the SIMBAD database, operated at CDS, Strasbourg, France (\url{http://simbad.u-strasbg.fr/Simbad}).

This work has made use of data from the European Space Agency (ESA) mission \gaia{} (\url{https://www.cosmos.esa.int/gaia}), processed by the \gaia{} Data Processing and Analysis Consortium (DPAC, \url{https://www.cosmos.esa.int/web/gaia/dpac/consortium}). Funding for the DPAC has been provided by national institutions, in particular the institutions participating in the \gaia{} Multilateral Agreement.

This research has benefited from the SpeX Prism spectral Libraries, maintained by Adam Burgasser at \url{https://cass.ucsd.edu/~ajb/browndwarfs/spexprism/index.html}. SPLAT is an collaborative project of research students in the UCSD Cool Star Lab, aimed at developing research through the building of spectral analysis tools. Contributors to SPLAT have included Christian Aganze, Jessica Birky, Daniella Bardalez Gagliuffi, Adam Burgasser (PI), Caleb Choban, Andrew Davis, Ivanna Escala, Joshua Hazlett, Carolina Herrara Hernandez, Elizabeth Moreno Hilario, Aishwarya Iyer, Yuhui Jin, Mike Lopez, Dorsa Majidi, Diego Octavio Talavera Maya, Alex Mendez, Gretel Mercado, Niana Mohammed, Johnny Parra, Maitrayee Sahi, Adrian Suarez, Melisa Tallis, Tomoki Tamiya, Chris Theissen, and Russell van Linge. This project has been supported by the National Aeronautics and Space Administration under Grant No. NNX15AI75G.

The Pan-STARRS1 Surveys (PS1) and the PS1 public science archive have been made possible through contributions by the Institute for Astronomy, the University of Hawaii, the Pan-STARRS Project Office, the Max-Planck Society and its participating institutes, the Max Planck Institute for Astronomy, Heidelberg and the Max Planck Institute for Extraterrestrial Physics, Garching, The Johns Hopkins University, Durham University, the University of Edinburgh, the Queen's University Belfast, the Harvard-Smithsonian Center for Astrophysics, the Las Cumbres Observatory Global Telescope Network Incorporated, the National Central University of Taiwan, the Space Telescope Science Institute, the National Aeronautics and Space Administration under Grant No. NNX08AR22G issued through the Planetary Science Division of the NASA Science Mission Directorate, the National Science Foundation Grant No. AST-1238877, the University of Maryland, Eotvos Lorand University (ELTE), the Los Alamos National Laboratory, and the Gordon and Betty Moore Foundation.

This publication makes use of data products from the Two Micron All Sky Survey, which is a joint project of the University of Massachusetts and the Infrared Processing and Analysis Center/California Institute of Technology, funded by the National Aeronautics and Space Administration and the National Science Foundation.

This publication makes use of data products from the Wide-field Infrared Survey Explorer, which is a joint project of the University of California, Los Angeles, and the Jet Propulsion Laboratory/California Institute of Technology, and NEOWISE, which is a project of the Jet Propulsion Laboratory/California Institute of Technology. WISE and NEOWISE are funded by the National Aeronautics and Space Administration.

We acknowledge the relevant open source packages used in our \textsc{Python} codes (\citealt{vanrossum1991}): \texttt{Astropy} (\citealt{astropy2013, astropy2018, astropy2022}), \texttt{Matplotlib} (\citealt{hunter2007}), \texttt{NumPy} (\citealt{harris2020}), \texttt{pandas} (\citealt{mckinney2010}), \texttt{rvfitter} (\citealt{cooper2022}), \texttt{SciPy} (\citealt{virtanen2020}), \texttt{specutils} (\citealt{specutils}), \splat{} (\citealt{burgasser2017}).

\section *{Data Availability}
The data underlying this article will be available in \href{https://vizier.cds.unistra.fr/viz-bin/VizieR}{CDS VizieR}, the \href{https://gucds.inaf.it}{GUCDS Data Browser}, and the \href{http://simple-bd-archive.org/}{SIMPLE Database}.

\bibliographystyle{mnras}
\bibliography{sources}
\nocite{*}

\appendix

\section{Observing Log}
\label{appendix:log}
The observation log for this work is presented in Table~\ref{table:obslog}.

\begin{table*}
    \caption{Observation log for the observations used for this work, showing details of gratings and exposure times for each observation made. The observation date and average airmass of each object are also noted.}
    \label{table:obslog}
    \centering
    \small
    \begin{tabular}{|c|c|c|c|c|c|c|c|c|}
	\hline
        Object & Observation Date & Instrument & Average & Exp Time & Telluric & Standard Exp Time & Standard & Standard \\
        Short Name & (UTC) & -- & Airmass & DIT (s) $\times$ NDIT & Standard & DIT (s) $\times$ NDIT & \textit{B} (mag) \footnotesize{$^{a}$} & \textit{V} (mag) \footnotesize{$^{a}$} \\
        \hline
        J0526$-$5026 & 2018 Apr 03 & ARCoIRIS & 1.62 & 240 $\times$ 4 & HD 36381 & 8 $\times$ 4 & 7.98 & 7.99 \\
        J0817$-$6155 & 2018 Apr 03 & ARCoIRIS & 1.18 & 45 $\times$ 4 & HD 62091 & 5 $\times$ 4 & 7.67 & 7.61 \\
        J1036$-$3441 & 2018 Apr 03 & ARCoIRIS & 1.01 & 300 $\times$ 4 & HD 92678 & 4 $\times$ 4 & 6.84 & 6.80 \\
        J1048$-$5254 & 2018 Apr 03 & ARCoIRIS & 1.16 & 90 $\times$ 4 & HD 95534 & 5 $\times$ 4 & 6.82 & 6.81 \\
        J1158$-$0008 & 2018 Apr 03 & ARCoIRIS & 1.18 & 300 $\times$ 4 & HD 97585 & 1.2 $\times$ 4 & 5.37 & 5.40 \\
        J1212$+$0206 & 2018 Apr 03 & ARCoIRIS & 1.18 & 300 $\times$ 4 & HD 97585 & 1.2 $\times$ 4 & 5.37 & 5.40 \\
        J1215$+$0042 & 2018 Apr 03 & ARCoIRIS & 1.26 & 240 $\times$ 4 & HD 111744 & 8 $\times$ 4 & 8.86 & 8.84 \\
        J1252$+$0347 & 2018 Apr 03 & ARCoIRIS & 1.21 & 300 $\times$ 4 & HD 97585 & 1.2 $\times$ 4 & 5.37 & 5.40 \\
        J1307$+$0246 & 2018 Apr 03 & ARCoIRIS & 1.01 & 180 $\times$ 4 & HD 109309 & 2 $\times$ 4 & 5.43 & 5.47 \\
        J1452$+$0931 & 2018 Apr 03 & ARCoIRIS & 1.30 & 240 $\times$ 4 & HD 126129 & 1.2 $\times$ 4 & 5.03 & 5.04 \\
        J1536$+$0646 & 2018 Apr 03 & ARCoIRIS & 1.27 & 300 $\times$ 4 & HD 140775 & 1.2 $\times$ 4 & 5.60 & 5.58 \\
        J1628$-$4652 & 2018 Apr 03 & ARCoIRIS & 1.04 & 300 $\times$ 4 & HD 146802 & 10 $\times$ 4 & 8.88 & 8.81 \\
        J1700$-$4048 & 2018 Apr 03 & ARCoIRIS & 1.03 & 300 $\times$ 4 & HD 151681 & 8 $\times$ 4 & 8.46 & 8.32 \\
        J0948$+$5300 & 2018 June 16 & SpeX SXD & 1.20 & 299.8 $\times$ 8 & HD 92245 & 19.92 $\times$ 6 & 6.02 & 6.05 \\
        J1150$-$2914 & 2018 June 16 & SpeX SXD & 1.57 & 299.8 $\times$ 8 & HD 110443 & 9.730 $\times$ 8 & 9.51 & 9.34 \\
        J1250$+$0455 & 2018 June 16 & SpeX SXD & 1.04 & 299.8 $\times$ 8 & HD 111744 & 19.92 $\times$ 8 & 8.86 & 8.81 \\
        J1441$+$4217 & 2018 June 16 & SpeX SXD & 1.09 & 299.8 $\times$ 6 & HD 111744 & 19.92 $\times$ 8 & 8.86 & 8.81 \\
        J1143$+$5324 & 2018 June 17 & SpeX Prism & 1.20 & 1.9 $\times$ 8 & HD 99966 & 1.853 $\times$ 8 & 7.33 & 7.39 \\
        J0508$+$3319 & 2020 Mar 04 & SpeX SXD & 1.06 & 119.5 $\times$ 8 & HD 35076 & 9.730 $\times$ 4 & 6.37 & 3.44 \\
        J0723$+$4622 & 2020 Mar 24 & SpeX SXD & 1.12 & 299.8 $\times$ 8 & HD 66824 & 4.634 $\times$ 4 & 6.30 & 6.33 \\
        J0811$+$1855 & 2020 Mar 24 & SpeX SXD & 1.00 & 119.5 $\times$ 8 & HD 74721 & 9.730 $\times$ 4 & 8.76 & 8.70 \\
        J0916$-$1121 & 2020 Mar 24 & SpeX SXD & 1.17 & 239.6 $\times$ 4 & HD 73687 & 4.634 $\times$ 4 & 6.66 & 6.64 \\
        J0942$-$2551 & 2020 Mar 24 & SpeX SXD & 1.43 & 299.8 $\times$ 16 & HD 98949 & 9.730 $\times$ 2 & 7.51 & 7.52 \\
        J1126$-$2706 & 2020 Mar 24 & SpeX SXD & 1.47 & 239.6 $\times$ 8 & HD 98949 & 9.730 $\times$ 2 & 7.51 & 7.52 \\
        J1152$+$5901 & 2020 Mar 24 & SpeX SXD & 1.29 & 299.8 $\times$ 12 & HD 73687 & 9.730 $\times$ 2 & 6.66 & 6.64 \\
        J1320$+$4238 & 2020 June 28 & SpeX SXD & 1.09 & 359.6 $\times$ 4 & HD 128039 & 19.92 $\times$ 4 & 9.50 & 9.28 \\
        J1423$+$5146 & 2020 June 28 & SpeX SXD & 1.18 & 299.8 $\times$ 4 & HD 128998 & 4.634 $\times$ 4 & 5.82 & 5.82 \\
        J1544$+$3301 & 2020 June 28 & SpeX SXD & 1.07 & 599.6 $\times$ 4 & HD 158261 & 4.634 $\times$ 4 & 5.93 & 5.93 \\
        J1654$-$3819 & 2020 June 28 & SpeX SXD & 1.92 & 299.8 $\times$ 4 & HD 154056 & 9.730 $\times$ 4 & 9.21 & 9.07 \\
        J1713$-$3952 & 2020 June 28 & SpeX SXD & 1.98 & 299.8 $\times$ 4 & HD 162620 & 4.634 $\times$ 4 & 11.02 & 9.71 \\
        J0808$+$3157 & 2021 Jan 02 & SpeX SXD & 1.02 & 359.6 $\times$ 4 & HD 71906 & 9.730 $\times$ 4 & 6.13 & 6.17 \\
        J0900$+$5205 & 2021 Jan 02 & SpeX SXD & 1.18 & 359.6 $\times$ 8 & HD 83869 & 9.730 $\times$ 4 & 6.36 & 6.35 \\
        J0911$+$1432 & 2021 Jan 02 & SpeX SXD & 1.02 & 359.6 $\times$ 10 & HD 80613 & 9.730 $\times$ 4 & 6.55 & 6.55 \\
        J1315$+$3232 & 2021 Jan 02 & SpeX SXD & 1.05 & 359.6 $\times$ 6 & HD 121781 & 9.730 $\times$ 4 & 9.04 & 8.94 \\
        J0515$+$0613 & 2021 Feb 01 & SpeX SXD & 1.05 & 359.6 $\times$ 8 & HD 40814 & 9.730 $\times$ 4 & 9.11 & 9.04 \\
        J0542$+$0041 & 2021 Feb 01 & SpeX SXD & 1.14 & 359.6 $\times$ 8 & HD 45357 & 9.730 $\times$ 4 & 6.71 & 6.68 \\
        J0832$+$3538 & 2021 Feb 21 & SpeX Prism & 1.04 & 119.5 $\times$ 6 & HD 82191 & 4.634 $\times$ 6 & 6.70 & 6.62 \\
        J0850$-$0318 & 2021 Feb 21 & SpeX Prism & 1.09 & 179.8 $\times$ 12 & HD 83535 & 4.634 $\times$ 6 & 7.34 & 7.21 \\
        J0941$+$3315A & 2021 Feb 21 & SpeX SXD & 1.03 & 299.8 $\times$ 6 & HD 89239 & 9.730 $\times$ 6 & 6.50 & 6.52 \\
        J1158$+$3817 & 2021 Feb 21 & SpeX SXD & 1.07 & 299.8 $\times$ 8 & HD 109615 & 9.730 $\times$ 4 & 7.25 & 7.28 \\
        J1243$+$6001 & 2021 Feb 01 & SpeX Prism & 1.57 & 179.8 $\times$ 28 & HD 116405 & 9.730 $\times$ 20 & 8.27 & 8.32 \\
        J1313$+$1404 & 2021 Feb 21 & SpeX SXD & 1.01 & 299.8 $\times$ 10 & HD 121880 & 9.730 $\times$ 4 & 7.65 & 7.58 \\
        J1420$+$3235 & 2021 Feb 21 & SpeX SXD & 1.02 & 299.8 $\times$ 4 & HD 127067 & 29.65 $\times$ 4 & 7.10 & 7.11 \\
        J1514$+$3547 & 2021 Feb 21 & SpeX SXD & 1.04 & 299.8 $\times$ 8 & HD 128039 & 19.92 $\times$ 4 & 9.50 & 9.28 \\
        J1544$-$0435 & 2021 Apr 08 & SpeX Prism & 1.10 & 119.5 $\times$ 8 & HD 148573 & 9.730 $\times$ 8 & 9.50 & 9.28 \\
        J1637$+$1813 & 2021 Apr 08 & SpeX Prism & 1.01 & 119.5 $\times$ 8 & HD 156653 & 9.730 $\times$ 8 & 6.00 & 5.99 \\
        J1646$-$2115 & 2021 Apr 08 & SpeX Prism & 1.33 & 119.5 $\times$ 8 & HD 157734 & 9.730 $\times$ 8 & 9.18 & 8.97 \\
        J1737$+$4705 & 2021 Apr 08 & SpeX SXD & 1.13 & 299.8 $\times$ 12 & HD 174366 & 9.730 $\times$ 12 & 6.76 & 6.71 \\
        J1847$-$3419 & 2021 June 03 & SpeX SXD & 1.72 & 299.8 $\times$ 8 & HD 182985 & 19.92 $\times$ 6 & 7.58 & 7.46 \\
        J1938$+$4321 & 2021 June 03 & SpeX SXD & 1.11 & 299.8 $\times$ 6 & HD 193594 & 19.92 $\times$ 6 & 7.80 & 7.78 \\
        J2019$+$2256 & 2021 June 03 & SpeX Prism & 1.04 & 149.7 $\times$ 6 & HD 201671 & 0.463 $\times$ 8 & 6.67 & 6.65 \\
	\hline
    \end{tabular}
    \begin{tabular}{p{\textwidth}}
         \footnotesize{$^{a}$} -- \textit{B} and \textit{V} magnitudes are taken from the Tycho-2 catalogue of \cite{hog2000}. 
    \end{tabular}
\end{table*}

\section{Non-UCD Objects}
\label{appendix:nonucd}
On several of the observing nights, conditions were too poor to be able to make observations of the UCD targets from the list of UCDs. In these cases, alternative objects were observed, so as to not waste observing time. The same steps were taken to analyse the spectra of these objects that as for the UCDs, albeit in not as much depth. As would be expected, all of the non-UCDs that were observed can be found in \gaia{}~DR3, and most of them can be found in SIMBAD with published classifications (shown in Table~\ref{table:nonucd}). Additionally, some of the UCDs were determined to have spectral classifications earlier than M7 so are presented in this appendix.

\subsection{Data Collection and Reduction}
\label{appendix:nonucd.data}
In addition to the UCDs presented in the main body of this work, we also observed some objects which are not classified as UCDs. Some of these objects were in the GUCDS candidate list, with their new spectroscopic classifications narrowly missing the M7 classification required to be considered a UCD, while their photometric classifications from literature may suggest that they are UCDs. Others of these non-UCDs are giant stars, which were observed when observing conditions were too poor to allow for UCD observations (e.g. too much cloud). Most of the observations of non-UCD targets were made using the prism configuration of the IRTF SpeX instrument, and only one of the non-UCDs was observed using the ARCoIRIS instrument at Blanco in Chile. Details of the observations made of the non-UCDs can be seen in Table~\ref{table:nonucd-obs}. As with the UCDs, the non-UCDs spectra were reduced with \textsc{Spextool}.

\subsection{Spectral Classification}
\label{appendix:nonucd.spt}
\subsubsection{Spectroscopic Classification using \splat}
\label{appendix:nonucd.spt.splat}
\splat{} can only classify the spectra of UCDs, so any non-UCDs cannot be classified with high confidence. Classification using \splat{} is carried out for sake of consistency and to give an indication of the strength of using \splat{} to classify objects outside the realm of UCDs. Again, both the Kirkpatrick method (\citealt{kirkpatrick2010}) and the whole spectrum are used for classification (Table~\ref{table:nonucd}).

\subsubsection{Photometric Classification}
\label{appendix:nonucd.spt.phot}
As with the UCDs, the photometric classification method outlined by \citealt{skrzypek2015} can be used to determine photometric spectral types. Making use of a template derived using the same photometric bands as for the UCDs (but instead for K and M-type non-UCDs, rather than UCDs), the non-UCDs can be classified photometrically. Table~\ref{table:nonucd} shows the photometric spectral classifications for the non-UCDs. Since \splat{} can only accurately classify UCDs, the photometric classifications are adopted for the non-UCDs.

\subsubsection{Comparison of Spectroscopic and Photometric Spectral Types}
\label{appendix:nonucd.spt.comp}
Comparing the spectroscopic and photometric spectral types can give an indication of the strength of \splat{} for classifying non-UCDs. Overall, we can see that the non-UCD spectral types given by \splat{} tend to agree for within $\pm$\,2.0 subtypes the objects with M-type photometric classifications; however, since M0 is the earliest spectral type that \splat{} can classify, the non-UCDs with K-type photometric classifications have vastly different spectroscopic classifications given by \splat. 

\begin{table*}
    \caption{Basic data for each of the non-UCD objects in this paper. Spectroscopic classifications (SpT) of the objects are as given by SPLAT, and photometric spectral types (PhT) are also shown. Where applicable, published classifications for each object are also shown. Objects marked with a dagger $\dagger$ are those observed with SpeX Prism, those marked with a double dagger $\ddagger$ are observed with SpeX SXD, and those marked with an asterisk * are observed with ARCoIRIS.}
    \label{table:nonucd}
    \centering
    \small
    \begin{tabular}{|c|c|c|c|c|c|c|}
	\hline
	Object & SIMBAD & SpT & PhT & Published & SIMBAD & \teff \\
        Short Name & Name & (this work) & (this work) & NIR Type & Spectral Type & (K) \\
	\hline
        J0547$+$0817 & \gaia{}~DR3 3335240017439310464 & M0 & M0 & -- & -- & 3400 $\pm$ 270 \\
        J0631$+$4129 & LP 205$-$44 & M5 & M5 & M6 V $^{[1]}$ & M5 $^{[2]}$ & 3500 $\pm$ 230 \\
        J0651$+$2712 & PM J06518$+$2712 & M0 & K7 & -- & K7e D $^{[3]}$ & 3500 $\pm$ 260 \\
        J0723$+$2024 & BD$+$20 1790 & M0 & K4 & -- & K5e D $^{[4]}$ & 3500 $\pm$ 135 \\
        J0752$+$1244 & PM J07529$+$1244 & M1 & M0 & -- & M0 $^{[5]}$ & 3500 $\pm$ 315 \\
        J0941$+$3315B & LP 260$-$43 & M5 & M5 & -- & M5 $^{[6]}$ & 3600 $\pm$ 290 \\
        J1022$+$4129 & * mu. UMa & M2 & M0 & -- & M0III B $^{[7]}$ & 3500 $\pm$ 110 \\
        J1103$+$3558 & HD 95735 & M3 & M3 & -- & M2$+$V B $^{[7,a]}$ & 3500 $\pm$ 95 \\
        J1128$+$4933 & V* HP UMa & M7 & M5 & -- & M6 D $^{[8]}$ & 1700 $\pm$ 250 \\
        J1154$+$3708 & BD$+$37 2228 & M7 & M4 & -- & M4:III D $^{[9]}$ & 3500 $\pm$ 125 \\
        J1215$+$3914 & LP 216$-$82 & M3 & M5 & -- & -- & 1500 $\pm$ 30 \\
        J1238$+$0659 & V* R Vir & M4 & M5 & -- & M3.5--7e B $^{[10]}$ & 3500 $\pm$ 190 \\
        J1345$+$1453 & HD 119850 & M0 & M0 & -- & M2V B $^{[7]}$ & 3500 $\pm$ 145 \\
        J1442$+$6603A & G 239$-$25 & M1 & M3 & -- & M3V C $^{[8]}$ & 3500 $\pm$ 260 \\
        J1516$+$2756 & \gaia{}~DR2 1271645228286891008 & M0 & K4 & -- & -- & 3500 $\pm$ 25 \\
        J1603$-$5046 & \gaia{}~DR3 5981801923255812608 & M7 & M6 & -- & -- & 1500 $\pm$ 20 \\
	\hline
    \end{tabular}
    \begin{tabular}{l l l l l}
        \textbf{References.} & \footnotesize{$^{[1]}$ -- \cite{newton2014}} & \footnotesize{$^{[2]}$ -- \cite{reid1995}} & \footnotesize{$^{[3]}$ -- \cite{bowler2019}} & \footnotesize{$^{[4]}$ -- \cite{reid2004}} \\
        
        & \footnotesize{$^{[5]}$ -- \cite{lepine2011}} & \footnotesize{$^{[6]}$ -- \cite{mason2001}} & \footnotesize{$^{[7]}$ -- \cite{keenan1989}} & \footnotesize{$^{[8]}$ -- \cite{stephenson1986}} \\
        
        & \footnotesize{$^{[9]}$} -- \cite{skiff2008} & \footnotesize{$^{[10]}$ -- \cite{keenan1974}} \\
    \end{tabular}
    \begin{tabular}{l l}
        \textbf{Notes.} & \footnotesize{$^{[a]}$ -- The $+$ symbol is used by \cite{keenan1989} to indicate a quarter sub-type (i.e. M2$+$ is indicative of a spectral type of M2.25).} \\
    \end{tabular}
\end{table*}

\begin{table*}
    \vspace{3ex}
    \captionof{table}{Observation log for the non-UCD observations, showing details of gratings and exposure times for each observation made. The observation date and average airmass of each object are also noted.}
    \label{table:nonucd-obs}
    \centering
    \small
    \begin{tabular}{|c|c|c|c|c|c|c|}
	\hline
	Object & Observation Date & Instrument & Average & Exp Time & Telluric & Standard Exp Time \\
	Short Name & (UTC) & & Airmass & DIT (s) $\times$ NDIT & Standard & DIT (s) $\times$ NDIT \\
	\hline
        J1603$-$5046 & 2018 Apr 03 & ARCoIRIS & 1.07 & 300 $\times$ 4 & HD 146802 & 10 $\times$ 4 \\
	J1442$+$6603A & 2018 June 16 & SpeX SXD & 1.63 & 59.8 $\times$ 4 & HD 128039 & 59.77 $\times$ 6 \\
        J1128$+$4933 & 2018 June 17 & SpeX SXD & 1.22 & 4.634 $\times$ 28 & HD 100417 & 19.92 $\times$ 8 \\
        J1154$+$3708 & 2018 June 17 & SpeX SXD & 1.08 & 4.634 $\times$ 10 & HD 100417 & 19.92 $\times$ 8 \\
	J1238$+$0659 & 2018 June 17 & SpeX SXD & 1.07 & 299.8 $\times$ 8 & HD 100417 & 19.92 $\times$ 8 \\
	J0547$+$0817 & 2021 Feb 01 & SpeX SXD & 1.28 & 359.6 $\times$ 10 & HD 46710 & 9.730 $\times$ 4 \\
	J0631$+$4129 & 2021 Feb 21 & SpeX Prism & 1.08 & 9.730 $\times$ 6 & HD 45105 & 9.730 $\times$ 6 \\
	J0651$+$2712 & 2021 Feb 21 & SpeX Prism & 1.01 & 9.730 $\times$ 8 & HD 46553 & 4.634 $\times$ 6 \\
	J0723$+$2024 & 2021 Feb 21 & SpeX Prism & 1.01 & 9.730 $\times$ 6 & HD 62510 & 9.730 $\times$ 8 \\
	J0752$+$1244 & 2021 Feb 21 & SpeX Prism & 1.01 & 9.730 $\times$ 8 & HD 67959 & 4.634 $\times$ 6 \\
	J0941$+$3315B & 2021 Feb 21 & SpeX SXD & 1.03 & 299.8 $\times$ 6 & HD 89239 & 9.730 $\times$ 6 \\
	J1215$+$3914 & 2021 Feb 21 & SpeX SXD & 1.06 & 299.8 $\times$ 4 & HD 116246 & 9.730 $\times$ 8 \\
	J1516$+$2756 & 2021 Feb 21 & SpeX SXD & 1.02 & 299.8 $\times$ 2 & HD 127067 & 29.65 $\times$ 4 \\
	J1103$+$3558 & 2021 May 06 & SpeX Prism & 1.14 & 29.7 $\times$ 4 & HD 103287 & 29.65 $\times$ 4 \\
	J1022$+$4129 & 2021 June 06 & SpeX Prism & 1.08 & 119.5 $\times$ 4 & HD 103287 & 29.65 $\times$ 4 \\
	J1345$+$1453 & 2021 June 06 & SpeX Prism & 1.01 & 59.8 $\times$ 12 & HD 130109 & 29.65 $\times$ 8 \\
	\hline
    \end{tabular}
\end{table*}

\begin{figure*}
    \includegraphics[width=\linewidth]{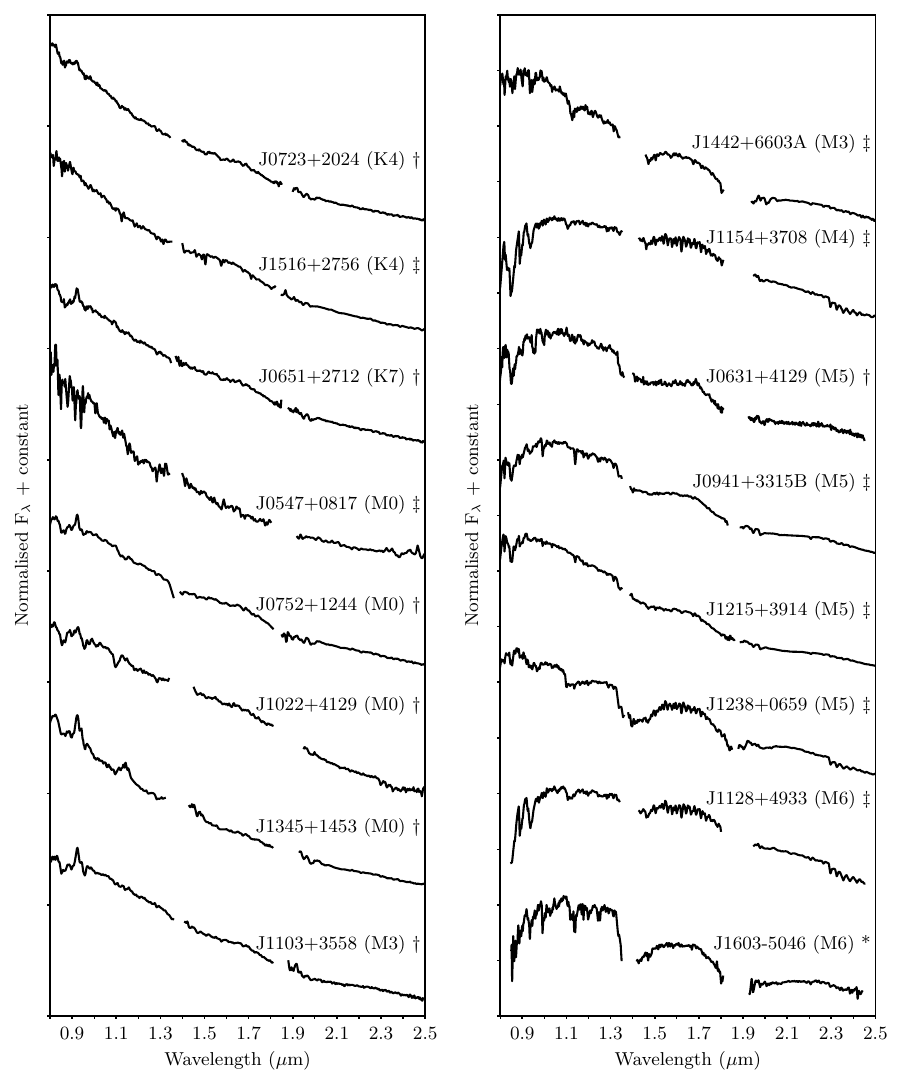}
    \caption{Stacked spectral plot of the non-UCDs observed, sorted by spectral type -- left: K3.5--K4.5, right: K8--M6.5.  Each spectrum is normalised at 1.27\,\um{} and vertically offset by equal flux increments. Noisy areas around the H$_2$O bands at ${\sim}$1.3\,\um{} and ${\sim}$1.9\,\um{} have been removed to make plots clearer. Objects marked with a dagger $\dagger$ are those observed with SpeX Prism, those marked with a double dagger $\ddagger$ are observed with SpeX SXD, and those marked with an asterisk * are observed with ARCoIRIS.}
    \label{figure:nonucd-spec}
\end{figure*}

\subsection{Spectral Analysis}
\label{appendix:nonucd.spec}
\subsubsection{Spectral Plots}
\label{appendix:nonucd.spec.plot}
Fig.~\ref{figure:nonucd-spec} shows the stacked spectral plots for the non-UCDs. Since these observations were made when conditions were too poor for the UCDs of interest to be observed, most of the objects were observed with the SpeX Prism configuration. Objects marked with a dagger $\dagger$ in Fig.~\ref{figure:nonucd-spec} are those observed with SpeX Prism, those marked with a double dagger $\ddagger$ are observed with SpeX SXD, and those marked with an asterisk * are observed with ARCoIRIS. As would be expected with observations of brighter objects, these spectra have better signal-to-noise than those of the UCDs even though the observing conditions were poor, resulting in clearer spectral plots. The shapes of the spectra follow the expected progression of spectral types K through M, and any spectra with features which appear unusual (such as those seen in J1103$+$3558) are due to differences in resolution of the instruments used to obtain the spectra.

\subsection{Colour-Magnitude Diagrams}
\label{appendix:nonucd.cmd}
\begin{figure}
    \centering
    \includegraphics[width=\linewidth]{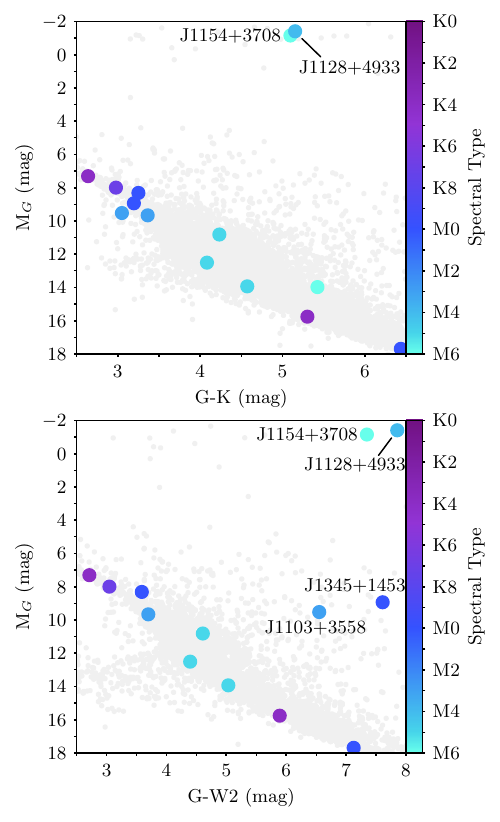}
    \caption{Top: CMD of M$_G$ versus \textit{G}$-$\textit{Ks} for the objects. Bottom: CMD of M$_G$ versus \textit{G}$-$\textit{W2} for the objects. The colour corresponds to the spectral type of the object: purple objects are earlier types and blue objects are later types. Median uncertainties are too small to plot. Grey points are objects from the GUCDS Master list, showing the expected distribution of UCDs.}
    \label{figure:nonucd-cmd}
\end{figure}

As is the case for the UCDs, we can plot CMDs for the non-UCDs (Fig.~\ref{figure:nonucd-cmd}). It is clear from these plots that none of the non-UCDs have unusual colours relative to their brightness, although there is no clear relation between spectral type and position on the CMD. Two of the non-UCDs (J1128$+$4933 and J1154$+$3708) are obviously giant stars, lying far above the main sequence on the giant branch of the CMD.

\section{Creating the Template for Photometric Classification}
\label{appendix:template}
We created a template of 2MASS \textit{J}, \textit{H} and \textit{Ks}, and CatWISE2020 \textit{W1} and \textit{W2} magnitudes using the lists of UCDs from \cite{gagne2014a} and \cite{gagne2014b}. Plotting the magnitudes for each of the UCDs in these lists against their spectral types and fitting a polynomial to each magnitude-spectral type relation allows use to create our magnitude template (Table~\ref{table:template}). The $\chi^2$ values for each object are in the same way as outlined by \cite{skrzypek2015}, and the spectral type corresponding to the smallest $\chi^2$ value is determined to be the best-fitting photometric spectral type.

\begin{table}
    \centering
    \caption{Template magnitudes used for photometric spectral classification of the UCDs.}
    \label{table:template}
    \begin{tabular}{c|c|c|c|c|c}
        \hline
        Spectral & \textit{J} & \textit{H} & \textit{Ks} & \textit{W1} & \textit{W2} \\
        Type & (mag) & (mag) & (mag) & (mag) & (mag) \\
        \hline
        M6 & 15.74 & 14.96 & 14.48 & 14.16 & 13.90 \\
        M7 & 15.76 & 14.96 & 14.47 & 14.13 & 13.87 \\
        M8 & 15.77 & 14.97 & 14.46 & 14.10 & 13.84 \\
        M9 & 15.80 & 14.98 & 14.46 & 14.08 & 13.81 \\
        L0 & 15.82 & 14.99 & 14.46 & 14.07 & 13.80 \\
        L1 & 15.85 & 15.01 & 14.48 & 14.07 & 13.78 \\
        L2 & 15.88 & 15.04 & 14.49 & 14.08 & 13.78 \\
        L3 & 15.91 & 15.07 & 14.51 & 14.10 & 13.78 \\
        L4 & 15.95 & 15.10 & 14.54 & 14.12 & 13.78 \\
        L5 & 15.99 & 15.14 & 14.57 & 14.16 & 13.79 \\
        L6 & 16.03 & 15.19 & 14.61 & 14.20 & 13.80 \\
        L7 & 16.07 & 15.24 & 14.66 & 14.25 & 13.83 \\
        L8 & 16.12 & 15.29 & 14.71 & 14.31 & 13.85 \\
        L9 & 16.17 & 15.35 & 14.76 & 14.39 & 13.88 \\
        T0 & 16.22 & 15.42 & 14.82 & 14.47 & 13.92 \\
        T1 & 16.27 & 15.49 & 14.89 & 14.55 & 13.96 \\
        T2 & 16.33 & 15.56 & 14.96 & 14.65 & 14.01 \\
        T3 & 16.39 & 15.64 & 15.04 & 14.76 & 14.06 \\
        T4 & 16.46 & 15.72 & 15.12 & 14.87 & 14.12 \\
        T5 & 16.52 & 15.81 & 15.21 & 15.00 & 14.18 \\
        T6 & 16.59 & 15.91 & 15.31 & 15.13 & 14.25 \\
        T7 & 16.66 & 16.01 & 15.41 & 15.27 & 14.33 \\
        T8 & 16.74 & 16.11 & 15.51 & 15.42 & 14.41 \\
        T9 & 16.81 & 16.22 & 15.62 & 15.58 & 14.49 \\
        \hline
    \end{tabular}
\end{table}

\section{Full Sample Data}
\label{appendix:fulldata}
Table~\ref{table:fulldata} shows the collated data for the full sample of UCDs presented in this work. A machine-readable version of this table is available in its entirety from CDS.

\begin{table*}
\centering
\caption{Collated data for all UCDs presented in this work with the first object as an example.}
\label{table:fulldata}
\begin{tabular}{llll}
    \hline
    Parameter & Unit & Comment & Example \\
    \hline
    short\_name & -- & Object Short Name & J0508$+$3319 \\
    source\_id & -- & \gaia~DR3 Source ID & 181724125038647040 \\
    ra & deg & Right ascension (ICRS, epoch 2016.0) & 77.228 \\
    ra\_error & mas & Uncertainty & -- \\
    dec & deg & Declination (ICRS, epoch 2016.0) & 33.32 \\
    dec\_error & mas & Uncertainty & -- \\
    simbad & -- & SIMBAD object name & 2MASS J05085506$+$3319272 \\
    parallax & mas & \gaia~DR3 parallax & 53.026 \\
    parallax\_error & mas & Uncertainty & 0.46 \\
    distance & pc & Parallactic distance from \gaia~DR3 parallax & 18.9 \\
    distance\_error & pc & Uncertainty & 0.2 \\
    spec\_distance & pc & Spectrophotometric distance & 22.0 \\
    spec\_distance\_error & pc & Uncertainty & 1.4 \\
    pmra & mas\,yr$^{-1}$ & \gaia~DR3 proper motion in RA & -215.533 \\
    pmra\_error & mas\,yr$^{-1}$ & Uncertainty & 0.634594 \\
    pmdec & mas\,yr$^{-1}$ & \gaia~DR3 proper motion in dec & -617.068 \\
    pmdec\_error & mas\,yr$^{-1}$ & Uncertainty & 0.427902 \\
    spt & -- & Spectroscopic spectral classification & L2 \\
    spt\_error & -- & Uncertainty & 0.5 \\
    spt\_published & -- & Spectroscopic spectral classification from literature & L2 \\
    spt\_refname & -- & ADS bibcode for SpT & 2016ApJS..224...36K \\
    pht & -- & Photometric spectral classification & L1 \\
    pht\_error & -- & Uncertainty & 2.5 \\
    pht\_published & -- & Photometric spectral classification from literature & -- \\
    pht\_refname & -- & ADS bibcode for PhT & -- \\
    vtan & \kms & Tangential velocity & 58.4276 \\
    vtan\_error & \kms & Uncertainty & 0.5122 \\
    rv & \kms & Radial velocity & 53.0 \\
    rv\_error & \kms & Uncertainty & 13.6 \\
    rv\_published & \kms & Radial velocity from literature & -- \\
    rv\_refname & -- & ADS bibcode for radial velocity & -- \\
    u & \kms & Space velocity U-component & -81.2 \\
    v & \kms & Space velocity V-component & -21.8 \\
    w & \kms & Space velocity W-component & -53.9 \\
    teff & K & Effective temperature & 2000 \\
    teff\_error & K & Uncertainty & 60 \\
    ruwe & -- & \gaia~DR3 re-normalised unit weight error value & 1.0803 \\
    companion & -- & \gaia~DR3 source ID of wide companion object & -- \\
    banyan & -- & \banyan{} young moving group & -- \\
    banyan\_prob & -- & Probability of group membership & -- \\
    gravity & -- & Surface gravity classification & FLD-G \\
    notes & -- & Young, thick disc, binary & Thick disc \\
    jmag\_2mass & mag & 2MASS J-band magnitude & 14.217 \\
    jmag\_2mass\_error & mag & Uncertainty & 0.032 \\
    hmag\_2mass & mag & 2MASS H-band magnitude & 13.242 \\
    hmag\_2mass\_error & mag & Uncertainty & 0.037 \\
    kmag\_2mass & mag & 2MASS K-band magnitude & 12.623 \\
    kmag\_2mass\_error & mag & Uncertainty & 0.028 \\
    w1mag\_wise & mag & CatWISE2020 W1-band magnitude & 12.133 \\
    w1mag\_wise\_error & mag & Uncertainty & 0.015 \\
    w2mag\_wise & mag & CatWISE2020 W2-band magnitude & 11.899 \\
    w2mag\_wise\_error & mag & Uncertainty & 0.014 \\
    gmag\_gaia & mag & \gaia~DR3 G-band magnitude & 19.141 \\
    gmag\_gaia\_error & mag & Uncertainty & 0.003 \\
    mg & mag & \gaia~DR3 G-band absolute magnitude & 17.763 \\
    mg\_error & mag & Uncertainty & 0.019 \\
    \hline
\end{tabular}
\end{table*}

\bsp	

\label{lastpage}
\end{document}